\documentclass[twocolumn]{aastex63}

\usepackage{amsmath,amssymb,natbib}
\usepackage{microtype}
\usepackage{hyperref}
\usepackage{graphicx}
\newcommand{\Mbar}{$\mathrm{\overline{M}}$}

\newcommand{\code}[1]{\texttt{#1}}

\newcommand{\johnny}[1]{#1}

\newcommand{\resp}[1]{#1}

\newcommand{\respp}[1]{#1}

\definecolor{citecolor}{rgb}{0,0.1,0.43}
\definecolor{linkcolor}{rgb}{0,0.1,0.43}

\hypersetup{breaklinks,colorlinks,urlcolor=citecolor,linkcolor=linkcolor,citecolor=citecolor}
\graphicspath{{figures/}}
 
\shorttitle{sbf distances to low-luminosity galaxies}
\shortauthors{Greco et al.}

\begin{document}\sloppy\sloppypar\raggedbottom\frenchspacing 

\title{Measuring distances to low-luminosity galaxies using surface brightness fluctuations}
\author[0000-0003-4970-2874]{Johnny P. Greco}

\altaffiliation{NSF Astronomy \& Astrophysics Postdoctoral Fellow}
\affiliation{Center for Cosmology and AstroParticle Physics (CCAPP), The Ohio State University, Columbus, OH 43210, USA}

\author[0000-0002-8282-9888]{Pieter van Dokkum}
\affiliation{Yale Center for Astronomy and Astrophysics, Yale University, New Haven, CT 06511, USA}

\author[0000-0002-1841-2252]{Shany Danieli}
\affiliation{Department of Physics, Yale University, New Haven, CT 06520, USA}
\affiliation{Yale Center for Astronomy and Astrophysics, Yale University, New Haven, CT 06511, USA}
\affiliation{Department of Astronomy, Yale University, New Haven, CT 06511, USA}

\author[0000-0002-5382-2898]{Scott G. Carlsten}
\affil{Department of Astrophysical Sciences, 4 Ivy Lane, Princeton University, Princeton, NJ 08544}

\author[0000-0002-1590-8551]{Charlie Conroy}
\affiliation{Harvard-Smithsonian Center for Astrophysics, 60 Garden Street, Cambridge, MA, USA}

\correspondingauthor{Johnny Greco}
\email{jgreco.astro@gmail.com}

\begin{abstract}
We present an in-depth study of surface brightness fluctuations (SBFs) in low-luminosity stellar systems. Using the MIST models, we compute theoretical predictions for absolute SBF magnitudes in the LSST, HST \resp{ACS/WFC}, and proposed \resp{Roman Space Telescope} filter systems. We compare our calculations to observed SBF\resp{-color relations of systems that span a wide range of age and metallicity}. Consistent with previous studies, we find that single-age population models show excellent agreement with \resp{observations} of low-mass galaxies with $0.5 \lesssim  g - i \lesssim  0.9$. For bluer galaxies, the observed relation is better fit by models with composite stellar populations. To study SBF recovery from low-luminosity systems, we perform detailed image simulations in which we inject fully populated model galaxies into deep ground-based images from real observations. \resp{Our simulations show that LSST} will provide data of sufficient quality and depth to measure SBF \resp{magnitudes} with precisions of \resp{${\sim}0.2$-0.5~mag in} ultra-faint $\left(\mathrm{10^4 \leq M_\star/M_\odot \leq 10^5}\right)$ and low-mass classical (M$_\star\leq10^7$~M$_\odot$) dwarf galaxies out to ${\sim}4$~Mpc and ${\sim}25$~Mpc, respectively, within the first few years of its deep-wide-fast survey. Many \resp{significant practical challenges and systematic uncertainties} remain, including an irreducible ``sampling scatter'' in the SBFs of ultra-faint dwarfs due to their undersampled stellar mass functions. We nonetheless conclude that SBFs in the new generation of wide-field imaging surveys have the potential to play a critical role in the efficient confirmation and characterization of dwarf galaxies in the nearby universe. 
\end{abstract}


\section{Introduction} \label{sec:intro}

Current and future generations of imaging surveys---which are simultaneously wide, deep, and sharp---will uncover thousands of diffuse dwarf galaxy candidates beyond the Local Group \citep[e.g.,][]{Muller-2017-lsb-Centaurus, Bennet-2017, Greco2018cat, Zaritsky2019, Carlsten2019-survey, Prole2019}. Reliable distances will be required to confirm the nature of these candidates and to study their numbers and physical properties as a function of environment. Unfortunately, measuring precise distances to low surface brightness (LSB) objects is extremely difficult and time consuming, especially for quenched spheroidal systems that lack emission lines for measuring redshifts \citep[e.g.,][]{van-Dokkum:2015ab, Kadowaki:2017aa, Gu-2018}.

\begin{figure*}[ht!]
    \centering
    \vspace{0.2cm}
    \includegraphics[width=\textwidth]{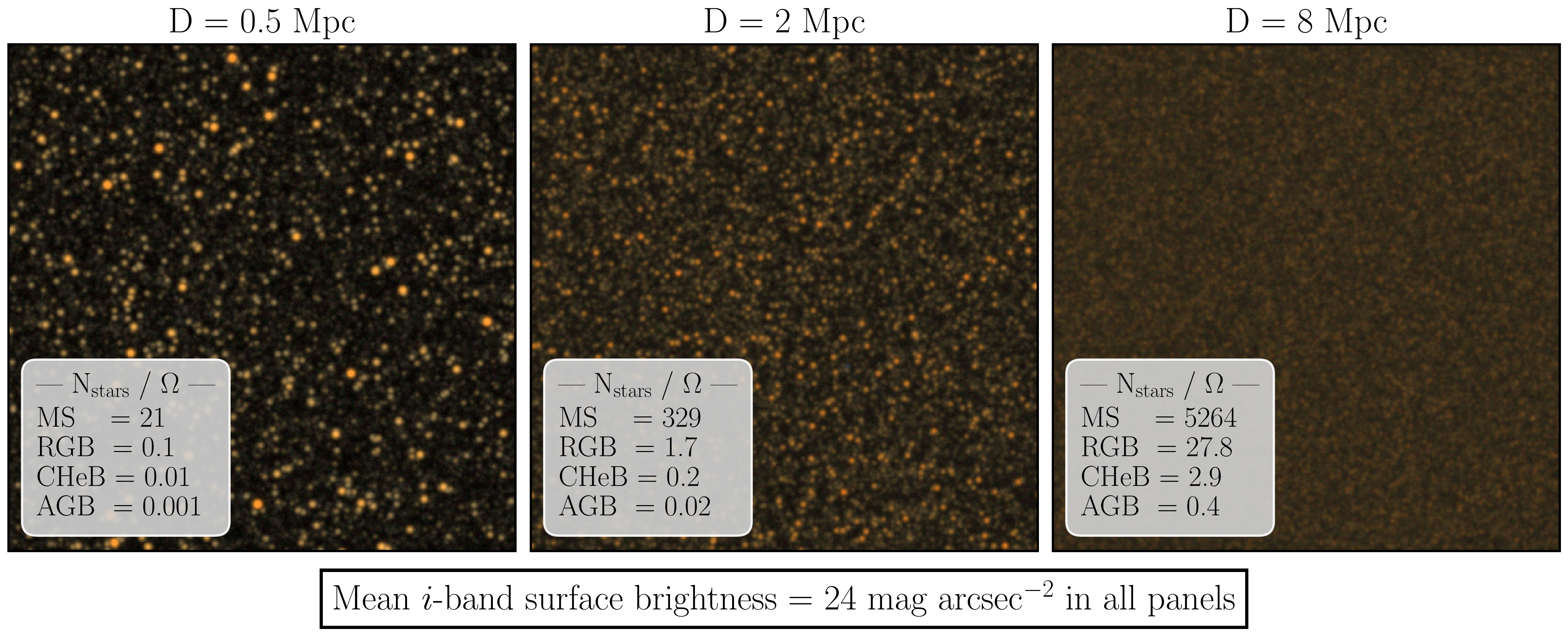}
    \caption{Illustration of the decreasing amplitude of SBFs as a function of distance, at fixed surface brightness. Each panel shows a $gri$-composite image of a simple stellar population of \resp{age 3~Gyr, metallicity $\mathrm{[Fe/H] = -1.5}$,} and mean $i$-band surface brightness 24~mag arcsec$^{-2}$. The simulations assume a ground-based observatory with 0\farcs6 seeing and a pixel scale of 0\farcs2 pixel$^{-1}$. The text box in each panel lists the number of stars per resolution element ($\Omega$) for stars on the main sequence (MS), red giant branch (RGB), core helium burning (CHeB) phase, and asymptotic giant branch (AGB). The images are 500 pixels on a side.}
    \vspace{0.2cm}
    \label{fig:intro}
\end{figure*}

Even if redshift measurements are feasible, there remains an enormous LSB discovery space within the Local Volume \citep[${\lesssim}10$~Mpc;][]{Danieli:2018aa}, where redshift-based distances are highly uncertain due to the local peculiar velocity field. It is, therefore, important and timely to study velocity-independent distance indicators in the ultra-LSB dwarf galaxy regime \citep[e.g.,][]{vD2018-DF2-distance, Trujillo:2019, Danieli2019-DF4}. After trigonometric parallax distances, robustly calibrated standard candles (and rulers) are the gold standard for measuring metric distances \citep{Jacoby-1992-distances, Beaton2018-old-pops}, the most important of which have been variable stars along the instability strip \citep{Leavitt-Law-1912, Freedman2001} and the tip of the red giant branch \citep{Lee-1993-TRGB, Madore-Freedman-1995}. While these methods have been essential for establishing the cosmic distance ladder, they require the detection of individual stars, which significantly limits the range of distances that they can probe from the ground. 

Surface brightness fluctuations (SBFs) provide a method for measuring distances to semi-resolved galaxies using imaging data alone, making it one of the most promising tools for confirming and studying dwarf galaxies with current and future wide-field imaging surveys. \citet{Tonry:1988} first established the theoretical framework for SBFs and demonstrated the feasibility of determining extragalactic distances through careful analysis of ground-based imaging. 

In essence, the SBF method takes advantage of the discrete nature of stellar systems. The stochasticity in the number of stars within each resolution element of a galaxy image leads to pixel-to-pixel brightness fluctuations, which have a {\it variance} that decreases as the square of the galaxy's distance. Figure~\ref{fig:intro} illustrates the appearance of SBFs as a function of distance for a simple stellar population of intermediate age and mean surface brightness 24~mag arcsec$^{-2}$. The simulations assume a ground-based observatory with 0\farcs6 seeing. The text box in each panel lists the number of stars per resolution element for various phases of stellar evolution. At 0.5~Mpc, individual giant stars are visible, with 0.11 red giant branch (RGB) stars per resolution element. In contrast, the same population at 8~Mpc has 27.8~RGB stars per resolution element, reducing the pixel-to-pixel variance in surface brightness by a factor of ${\sim}$250.

In the decades that followed the seminal work of \citet{Tonry:1988}, enormous effort has been devoted to developing the SBF method and applying it to early-type systems and spiral bulges from the ground \citep[e.g.,][]{Tonry:1990, Tonry:1997, Tonry:2001aa} and space \citep[e.g.,][]{Cantiello:2007aa, Blakeslee:2009, Blakeslee:2010aa, Jensen:2015, Cantiello-GW-source:2018}. There has also been considerable work on SBFs in early-type dwarfs \citep[e.g.,][]{Jerjen:1998, Jerjen:2000a, Jerjen:2001, Mieske:2003, Cantiello:2018aa}. More recently, SBF distance measurements have been obtained for ultra-LSB dwarfs using both space- and ground-based imaging \citep{Cohen:2018, vD2018-DF2-distance, Blakeslee-DF2:2018, Carlsten:2019aa}. 

\resp{A challenging but also potentially powerful feature of this method is that} the amplitude of luminosity fluctuations in a galaxy depends sensitively on its underlying stellar population. This effect is generally accounted for using the correlation between the SBF signal and integrated color \citep{Tonry-1989}---the so-called fluctuation-color relation. Decades of effort have been devoted to calibrating this relation \resp{using both stellar population synthesis \citep[e.g.,][]{Worthey:1993, Buzzoni:1993, Liu:2000, Cantiello:2003aa} and empirical methods \citep[e.g.,][]{Tonry:1997, Blakeslee:2009, Cantiello:2018aa}.} This dependence on stellar population parameters can also be seen as a strength of the SBF method, as it provides additional stellar population information that complements integrated light measurements \citep[e.g.,][]{Cantiello:2007aa, Lee-Worthey-Blakeslee-2010, Conroy-2016-pcmd, Cook:2019}.  

Although SBFs have most often been measured in early-type systems in dense environments, much bluer systems are also observed to follow a fluctuation-color relation \citep[e.g.,][]{Jerjen:2001, Rekola:2005}. The challenge with low-mass blue systems is that they are generally young and actively forming stars, and since young stars tend to cluster, such systems may break the fundamental assumption of the SBF method that the stars are Poisson distributed. Nevertheless, if regions in a blue galaxy can be isolated that are apparently free of star-forming lumps, it is still possible to use SBFs to constrain its distance and stellar population. Indeed, \citet{Carlsten:2019aa} used archival Canada-France-Hawaii Telescope (CFHT) imaging of dwarf galaxies with tip of the RGB distances to extend the fluctuation-color relation to dwarf galaxies with colors as blue as $g-i\sim0.3$. 

With new and upcoming ground-based imaging surveys such as the Dark Energy Survey \citep{DES:2016aa}, the Hyper Suprime-Cam Subaru Strategic Program \citep{Aihara:2018aa}, and ultimately the Legacy Survey of Space and Time \citep[LSST;][]{LSST-ref-design-2019} with the Rubin Observatory, we are entering a new era for ground-based SBF measurements. From space, \resp{the Roman Space Telescope \citep[RST;][]{Spergel:2015}} will provide HST-quality imaging in the infrared over large areas of the sky, significantly increasing the number of systems for which it will be possible to measure accurate distances and study stellar populations in both the resolved and semi-resolved regimes. 

In this work, we study SBFs \resp{with stellar population modeling and numerical simulations of astronomical images}, with a particular emphasis on low-luminosity stellar systems. The paper is organized as follows. In Section~\ref{sec:models}, we describe our procedure for generating synthetic stellar populations and modeling their luminosity fluctuations, and we present new theoretical predictions for the fluctuation-color relation for simple stellar populations (SSPs) in the LSST, HST, and RST filter systems. In Section~\ref{sec:compare-with-obs}, we compare our models with observations of dwarf galaxies that have independently-measured distances. In Section~\ref{sec:sims}, we describe our image simulations, which we use in Section~\ref{sec:sbf-in-pratice} to study the feasibility of measuring SBFs in low-luminosity galaxies with LSST-like data. \johnny{In Section~\ref{sec:D_lim}, we present the limiting SBF distance, as a function of galaxy stellar mass, that can be reached using images that are comparable to what is expected for 2-year LSST stacks.} Finally, we conclude with a brief summary in Section~\ref{sec:summary}.

\section{Modeling Luminosity Fluctuations} \label{sec:models}

\begin{figure*}[ht!]
    \centering
    \includegraphics[width=\textwidth]{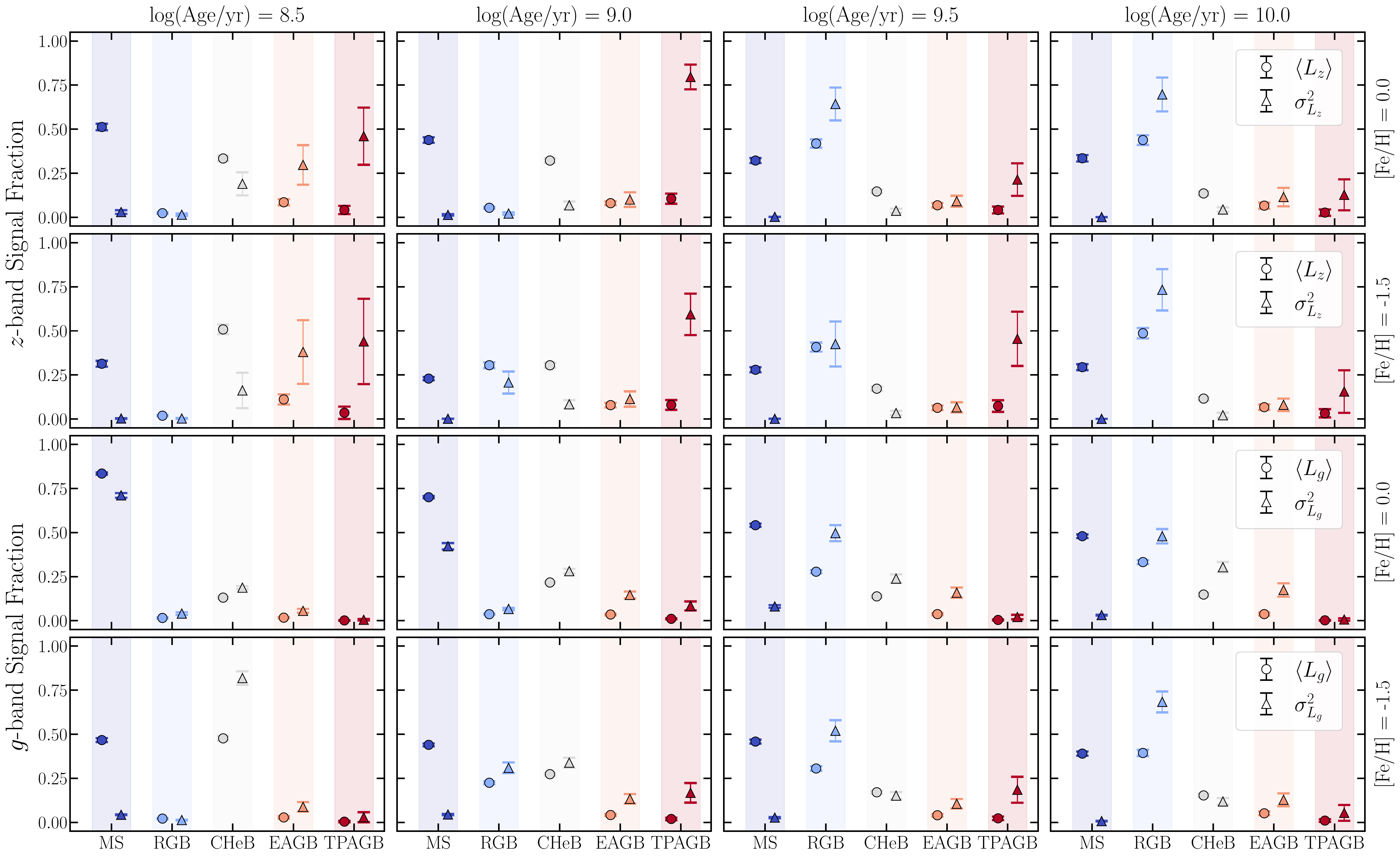}
    \caption{Fraction of the \resp{luminosity} (circles) and the \resp{luminosity variance} (triangles) contributed by the main sequence (MS), red giant branch (RGB), core helium burning stars (CHeB), early asymptotic giant branch (EAGB), and the thermally pulsating asymptotic giant branch (TPAGB)  \johnny{for a $10^5$~M$_\odot$ SSP}. The age of the SSP increases from left to right and is indicated at the top of each column. The metallicity of each row is indicated on the right. The top (bottom) two rows show the signal fractions in the LSST $z$-band ($g$-band). The error bars show the $1\sigma$ width of 1000 realizations of the SSP, where the median values are consistent with an SSP that fully samples the mass function (see Section~\ref{sec:imf-sample}). For a $10^6$~M$_\odot$ SSP, all error bars \johnny{would be comparable to the symbol sizes}. The phase of stellar evolution that dominates the \resp{each signal} is a strong function of age, metallicity, and observation bandpass. The RGB dominates the \resp{luminosity variance} in old populations regardless of metallicity, but for young to intermediate ages, the dominant source of fluctuations varies strongly with metallicity and bandpass, spanning all phases of stellar evolution. See Section~\ref{sec:phases} for a note about how we label stellar evolutionary phases.}
    \label{fig:sig-frac}
    \vspace{0.2cm}
\end{figure*}

\begin{figure}[th!]
    \centering
    \includegraphics[width=\columnwidth]{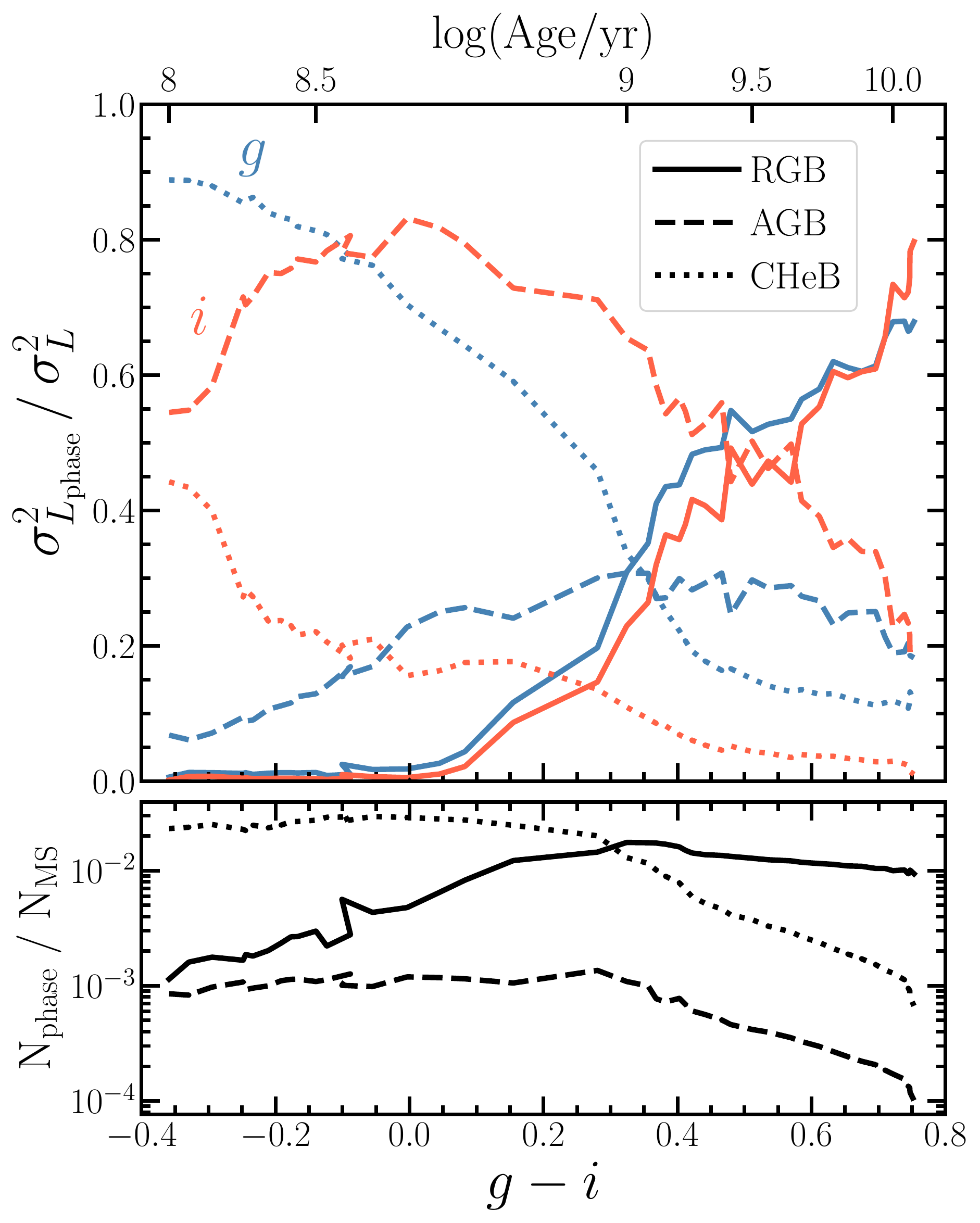}
    \caption{{\it Top}: Fraction of LSST $g$- (blue) and $i$-band (orange) \resp{luminosity variance} contributed by RGB, AGB, and CHeB stars as a function of $g-i$ color and age. See Section~\ref{sec:phases} for a note about how we label stellar evolutionary phases. Calculations are for an SSP with [Fe/H] = $-1.5$. With the metallicity fixed, the ages indicated by the ticks on the top axis map directly onto the color on the bottom axis. {\it Bottom}: The ratio of the number of stars in the evolved phases shown in the top panel to the number of main sequence stars.}
    \label{fig:phase-sbf-color}
\end{figure}

\subsection{Stellar Population Generation} \label{sec:pop-generation}

We adopt the stellar isochrones from the Modules for Experiments in Stellar Astrophysics (MESA; \citealt{Paxton:2011, Paxton:2013, Paxton:2015}) Isochrones and Stellar Tracks (MIST) project\footnote{\url{http://waps.cfa.harvard.edu/MIST}} \citep{Choi:2016, Dotter:2016}. Specifically, we use version 1.2 of the rotating models with $v/v_\mathrm{crit} = 0.4$. We restrict our calculations to age and metallicity ranges of $8 \leq \log(\mathrm{Age/yr}) \leq10.15$ and $-2.5 \leq \mathrm{[Fe/H]} \leq 0$, respectively. We are currently limited to models with solar-scaled abundance ratios but plan to explore SBFs with non-solar-scaled models in the future. 

Since rare, luminous stars dominate the SBF signal (see Section~\ref{sec:sbf-signal}), the treatment of post-main sequence evolution is especially important for modeling luminosity fluctuations \citep[e.g.,][]{Cantiello:2003aa}. In MIST, \resp{mass loss in stars with masses below 10~M$_\odot$ is treated with the prescription of \citet{Reimers:1975} along the RGB and that of \citet{Bloecker:1995} along the asymptotic giant branch (AGB). The free parameters of these prescriptions are tuned to match observational constraints, including the initial–final stellar mass relation, the AGB luminosity function in the Magellanic Clouds, and asteroseismic measurements from cluster members in the Kepler fields. We refer the reader to \citet{Paxton:2011} and \citet{Choi:2016} for more details about MIST's treatment of post-main sequence evolution.}

To generate single-burst stellar populations, we assume the initial mass function from \citet{Kroupa:2001}, with a minimum mass of 0.1~M$_\odot$. \resp{The maximum masses are set by the MIST isochrones for each combination of age and metallicity. For reference, the maximum mass of an SSP with $\mathrm{[Fe/H] = -1}$ is 5.1, 2.0, and 0.9~M$_\odot$ at ages of 100~Myr, 1~Gyr, and 10~Gyr, respectively. For the range of metallicities presented in this work, the maximum masses vary with respect to these values by ${\sim}5$-15\%. Note that the choice of mass function is not crucial, as previous work has shown that the SBF signal is relatively insensitive to it \citep{Worthey:1993, Blakeslee:2001aa, Cantiello:2003aa}.} Given a set of initial masses, we then linearly interpolate the MIST isochrones to predict stellar magnitudes. 

For our fiducial calculations, we adopt the filter systems of LSST $ugrizy$, HST ACS/WFC, and RST, which represent the most important current and future observatories for measuring SBFs from the ground and space. \resp{To compare our models to observations, we additionally perform calculations in the Two Micron All Sky Survey (2MASS) $J\!H\!K_\mathrm{s}$ \citep{Cohen-2MASS:2003}, $U\!B\!V\!R\!I$ \citep{Bessell:2012}, and CFHT\footnote{\url{http://www.cfht.hawaii.edu/Instruments/Imaging/Megacam/specsinformation.html}} filter systems. Throughout this work, $U\!B\!V\!R\!I\!J\!H\!K_\mathrm{s}$ magnitudes are presented in the Vega system, and all other magnitudes are in the AB system \citep{Oke:1983aa}.}

\subsection{Identifying Stellar Evolutionary Phases}\label{sec:phases}

Throughout this work, we label phases of stellar evolution according to the MIST primary equivalent evolutionary phases \citep[EEPs;][]{Dotter:2016, Choi:2016}, which are physically-defined reference points that can be identified across a full set of stellar evolution tracks. While this is very useful from a computational point of view, it means that in some cases our terminology will differ from standard nomenclature. 

For example, we classify RGB stars as having a MIST EEP number between the terminal age main sequence ($\mathrm{EEP = 454}$) and the tip of the RGB ($\mathrm{EEP = 605}$). Physically, this corresponds to the evolutionary phase between hydrogen exhaustion and core helium burning (CHeB), regardless of initial stellar mass. This makes it possible to identify an RGB phase for high-mass stars, which do not go through a ``red giant'' phase in the traditional sense. Additionally, the subgiant branch is part of the RGB in this framework. Similarly, the AGB corresponds to the post-CHeB evolutionary phase. Unless explicitly noted, we use the term AGB to refer to the combination of the early and thermally-pulsating phases of the AGB.

\begin{figure*}[ht!]
    \centering
    \includegraphics[width=\textwidth]{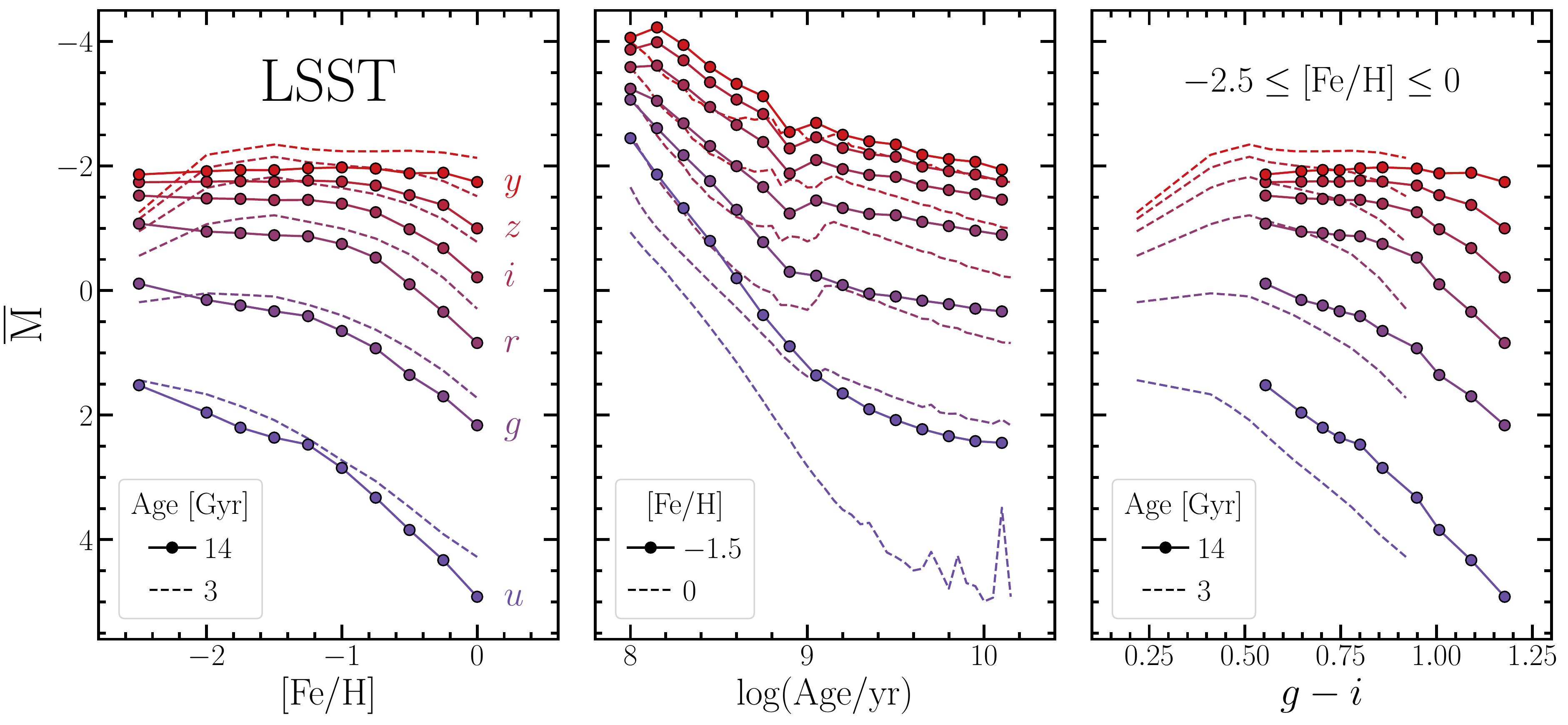}
    \caption{Absolute SBF magnitudes $\left(\mathrm{\overline{M}}\right)$ in the LSST $ugrizy$ filter system as a function of metallicity (left), age (middle), and $g-i$ color (right). All calculations assume simple stellar populations (SSPs) with  the indicated ages and metallicities. The SSPs span an age and metallicity range of $8 \leq \log(\mathrm{Age/yr}) \leq 10.15$ and $-2.5 \leq [\mathrm{Fe/H}] \leq 0$, respectively. In general, \Mbar\ becomes fainter with decreasing wavelength, and SBFs from older SSPs in redder bandpasses have a weaker dependence on metallicity, making \johnny{them best suited as} distance indicators. In contrast, $u$-band SBF is an excellent metallicity tracer, with little age dependence for ages $\gtrsim3$~Gyr, though its intrinsic faintness and the $u$-band's poor filter throughput make it difficult to observe in practice.}
    \label{fig:lsst-sbf-color}
\end{figure*}

\begin{figure*}[ht!]
    \centering
    \includegraphics[width=\textwidth]{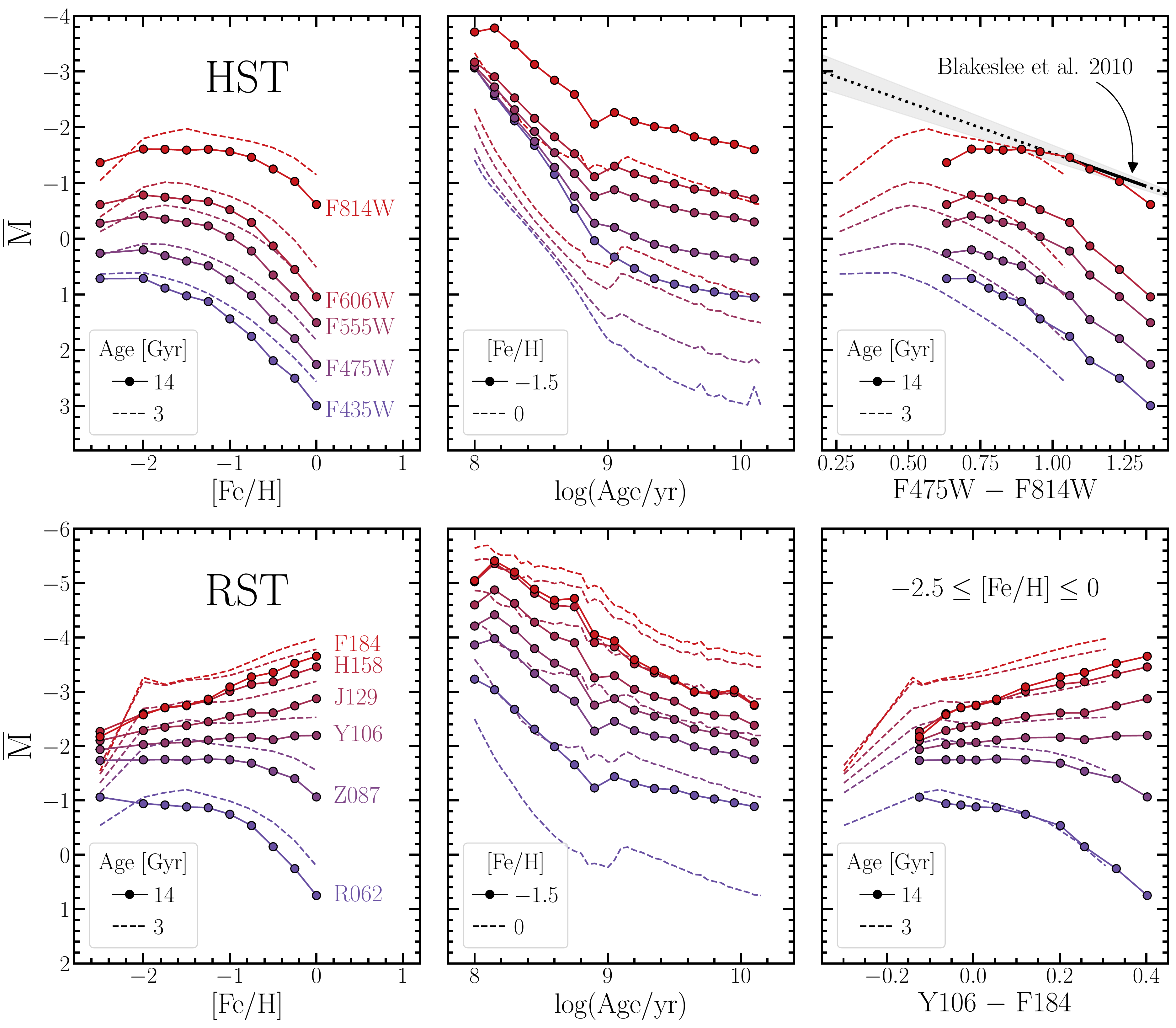}
    \caption{Same as Figure~\ref{fig:lsst-sbf-color} except for the HST ACS/WFC (top row) and RST (bottom row) filter systems. In the right panel of the top row, the solid black line shows the empirical calibration of from \citet{Blakeslee:2010aa}; the gray shaded region indicates the $1\sigma$ uncertainties in the fit, and the dotted black line is an extrapolation of the calibration.}
    \label{fig:space-sbf-color}
\end{figure*}

\subsection{The SBF Signal}\label{sec:sbf-signal}

The SBF signal is traditionally quantified by the ratio of the second moment to the first moment of the stellar luminosity function \citep{Tonry:1988}: 
\begin{equation}\label{eqn:Lbar}
    \overline{\mathrm{L}} \equiv \frac{\sum_j n_j L_j^2}{\sum_j n_j L_j} \equiv \frac{\sigma_L^2}{\langle L\rangle}, 
\end{equation}
where $n_j$ is the number of stars of type $j$ with luminosity $L_j$, \resp{$\sigma_L^2$ is the luminosity variance, and $\langle ...\rangle$ denotes the expectation value.} Equivalently, $\overline{\mathrm{L}}$ may be viewed as the luminosity-weighted mean luminosity of a population of stars. The corresponding absolute and apparent fluctuation magnitudes of  $\overline{\mathrm{L}}$ are denoted as $\overline{\mathrm{M}}$ and $\overline{m}$, respectively. As noted by \citet{Blakeslee:2001aa}, $\overline{\mathrm{L}}$ is well-defined for {\it any} population of stars in any bandpass. 

As is clear from Equation (\ref{eqn:Lbar}), the SBF signal is a function of the underlying stellar population. In general, the most luminous (and hence rare) stars contribute most to the fluctuation signal, and the specific stellar phase that dominates is a strong function of age, metallicity, and the observation bandpass \citep[e.g.,][]{Liu:2000}. \resp{We illustrate this point in Figure~\ref{fig:sig-frac}, which shows the fraction of $\langle L \rangle$ and $\sigma_L^2$ in $g$- and $z$-band that is contributed by different phases of stellar evolution. Note that the signal fractions presented here are subject to large model uncertainties and are primarily meant to provide an approximate sense of the dominant stellar contributors to these two important observables, as predicted by the MIST models.}

\resp{Each panel of Figure~\ref{fig:sig-frac} summarizes the signal fractions from 1000 realizations of an SSP of stellar mass $10^5~M_\odot$ with the indicated age (top of each column) and metallicity (far right of each row). The error bars indicate the $1\sigma$ spread of the realizations, which is due to incompletely sampling the mass function (see Section~\ref{sec:imf-sample})}. The median values of the data points are consistent with an SSP that fully samples the mass function. For a $10^6$~M$_\odot$ SSP, all error bars are comparable in size to the symbol sizes.

The \resp{$z$-band luminosity variance} is always dominated by post-main sequence stars. In contrast, \resp{${\sim}25$-50\% of the luminosity in this bandpass} originates from the main sequence (MS), regardless of age and metallicity. RGB stars are often named as the dominant source of SBFs, but this is only true for stellar populations with $\log(\mathrm{Age/yr}) \gtrsim 9.5$. At younger ages, AGB stars are important contributors to the SBF signal, with significant stochasticity in their contribution to low-mass stellar systems due to the short-lived and luminous nature of this phase of stellar evolution. The age at which the RGB begins to contribute SBFs is a function of metallicity, with RGB stars in more metal poor systems contributing significant SBFs at younger ages. 

In the bottom two rows of Figure~\ref{fig:sig-frac}, the $g$-band signal fractions show very different behavior from those of the $z$-band, particularly \resp{in young SSPs}. For ages $\log(\mathrm{Age/yr}) \lesssim 9.5$, the phases that dominate the $g$-band signal fractions are highly sensitive to metallicity. This is especially clear for the MS in the youngest age column, where the MS contributes ${\gtrsim}70\%$ \resp{of the luminosity variance} at high metallicity and ${\sim}0\%$ at low metallicity. At 1~Gyr, those fractions become ${\gtrsim}40\%$ at high metallicity and ${\sim}0\%$ at low metallicity. At [Fe/H] $ = -1.5$, CHeB stars produce ${\gtrsim}80\%$ of the \resp{luminosity variance} and ${\sim}50\%$ of the \resp{luminosity} in very young SSPs. The reason for this trend with metallicity is that the CHeB stars become hotter and more luminous as the metal content decreases due a lower Rosseland mean opacity \citep[e.g.,][]{Choi:2016}, leading to their larger contribution to both the integrated flux and SBF signals. 

In contrast to the $z$-band, AGB stars generally do not contribute much to the $g$-band signals, leading to far less scatter between SSP realizations in this bluer band. Similar to the $z$-band, the RGB dominates \resp{the $g$-band luminosity variance} at the oldest ages, though CHeB stars become important contributors in old SSPs with high metallicity.

For a metal-poor stellar population, at what age --- or equivalently, color --- does the RGB become a subdominant contributor to the SBF signal? At this transition color, we might expect an inflection point in the fluctuation-color relation. We explore this question in the top panel of Figure~\ref{fig:phase-sbf-color}, which shows the LSST $g$- and $i$-band \resp{luminosity variance} contributed by RGB, AGB (which includes both early and thermally-pulsating AGB stars), and CHeB stars for an SSP with [Fe/H] = $-1.5$, plotted as a function of $g-i$ color and age. The bottom panel shows the relative number of stars in each of the phases shown in the top panel compared to the number of main sequence stars. For colors redward of $g-i \sim 0.5$, RGB stars dominate the fluctuations. In contrast, \resp{fluctuations} from bluer populations begin to be dominated by AGB and CHeB stars, with the relative fractions depending on bandpass and metallicity. The $g$-band fluctuation magnitude in particular is an excellent probe of CHeB stars for SSPs with $g-i < 0.4$.

Figure~\ref{fig:sig-frac} and \ref{fig:phase-sbf-color} illustrate the well-known idea that, in principle, it is possible to use SBFs to isolate and study specific phases of stellar evolution in semi-resolved galaxies \citep[e.g.,][]{Liu:2000, Raimondo:2005, 2010MNRAS.403.1213G, Lee-Worthey-Blakeslee-2010, vD:2014, Conroy-2016-pcmd, Mitzkus:2018, Cook:2019}, provided the \resp{distance to the system is known and it} is relatively simple both in terms of its stellar population and morphology.

\subsection{The Fluctuation-Color Relation}\label{sec:sbf-color-relation}

Here, we present the MIST predictions for absolute fluctuation magnitudes (\Mbar) and their dependence on SSP parameters for ground- (LSST) and space-based (HST and RST) observatories. In order to convert apparent fluctuation magnitudes into absolute units, information about the underlying stellar population is necessary. This is usually parametrized using galaxy colors, since for old SSPs ($\gtrsim$5~Gyr), \Mbar\ is an approximately linear or constant function of integrated color, depending on the bandpass \citep[e.g.,][]{Tonry:1991a, Blakeslee:2001aa}. The ``fluctuation-color relation'' must be calibrated empirically with systems of known distance \citep[e.g.,][]{Tonry:1997, Blakeslee:2010aa, Jerjen:2001, Carlsten:2019aa} or theoretically using stellar populations synthesis \citep[e.g.,][]{Worthey:1993, Liu:2000, Cantiello:2003aa, Raimondo:2005}. 
 
In Figure~\ref{fig:lsst-sbf-color}, we show absolute fluctuation magnitudes in the LSST $ugrizy$ filter system as a function of metallicity, age, and $g-i$ color. \johnny{The data for this figure are in Table~\ref{tab:lsst}.} As previous studies have shown, the fluctuation magnitude generally brightens with increasing bandpass wavelength, regardless of the stellar population parameters. This trend is due to the increasing contribution of giant stars to the fluctuation signal in redder bandpasses (see Figure~\ref{fig:sig-frac}). Consistent with observations, the signal becomes fainter as the metallicity increases for wavelengths blueward of ${\sim}1~\mu$m, and the amplitude of this trend decreases with increasing wavelength. As first predicted by \citet{Worthey:1993}, the SBF signal from old SSPs in the $y$-band, which falls between ${\sim}$0.8~$\mu$m and 1.2~$\mu$m, is roughly independent of metallicity (color), making it a promising distance indicator. 

At the other end of the spectrum, the $u$-band has a steep metallicity dependence and\resp{, similar to the other filters,} a weak age dependence for SSPs that are older than ${\sim}3$~Gyr. Although SBFs in the $u$-band are exceedingly difficult to observe in practice due to their intrinsic faintness and the $u$-band's poor filter throughput, they have great potential as a stellar metallicity tracer, as well as a general probe of stellar populations. This was previously pointed out by \citet{Cantiello:2003aa} for the Johnson-Cousins $U$- and $B$-band. 

In the top panel of Figure~\ref{fig:space-sbf-color}, we show absolute fluctuation magnitudes in the HST ACS/WFC filter system as a function of metallicity, age, and F475W $-$ F814W color. The ranges of age and metallicity are the same as in Figure~\ref{fig:lsst-sbf-color}. The trends between bandpass and stellar population parameters are qualitatively similar to the LSST predictions. Extensive work has been done on empirically and theoretically calibrating the fluctuation-color relation in the HST filter system, with a particular focus on stellar populations that are similar to early-type galaxies in groups and clusters \cite[e.g.,][]{Blakeslee:2001aa, Cantiello:2003aa}. 

The solid black line in the top-right panel shows the F814W SBF vs. $\mathrm{F475W - F814W}$ color empirical calibration from \citet{Blakeslee:2010aa}, which is approximately linear in the color range $1.06 < \mathrm{F475W - F814W} < 1.32$. The dotted black line shows the linear extrapolation of this calibration, with the 1$\sigma$ uncertainties indicated by the gray shaded region. The zero point of the MIST relation is in good agreement with the empirical calibration, but the slope of the model prediction is slightly steeper within this color range --- a qualitatively similar discrepancy is seen in the model comparison in Figure~6 of \citet{Blakeslee:2010aa}. 

With the resolution of HST but $100\times$ the field of view, RST has the potential to become our most powerful SBF observatory. Importantly, RST's proposed infrared filter set has four filters in the wavelength range 1 - 2~$\mu$m, and SBFs are much brighter at infrared wavelengths than at optical wavelengths \citep{Worthey:1993}. Infrared SBFs have long been recognized as a way to extend the reach of the SBF method from the ground \citep{Jensen:1998} and space \citep{Jensen:2003}. \johnny{For bright elliptical galaxies, \citet{Jensen:2015} showed that it is possible to measure ${\sim}$5\% distances out to ${\sim}80$~Mpc using a single-orbit observation with the Wide Field Camera 3 Infrared Channel on HST.}

In the bottom panel of Figure~\ref{fig:space-sbf-color}, we show the SBF predictions for the proposed RST filter set. As expected from previous work, the fluctuation magnitude {\it brightens} with increasing metallicity for wavelengths ${\gtrsim}1~\mu$m. Similar to the LSST's $y$-band, the Y106 SBF signal is nearly independent of metallicity for old SSPs. For the reddest filters, the fluctuation magnitude can be as much as ${\sim}$4-5 magnitudes brighter than the corresponding optical signal. However, the metallicity dependence steadily steepens for filters redward of 1~$\mu$m, which is a challenge for calibrating infrared SBFs as a distance indicator. We provide HST and RST fluctuation magnitudes and mean magnitudes for color calculations in Table~\ref{tab:hst} and Table~\ref{tab:wfirst}, respectively.

\begin{figure*}[ht!]
    \centering
    \includegraphics[width=\textwidth]{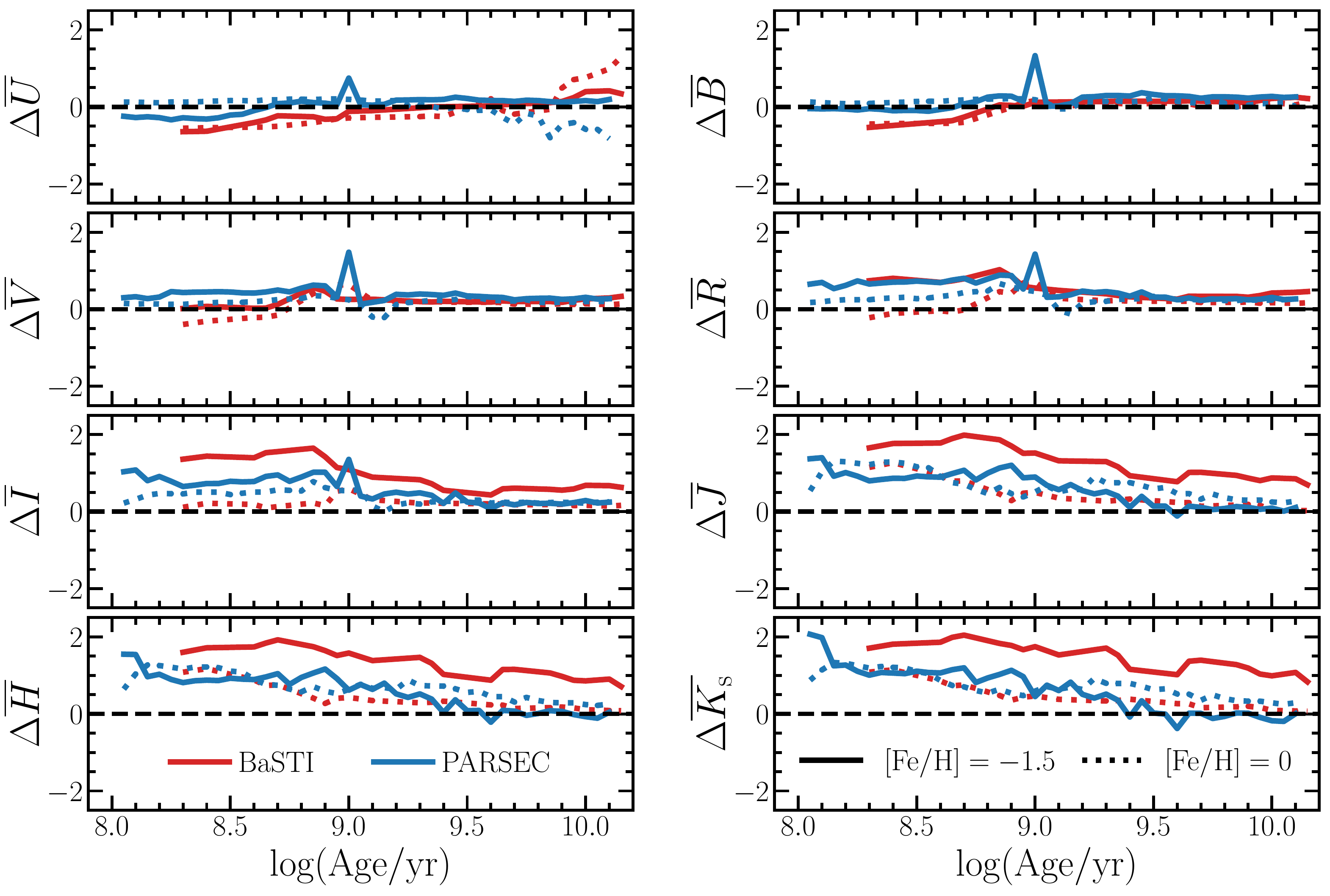}
    \caption{\resp{Comparison of $U\!B\!V\!R\!I\!J\!H\!K_s$ SBF predictions from different isochrone models,} plotted as a function of SSP age, assuming SSPs with $\mathrm{[Fe/H] = -1.5}$ (solid lines) and $\mathrm{[Fe/H] = 0}$ (dotted lines). Each panel shows the difference $\delta\overline{\mathrm{M}}_x \equiv \mathrm{\overline{M}_{MIST}} - \mathrm{\overline{M}_{compare}}$, where $\mathrm{\overline{M}_{MIST}}$ is the MIST prediction, $x$ is the bandpass, and ``compare'' refers to BaSTI (red lines) and PARSEC (blue lines)---the comparison isochrones. All calculations were consistently performed using the Flexible Stellar Population Synthesis software package. \resp{The agreement between all models is generally better at older SSP ages, as expected, and MIST is generally in better agreement with the PARSEC isochrone calculations. For the low-metallicity SSPs with ages older than ${\sim}$3~Gyr, PARSEC and MIST agree at the ${\sim}10$-30\% level in the optical and at the ${\sim}5$\% level in the near-infrared. The overall level of disagreement between these models highlights that the choice of isochrone remains as a major uncertainty in SBF calculations.}}
    \label{fig:fsps-compare-sbf}
\end{figure*}

\begin{figure*}[ht!]
    \centering
    \includegraphics[width=\textwidth]{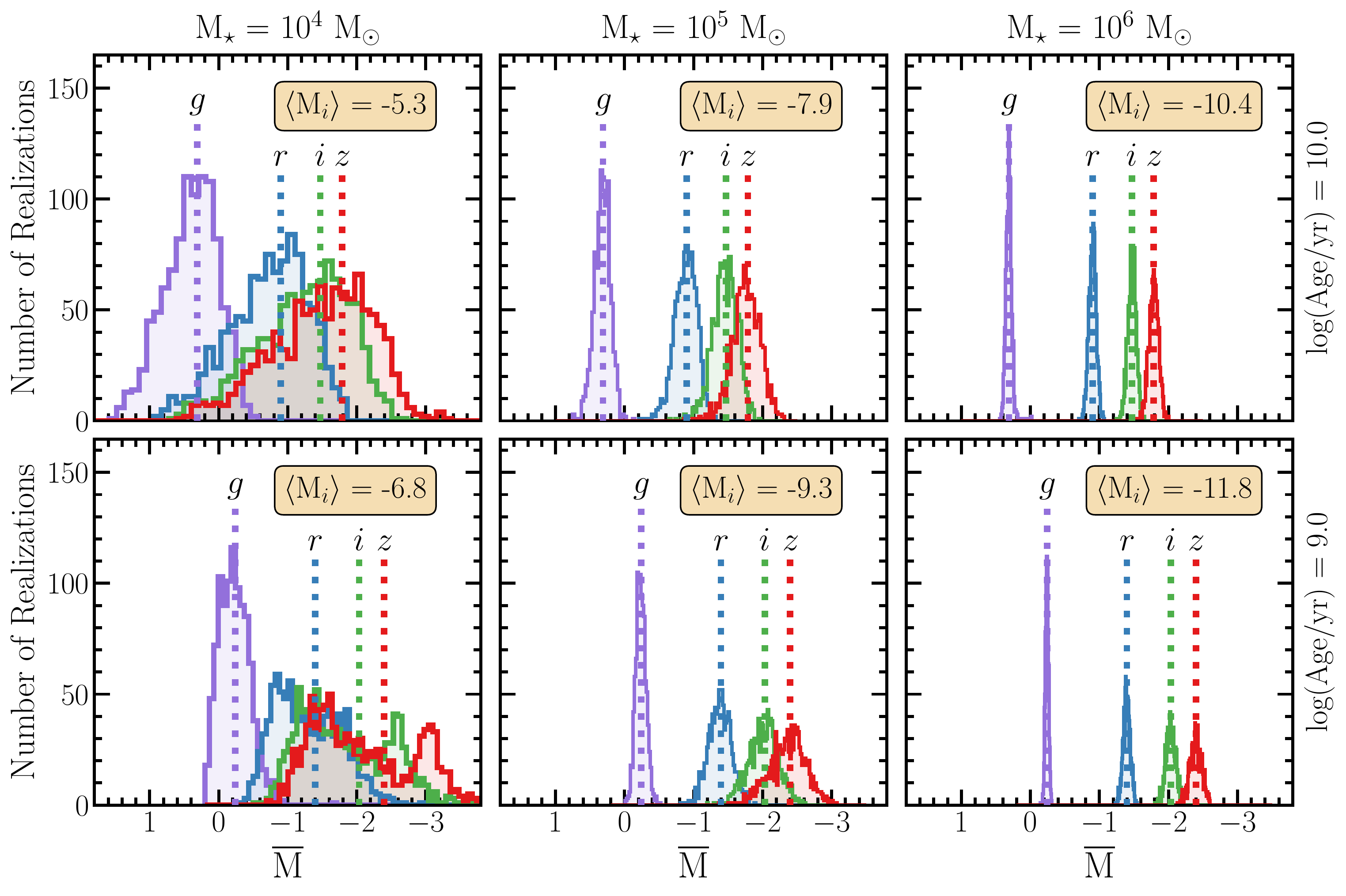}
    \caption{Fluctuation magnitudes in LSST $griz$ for simple stellar populations with masses 10$^4$ (left), 10$^5$ (middle), and 10$^6~M_\odot$ (right). We omit $u$ and $y$ for clarity; they follow the same trend, where the scatter decreases with bandpass wavelength. We generate 1000 realizations of each population, with a fixed metallicity of $\mathrm{[Fe/H]} = -1.5$ and for ages of 10 (top) and 1 Gyr (bottom). The vertical dotted lines show the asymptotic fluctuation signal for a stellar population that fully samples the mass function, with the associated bandpass indicated at the top of each line. In the ultra-faint stellar mass regime (${\lesssim}10^5~M_\odot$), the SBF signal is sensitive to the number of AGB stars, leading to the bimodal distributions in the bottom left. \resp{In each panel, the text box indicates the mean $i$-band absolute magnitude of the SSP realizations, which may be compared with Figure~\ref{fig:absmag-scatter}.}}
    \label{fig:m_rs}
\end{figure*}

\begin{figure}[ht!]
    \centering
    \includegraphics[width=\columnwidth]{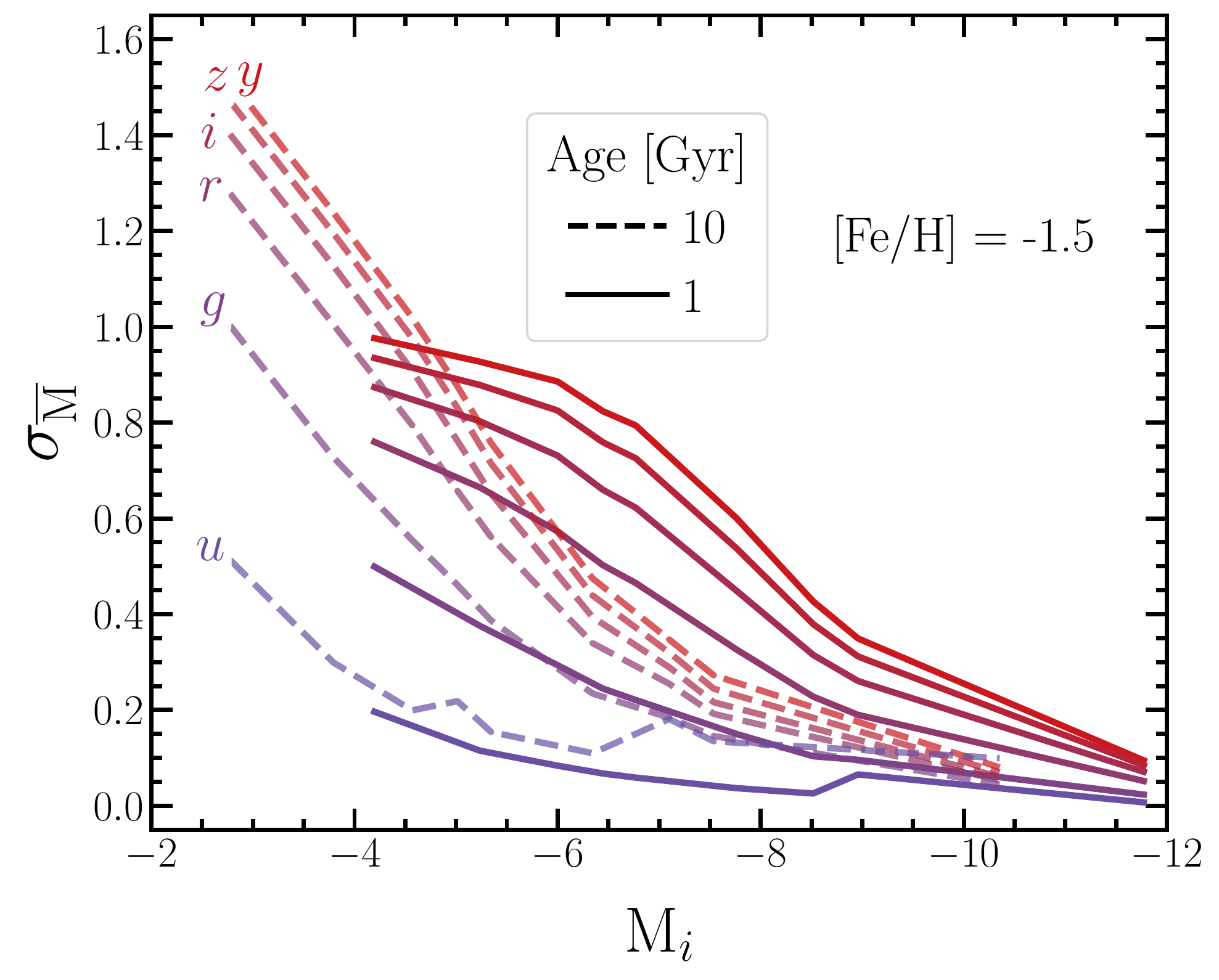}
    \caption{\johnny{Sampling scatter} in the distribution of LSST $ugrizy$ fluctuation magnitudes as a function of the absolute magnitude of the stellar population. Solid lines are for an SSP of age 1~Gyr, and dashed lines are for an SSP of age 10~Gyr. Both SSPs have a metallicity of $\mathrm{[Fe/H]}=-1.5$. The scatter increases with decreasing luminosity due to undersampling of the stellar mass function, and it increases with bandpass wavelength due to the increasing contribution of rare, luminous stars.}
    \label{fig:absmag-scatter}
\end{figure}

\subsection{Isochrone Comparison}

Many theoretical SBF predictions exist in the literature, which make different assumptions about important and often uncertain aspects of stellar evolution such as the treatment of mass loss along the RGB and AGB \citep[e.g.,][]{Blakeslee:2001aa, Cantiello:2003aa, Raimondo:2005, Raimondo:2009, 2010MNRAS.403.1213G}. However, the large number of assumptions and input parameters required to run stellar population synthesis codes \citep[e.g.,][]{Conroy-SPS-review-2013} makes it difficult to directly compare different model predictions. 

To compare our results with SBF predictions based on different isochrones, we use the Flexible Stellar Population Synthesis software package \citep{Conroy:2009aa, Conroy-Gunn-2010} to consistently calculate fluctuation magnitudes based on the MIST, BaSTI \citep{BaSTI-2004, Cordier-2007-BaSTI}, and PARSEC \citep{PARSEC-2012} isochrones. By holding all other model inputs fixed (e.g., initial mass function, spectral library, and filter throughput functions), we isolate the effect that varying the isochrones has on the SBF signal.

For each isochrone set, we calculate SBF predictions as a function of SSP age in \resp{standard $U\!B\!V\!R\!I\!J\!H\!K_s$  filters to cover wavelengths from the optical to near-infrared}. In Figure~\ref{fig:fsps-compare-sbf}, we show the difference in fluctuation magnitudes $\delta\overline{\mathrm{M}}_x \equiv \mathrm{\overline{M}_{MIST}} - \mathrm{\overline{M}_{compare}}$, where $x$ is the bandpass and ``compare'' refers to the BaSTI (red lines) and PARSEC (blue lines) isochrones,  the comparison isochrones. We show results for SSPs with $\mathrm{[Fe/H] = -1.5}$ (solid lines) and $\mathrm{[Fe/H] = 0}$ (dotted lines). 

The agreement between MIST and the comparison isochrones is generally better for older ages, as expected given the large uncertainties and timescales associated with the flux contributed by short-lived massive stars. For ages older than 3~Gyr, MIST differs with PARSEC at the ${\sim}$10-30\% level \resp{in bandpasses blueward of ${\sim}1~\mu$m}. The disagreement with BaSTI over this same wavelength and age range is much larger, with differences at \resp{the ${\sim}15$-90\% level}, depending on bandpass. In both comparisons, the agreement is generally better in the bluer bandpasses, with the exception of the $U$-band at solar metallicity near 10~Gyr. \resp{At near-infrared wavelengths, there is excellent agreement between MIST and PARSEC, with differences at the ${\sim}5$\% level for low-metallicity SSPs with ages $\gtrsim$3~Gyr. In contrast, the low-metallicity BaSTI near-infrared SBF magnitudes are typically offset by nearly 1 mag. That the largest discrepancies are in the near-infrared is not too surprising, since the uncertain AGB phase dominates the flux at these red wavelengths.}

\johnny{The overall level of disagreement in Figure~\ref{fig:fsps-compare-sbf} is concerning, though it is in line with known isochrone-driven differences between evolutionary calculations (e.g., see Figure 33-34 in \citealt{Choi:2016}). The great challenge is that clean, simple calibration data are scarce, especially for ages that are younger than globular clusters. For predicting SBFs, the likely most important differences between the isochrones are their very different treatments of AGB stars and stellar rotation in higher mass stars, as well as their different calibrations of convective overshooting and mass-loss. For additional isochrone comparisons, we refer the reader to \citet{Choi:2016}.}

\subsection{Undersampling the Mass Function} \label{sec:imf-sample} 

Theoretical predictions for fluctuation magnitudes generally assume the mass function has been fully populated, an assumption that is easily satisfied in giant elliptical galaxies but is likely false in low-luminosity stellar systems such as star clusters and low-mass dwarf galaxies \respp{\citep[e.g.,][]{Santos:1997, Raimondo:2005}}. For systems with stellar masses ${\lesssim}10^6~M_\odot$, stochasticity in the number of stars in rare, luminous evolutionary phases leads to \johnny{``sampling scatter''} in the fluctuation signal. Furthermore, SBFs are often measured within a subset of a galaxy's pixels, since some regions of the image might not be suitable for SBF measurements (e.g., due to foreground/background objects, star-forming knots, and/or dust lanes); thus, sampling of the mass function \resp{may be an} important consideration for \resp{more massive dwarf galaxies, depending on the fraction of their surface area that must be masked.}

In Figure~\ref{fig:m_rs}, we show fluctuation magnitudes in LSST $griz$ filters for SSPs with masses of 10$^4$ (left), 10$^5$ (middle), and 10$^6~M_\odot$ (right). The calculations assume a metallicity of $\mathrm{[Fe/H]} = -1.5$ and ages of 1~Gyr (bottom row) and 10~Gyr (top row). The dotted lines indicate \Mbar\ for a fully-sampled mass function. As expected, the \johnny{sampling scatter} increases with decreasing stellar mass. Interestingly, at fixed stellar mass, the \johnny{sampling scatter} is lower for bluer bandpasses, with little to no bias in the $g$-band across this range of SSP parameters. \respp{A similar effect is also apparent in the models of \citet{Raimondo:2005}, as can be seen in their Figure~2.} 

The lower \johnny{sampling scatter} in blue compared to red bandpasses is due to the ${\sim}1$-6 mag fainter flux from rare AGB stars, as well as the increased signal contribution from comparatively less rare CHeB stars. Therefore, for very low-mass galaxies, SBFs in bluer bands such as $u$ or $g$ \resp{have potential as distance indicators}, though this comes at the cost of much fainter SBF magnitudes and a steeper metallicity dependence than in redder bandpasses (see Section~\ref{sec:sbf-color-relation}). \resp{The stronger dependence of blue bandpasses on stellar population parameters (see Figure~\ref{fig:lsst-sbf-color}) means a robust SBF-color calibration and accurate color measurements would be required to take advantage of the smaller sampling scatter.} 

The strongly non-Gaussian shapes of the distributions in Figure~\ref{fig:m_rs}, particularly those in the bottom row, are primarily driven by the number of thermally-pulsating AGB stars. Note, for example, the bottom-left panel, for which the distributions of the red bands ($riz$) are bimodal, with one peak corresponding to zero thermally-pulsating AGB stars and the other corresponding to a non-zero number. The better-behaved distributions in the top row can be understood from inspection of the second and fourth rows of Figure~\ref{fig:sig-frac}. For an SSP of age 1~Gyr, the thermally-pulsating AGB dominates the SBF signal in the $z$-band, whereas the RGB and CHeB phase dominate it in the $g$-band, and when the SSP reaches 10~Gyr in age, the RGB dominates both bandpasses, leading to more stability in the predicted fluctuation signal. 

In practice, the \resp{stellar mass (or absolute magnitude)} within an SBF measurement aperture determines the amount of scatter in \Mbar. In Figure~\ref{fig:absmag-scatter}, we show the \johnny{sampling scatter} in LSST $ugrizy$ fluctuation magnitudes as a function $i$-band absolute magnitude for a metal poor SSP of age 1 Gyr (solid lines) and 10 Gyr (dashed lines). \resp{Denoted as $\sigma_\mathrm{\overline{M}}$, we quantify the sampling scatter as the standard deviation of the model absolute SBF magnitudes.} At 10~Gyr, the scatter ranges from $\lesssim0.1$ at M$_i=-10$ (which corresponds to M$_\star\sim10^6$~M$_\odot$) to larger than a magnitude for the lowest mass systems. In red bandpasses, the scatter from young to intermediate-age populations is generally larger by ${\sim}0.3$-0.5~mag compared to old populations, except for the lowest luminosities, where $\sigma_\mathrm{\overline{M}}\gtrsim1$~mag for older populations and the distribution of \Mbar\ is bimodal in younger populations (see Figure~\ref{fig:m_rs}). 

In massive elliptical galaxies, the \johnny{``cosmic scatter''} in the fluctuation-color relation\footnote{Specifically, the HST ACS/WFC F850LP fluctuation vs. $\mathrm{F475W - F850LP}$ color relation for $\mathrm{F475W - F850LP>1}$.} \johnny{due to intrinsic stellar population variations} was measured by \citet{Blakeslee:2009} to be ${\sim}0.06$~mag. \johnny{Intrinsic scatter (whether due to sampling or stellar population effects) has yet to be measured} in low-mass galaxies with M$_\star\sim10^6$-$10^7$~M$_\odot$. The typical observational uncertainties in SBF measurements of such systems is on the order of 0.3~mag \citep{Cohen:2018, Carlsten:2019aa}, which is larger than the predicted \johnny{sampling scatter} by ${\sim}0.2$~mag.

\section{Comparison to Observations}\label{sec:compare-with-obs}

\subsection{\resp{Optical SBFs in Dwarf Galaxies}}

In this section, we compare MIST \resp{optical} fluctuation-color relations to observations of dwarf galaxies, which have bluer colors and lower luminosities than galaxies used for most previous comparisons between theory and observations. Our comparison data come from the CFHT \citep {Carlsten:2019aa} and HST \citep{Cohen:2018}. Importantly, all the galaxies in this comparison sample have independently measured distances based on the tip of the RGB method, which we use to directly convert apparent fluctuation magnitudes into absolute units. 

\johnny{The tip of the RGB distance measurements from \citet{Carlsten:2019aa} were mostly drawn from the nearby galaxy catalog of \citet{Karachentsev:2013}, with additional distances taken from \citet{Cohen:2018} and \citet{Martinez-Delgado:2018}. The characteristic uncertainty in determining the tip of the RGB is ${\sim}0.1$~mag or 5\% in distance \citep[e.g.,][]{Danieli:2017, Cohen:2018, Beaton2018-old-pops, Jang:2018}.}

\subsubsection{Models with Simple Stellar Populations}

\begin{figure*}[ht!]
    \centering
    \includegraphics[width=\textwidth]{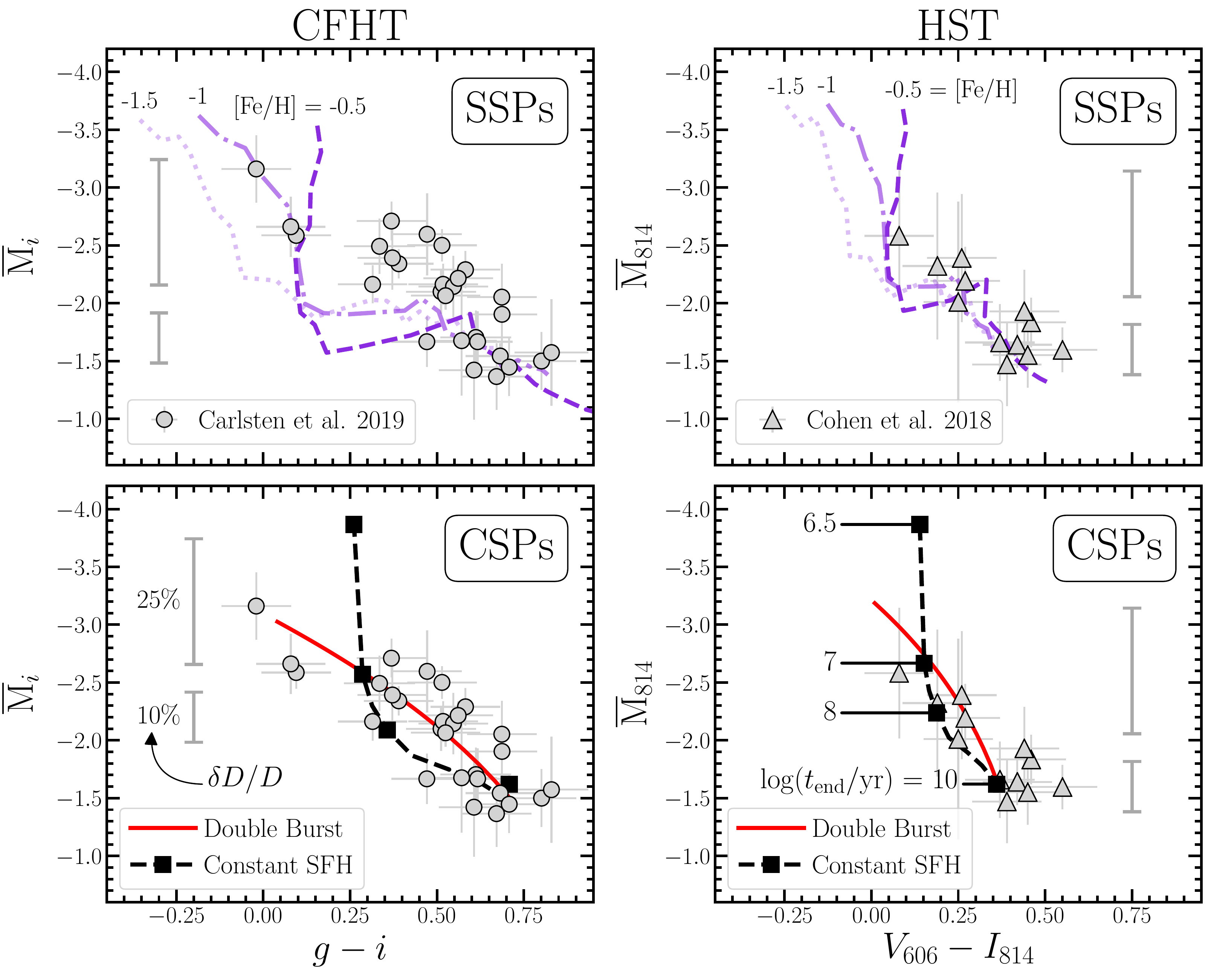}
    \caption{Comparison of MIST fluctuation-color relations with observations of \resp{relatively blue} low-luminosity systems from CFHT \citep[left column;][]{Carlsten:2019aa} and HST \citep[right column;][]{Cohen:2018}. In the top row, we show predictions for simple stellar populations (SSPs) with the indicated metallicities and ages spanning 100 Myr to 12.5 Gyr. In the bottom row, we show predictions for composite stellar populations (CSPs). The red lines show our double-burst model, which is described by Equation~(\ref{eqn:double-burst}). The dashed black lines show the predictions for a stellar population that has undergone constant star formation, where we vary the time since the star formation ended ($t_\mathrm{end}$). We indicate some values of $t_\mathrm{end}$ with the squares in the bottom-right panel. For reference, the two gray error bars without data points in each panel show the SBF magnitude uncertainty associated with fractional distance errors of 10\% and 25\%. Red and old stellar populations are well-described by SSPs, but for the bluest and youngest galaxies, \johnny{the double-burst model shows better agreement than single-age populations.}}
    \label{fig:compare-with-obs}
\end{figure*}

\johnny{In the top row of Figure~\ref{fig:compare-with-obs}, we compare MIST predictions for SSPs with the observed CFHT \Mbar$_i\ \mathrm{vs.}\ (g-i)$ and HST \Mbar$_{814}\ \mathrm{vs.}\ (V_{606} - I_{814})$ relations.} It is well-established that theoretical SSP colors and SBF magnitudes are consistent with the observed fluctuation-color relation for old, red galaxies \citep[e.g.,][]{Cantiello:2018aa}, as can be seen in our comparison with the empirical calibration of \citet{Blakeslee:2010aa} in top-right panel of Figure~\ref{fig:space-sbf-color}. However, the extent to which models agree with observations of bluer and lower-luminosity systems --- a regime that will become increasingly more relevant as the next generation of wide-field surveys come online --- remains an open question. While resolved-star observations of nearby dwarf galaxies reveal that they have generally undergone complex star formation histories \citep{McQuinn:2010-I, McQuinn:2010, Weisz:2011, Weisz:2014}, SSPs are still a useful point of comparison, since they are easy to interpret, and we might expect them to become more accurate as we push to the lowest-luminosity systems.

The purple lines in Figure~\ref{fig:compare-with-obs} show the MIST predictions for ages spanning 100~Myr to 12.5~Gyr. We show predictions for SSPs with metallicities of $\mathrm{[Fe/H]} = -1.5,\ -1,$ and $-0.5$; higher metallicities are unlikely for such low-luminosity systems \citep{Kirby:2013aa} and lead to predictions that are less consistent with the data. For reference, the two gray error bars without data points in each panel show the SBF magnitude uncertainty associated with fractional distance errors of 10\% and 25\%. 

Consistent with \citet{Cantiello:2018aa}, we find that the MIST SSPs are in excellent agreement with the CFHT observations for colors $g-i \gtrsim 0.5$. We also see that the bluest galaxies observed by \citet{Carlsten:2019aa} are consistent with very young SSPs with $\mathrm{[Fe/H]} = -1$, but there are currently only three galaxies with published data in this regime. Intriguingly, there is a region of parameter space near $g-i \sim 0.4$ that is inconsistent with pure SSPs --- this color is also near the transition color where AGB stars begin to dominate the $i$-band fluctuation signal in metal poor systems (Figure~\ref{fig:phase-sbf-color}). 

For the comparison with HST, the data are fully consistent with the SSP predictions, though the observational uncertainties are large and the fluctuation-color relation is steep, making it difficult to distinguish between different model predictions. \resp{Note the \citet{Cohen:2018} objects are predominantly quenched spheroidal systems in group environments, whereas the \citet{Carlsten:2019aa} objects span group and field environments, with the bluest systems being in the field. We, therefore, expect the stellar populations of the CFHT sample (left column of Figure~\ref{fig:compare-with-obs}) to be more diverse than the HST sample (right column). In fact, 9 of the \citet{Cohen:2018} objects are included in the \citet{Carlsten:2019aa} sample, 8 of which have $g-i\gtrsim0.5$ and $i$-band SBFs that are consistent with old, metal-poor SSPs.}

\subsubsection{Models with Composite Stellar Populations}

\resp{In the top-left panel of Figure~\ref{fig:compare-with-obs}, the SSP models predict fluctuation magnitudes that are too \resp{faint} near $g-i \sim 0.4$. While there are large model uncertainties in this color regime (e.g., the contribution from AGB stars), this discrepancy may suggest that composite stellar populations (CSPs) are needed to explain the data. This hypothesis is consistent with the visual appearance of the blue ($g-i\lesssim0.5$) galaxies from the CFHT sample, most of which have irregular morphologies and show clear evidence of the presence of a young stellar population (e.g., ultra-violet emission and bright blue knots, which must be masked for the SBF measurement) embedded within an older, redder stellar population.} 

We explore this CSP scenario using a simple ``double-burst'' model composed of two SSPs. The primary SSP is old with an age of 12.5~Gyr and [Fe/H] = $-1.5$, and the secondary SSP is very young with an age of 200~Myr and [Fe/H] = $-1$. We then vary the contribution from the secondary population from 0 - 50\% to generate a sequence of double-burst models. The total fluctuation luminosity is calculated as
\begin{equation}\label{eqn:double-burst}
	\overline{L}	 = \frac{ \beta \sum_j n_j\,L^2_{j,\,\mathrm{young}} + (1 - \beta)\,\sum_k n_k\,L^2_{k,\,\mathrm{old}}}{\beta \sum_j n_j\,L_{j,\,\mathrm{young}} + (1 - \beta)\,\sum_k n_k\,L_{k,\,\mathrm{old}}},
\end{equation}
where the old and young subscripts refer to the two subpopulations, and $\beta \in [0,\, 0.5]$ is the fraction of stars contributed by a second instantaneous burst of star formation  --- in other words, $\beta$ parameterizes the scale of the second burst of star formation compared to the stellar population that formed at early times.  

The double-burst models are indicated by the red lines in the bottom row of Figure~\ref{fig:compare-with-obs}. The agreement with the observed CFHT relation is very good across the entire color range, and it is generally much better than that of \johnny{pure SSPs. Although dwarf galaxy formation histories are certainly more complex and diverse than the double-burst scenario, this simple model nevertheless demonstrates that composite populations can reproduce the observed fluctuation-color relation of dwarf galaxies.}

\johnny{To explore the sensitivity of the theoretical predictions to star formation history, we generated a set of constant star formation models} using \code{PCMDPy} \citep{Cook:2019}. The dashed black lines in the bottom row of Figure~\ref{fig:compare-with-obs} show the predictions for a stellar population that has undergone constant star formation, where we vary the time since the star formation ended ($t_\mathrm{end}$). The black squares indicate where the star formation ended $10^{10},\ 10^8,\ 10^7,$ and $10^{6.5}$ years ago. For all the constant star formation models, we assume a metallicity of $\mathrm{[Fe/H] = -1.5}$; we verified that our results are not significantly changed if we instead assume a Gaussian metallicity distribution \johnny{centered at $\mathrm{[Fe/H] = -1.5}$ with a width of 0.2~dex.}

For the models considered here, the overall shape of the observed CFHT relation is best matched by the double-burst model, particularly in the color range $0.25 \lesssim g-i \lesssim 0.65$. There are, however, regions of parameter space that are degenerate between the double-burst, constant star formation, and SSP models, especially given the relatively large observational uncertainties. This degeneracy between different star formation histories is encouraging for measuring SBF distances to blue stellar populations. On the other hand, it limits the stellar population information content that can be extracted from such systems using luminosity fluctuations.

Composite stellar populations have previously been invoked to explain SBF observations \citep[e.g.,][]{Tonry:1990, Liu:2002}. For more massive early-type systems, three-component models from \citet{Blakeslee:2001aa} reproduce the observed $I$-band SBF slope and are consistent with infrared SBF observations \citep{Jensen:2003}. For dwarf spheroidals, \citet{Jerjen:2000a} adopted a two-component model composed of an old, metal poor population and an intermediate age population of solar metallicity that contributes 10-30\% of the signal. These authors identified two ``branches'' --- one linear branch and one curved branch --- in the observed and theoretical fluctuation-color planes. They suggest the curved branch is due to very old, metal poor populations, and the linear branch requires increasing fractions of younger stars. 

The dwarf spheroidal sample of \citet{Jerjen:2000a} spanned the color range $1<B-R<1.3$, which for a metal-poor SSP of age 10~Gyr, corresponds to approximately $0.6<g-i<0.9$ in the CFHT filter system. While we are exploring much bluer colors and younger stellar populations than \citet{Jerjen:2000a}, our double-burst model results are also consistent with the fluctuation-color relation for dwarf galaxies consisting of two branches. The first branch is linear and is well-described by old SSPs (not shown in Figure~\ref{fig:compare-with-obs}), and the second branch curves away from an extrapolation of the linear branch towards fainter SBF magnitudes and is associated with composite stellar populations.

\begin{figure}[t!]
    \centering
    \includegraphics[width=\columnwidth]{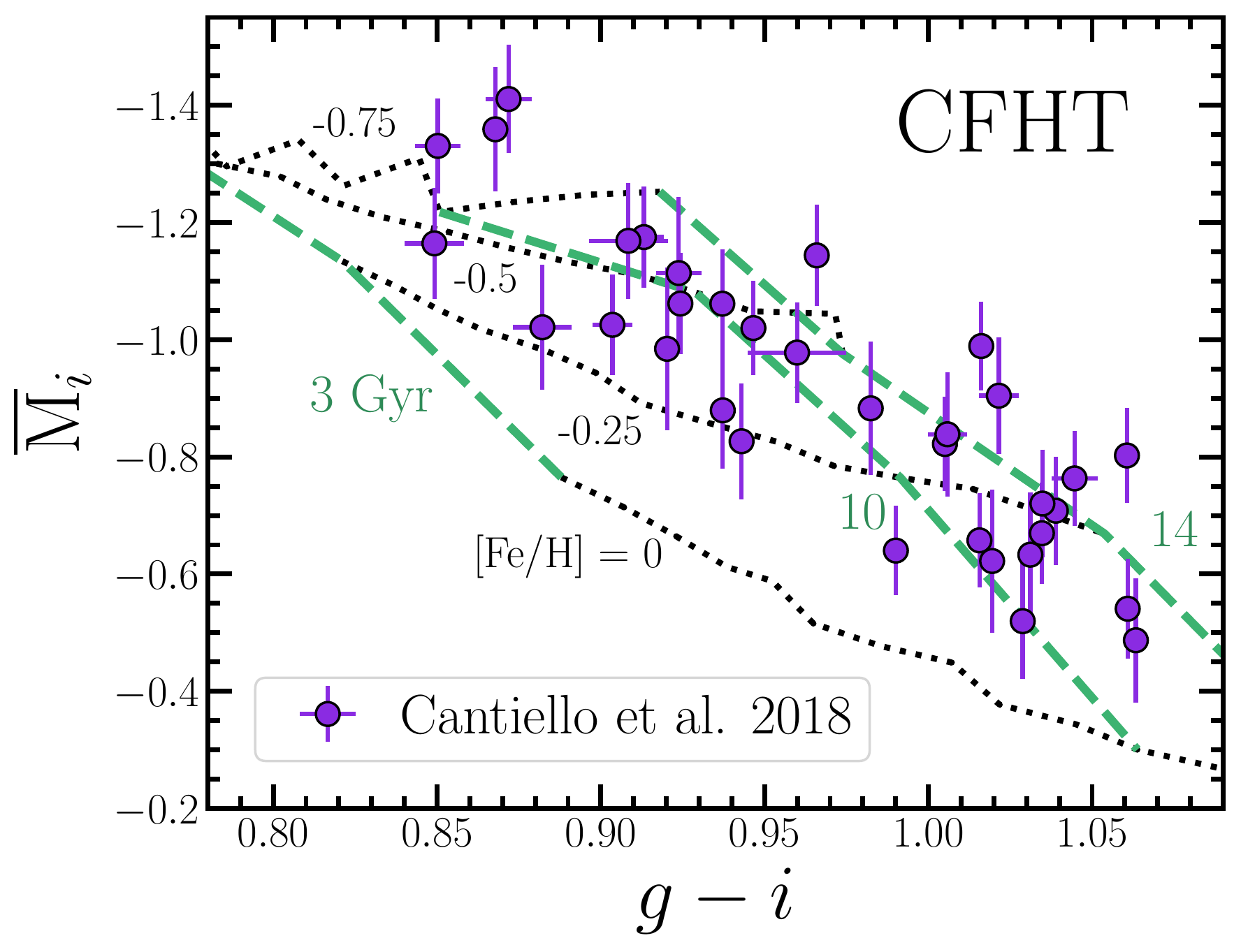}
    \caption{\resp{Comparison of MIST simple stellar population models (lines) with the observed CFHT $i$-band SBF vs. $g-i$ color relation for luminous red galaxies in the Virgo cluster \citep[purple circles;][]{Cantiello:2018aa}. Dotted black lines show models of constant metallicity (with $\mathrm{3\ Gyr \leq Age \leq 14\ Gyr}$), and dashed green lines show models of constant age (with $\mathrm{-0.75 \leq [Fe/H] \leq 0}$). The model parameters are indicated in the figure. While the MIST models show good agreement with the data, the inferred metallicities are lower than expected for this galaxy sample. We speculate this issue is related to the RGB temperature in high-metallicity MIST models, which will be addressed in future work.}}
    \label{fig:old-pops-compare}
\end{figure}

\begin{figure*}[ht!]
    \centering
    \includegraphics[width=\textwidth]{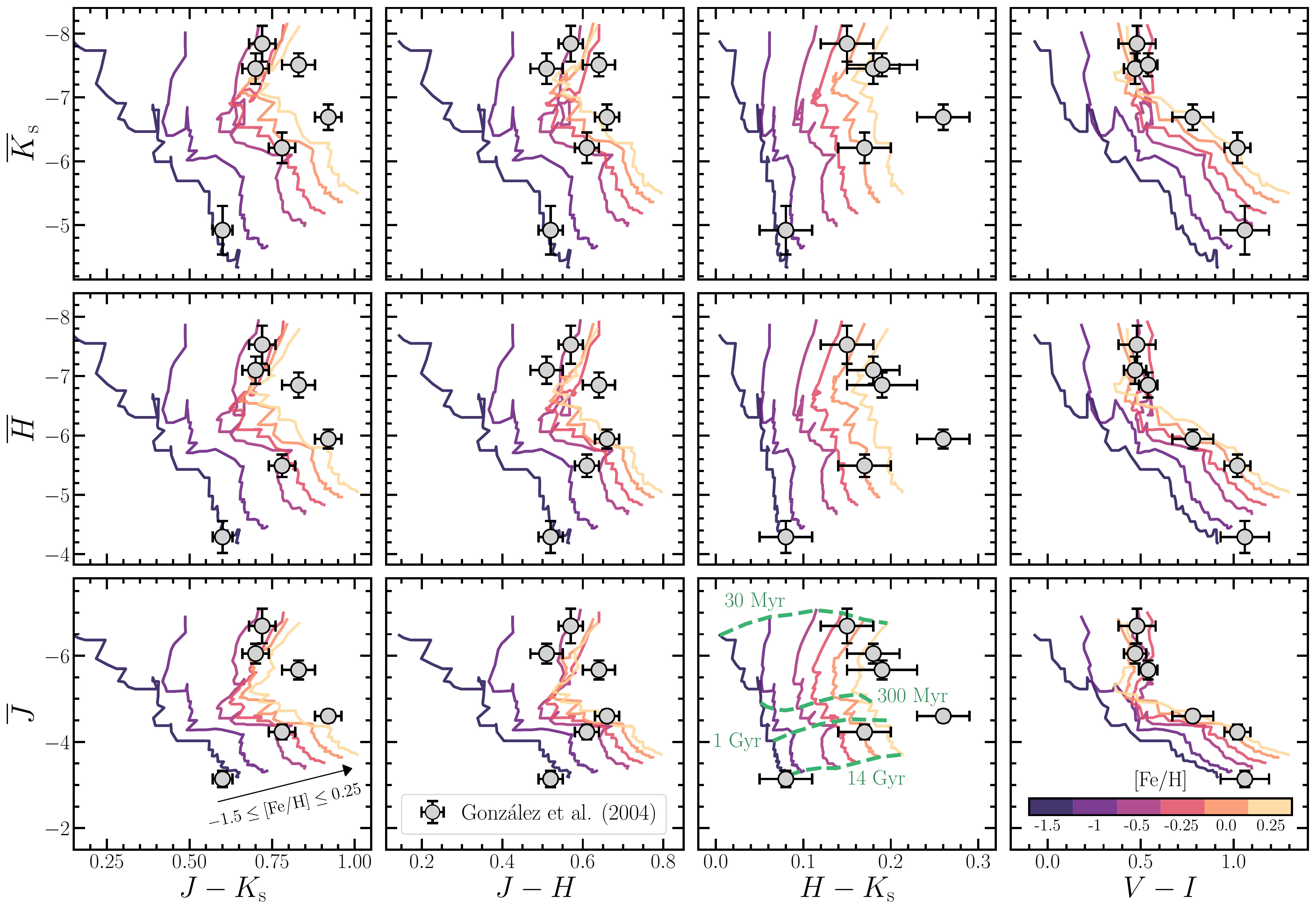}
    \caption{\resp{MIST near-infrared SBF-color relations compared with stacked 2MASS observations of star clusters in the Magellanic Clouds. The infrared SBF and color data are from \citet{Gonzalez:2004}, and the optical color data are from the compilation of \citet{Gonzalez:2005}. The legend, color bar, and general trends of the annotations in the bottom row apply to all panels. The stacked star clusters have previously estimated ages between ${\sim}10$~Myr and 14~Gyr and a mean metallicity of ${\sim}0.3$-$0.5\times$~solar \citep{Cohen:1982, Gonzalez:2004}. The infrared SBF-color relations disentangle the age-metallicity degeneracy, particularly the relations with $H-K_\mathrm{s}$ colors. However, while the agreement between the data and the model infrared colors are satisfactory for very young and old populations, the models are ${\sim}0.05$-$0.1$~mag bluer than the observations of intermediate-age star clusters, an effect that has also been noticed in other stellar population synthesis models \citep[e.g.,][]{Conroy-Gunn-2010}.}}
    \label{fig:compare-IR}
\end{figure*}

\subsection{\resp{Optical SBFs in Luminous Red Galaxies}}

\resp{Although our primary focus is SBFs in low-luminosity galaxies, it is still useful to compare our models with SBF observations of more luminous systems, which tend to be redder and more metal-rich. In Figure~\ref{fig:old-pops-compare}, we compare the MIST models with the observed CFHT $i$-band vs. $g-i$ color relation for luminous red galaxies in the Virgo cluster \citep{Cantiello:2018aa}. To convert the apparent fluctuation magnitudes to absolute units, we use the SBF distance measurements from \citet{Mei:2007}, which were derived from high-quality HST data from the ACS Virgo Cluster Survey \citep{Cote:2004}. The magnitudes have been corrected for Galactic extinction using the $E(B-V)$ values from the dust map of \citet{Schlegel:1998aa} and the recalibration from \citet{Schlafly:2011aa}.
}

\resp{In Figure~\ref{fig:old-pops-compare}, dotted black lines show models of constant metallicity, and dashed green lines show models of constant stellar population age. The MIST models show good agreement with the observations, with ages that are consistent with expectations for this sample. However, the MIST metallicities are lower than those inferred by other models \citep[e.g.,][]{Cantiello:2003aa, Raimondo:2005} by ${\sim}0.5$~dex. We suspect that this discrepancy is related to the RGB temperature in high-metallicity MIST models being ${\sim}50$-100~K cooler than other widely used models \citep[see Figure 16 of][]{Choi:2016}. This issue will be addressed in a future version of MIST. We note that the models are in better agreement at the metallicities that are expected for low-luminosity systems ($-2 \lesssim \mathrm{[Fe/H]} \lesssim -1$).}

\subsection{\resp{Near-Infrared SBFs in Star Clusters}}
\resp{As shown in previous work \citep[e.g.,][]{Worthey:1993, Jensen:1998} and Section~\ref{sec:sbf-color-relation}, the large luminosities and red colors of giant stars makes SBFs much more luminous in the infrared than the optical. Infrared observations  thus provide a powerful test for theoretical SBF calculations. In Figure~\ref{fig:compare-IR}, we compare our models with 2MASS observations of star clusters in the Large and Small Magellanic Clouds. \citet{Gonzalez:2004} carried out the SBF and color measurements by stacking the star clusters in bins of age to reduce the impact of stochastic stellar population effects (e.g., Section~\ref{sec:imf-sample}). We show SBF-color relations using both near-infrared (first three columns) and optical (rightmost column; \citealt{Gonzalez:2005}) colors. The star clusters span a wide age range of ${\sim}10$~Myr to 14~Gyr, with a low average metallicity of ${\sim}0.3$-$0.5\times$~solar \citep{Cohen:1982}, making this an excellent comparison data set for our models. }

\resp{For the infrared SBF vs. $V-I$ relations in the rightmost panel, MIST is in excellent agreement with the observations. The inferred SSP parameters are consistent with expectations based on previous work \citep[e.g.,][]{Cohen:1982, Gonzalez:2004}---the oldest star clusters have an average metallicity of $\mathrm{[Fe/H]\approx-1.5}$, and younger clusters have an average metallicity of roughly  half the solar value. For near-infrared colors, however, the results are mixed. MIST is in good agreement with observations of the youngest and oldest star clusters, but its predicted near-infrared colors are ${\sim}0.05$-$0.1$~mag bluer than observations of intermediate-age star clusters. This discrepancy is particularly clear in the third column, which shows $H-K_\mathrm{s}$ colors.}

\resp{One possible explanation for the near-infrared color offset may be that MIST under-predicts the number AGB stars in SSPs with the parameters considered here. \citet{Choi:2016} showed that MIST does indeed under-predict the number of AGB stars in the Large Magellanic Cloud, but its predictions are in relatively good agreement with the observed AGB population of the Small Magellanic Cloud. This issue will be explored in more detail in a future version of MIST. For now, we caution that---at intermediate SSP ages---the near-infrared colors presented in this work are likely too blue, and if the reason is related to under-predicting the number of AGB stars, this will impact our calculations of the SBF contributions from AGB stars in Section~\ref{sec:sbf-signal}.}

\resp{The near-infrared color discrepancies notwithstanding, it is worth noting the clear separation of age and metallicity in near-infrared SBF vs. $H-K_\mathrm{s}$ parameter space, where the SBF magnitudes depend mostly on metallicity and the colors depend mostly on age. The potential of near-infrared SBF measurements to break the age-metallicity degeneracy is well-known \citep[e.g.,][]{Blakeslee:2001aa}, and such measurements will likely play a major role in extragalactic studies in the RST era.}

\section{Image Simulations} \label{sec:sims}

To study the measurement of SBFs in practice, we use image simulations in which dwarf galaxies are built star-by-star from the faint end of main sequence to the most luminous phases of stellar evolution. \citet{Tonry:1988} used conceptually similar image simulations \citep{Bahcall:1988} to demonstrate the efficacy of their novel method, though their simulations were necessarily limited. More recently, SBFs have generally been simulated by injecting Poisson variance into mock galaxies with smooth surface brightness distributions \citep[e.g.,][]{Mieske:2003, Mei:2001sims, Mei2005sbfsims, Cantiello:2007aa, Carlsten:2019aa}.

Since we are interested in low-luminosity galaxies composed of ${\sim}10^5$ - 10$^8$ stars, it is computationally feasible to build a large number of synthetic galaxy images one star at a time. This means that SBFs in our simulations are exactly analogous to real observations. Furthermore, by selectively injecting stars, we can straightforwardly disentangle the SBF and luminosity signals from various phases of stellar evolution and use our mock images to build intuition for how stellar populations appear in real observations with arbitrary galaxy properties and observing conditions. In this work, we focus on ground-based observing conditions that are similar to what is expected to be typical for LSST.

\subsection{ArtPop} \label{sec:artpop}

We create our image simulations using an updated version of the code \code{ArtPop}, which was introduced in \citet{Danieli:2018aa}. \johnny{An expanded and generalized implementation of the code will be made publicly available} in the near future (J. Greco \& S. Danieli, in preparation). Here, we give a brief overview of relevant aspects of the code.

\code{ArtPop} takes the following parameters as input: the total stellar mass or number of stars to be simulated, stellar population \johnny{and galaxy morphology parameters}, the distance to the stellar system, the pixel scale, the photometric system, the point-spread function (PSF), \johnny{and various other parameters that control the noise and count level in the image.} 

Given a set of observational parameters, \code{Artpop} calculates stellar magnitudes based on the MIST isochrones using the procedure described in Section~\ref{sec:pop-generation}. Stellar magnitudes in bandpass $x$ are converted to photon counts using the analytic expression 
\begin{equation}\label{eqn:counts}
	C_x = A\,\frac{t_\mathrm{exp}}{h}\frac{\Delta \lambda_x}{\lambda_{\mathrm{eff,}\,x}}10^{-0.4\,(m_x\,+\, 48.6)},
\end{equation}
where $A$ is the collecting area of the telescope, $t_\mathrm{exp}$ is the exposure time, $h$ is the Planck constant, $\Delta \lambda_x$ and $\lambda_{\mathrm{eff,}\,x}$ are the width and effective wavelength of bandpass $x$, and $m_x$ is the stellar AB  magnitude in bandpass $x$. 

\code{Artpop} then injects the stars into an image according to a user-specified spatial distribution and convolves with the PSF. Finally, noise is added to the image. For a \johnny{fully artificial} observation, Poisson noise is generated from the combined counts of the source and sky, and the read noise is assumed to be Gaussian. If \johnny{instead} the galaxy is injected into a real image, Poisson noise is optionally generated from the source counts before converting into the image flux units, provided the necessary parameters for Equation~(\ref{eqn:counts}) are given as input.

\begin{figure*}
	\centering
	\gridline{\fig{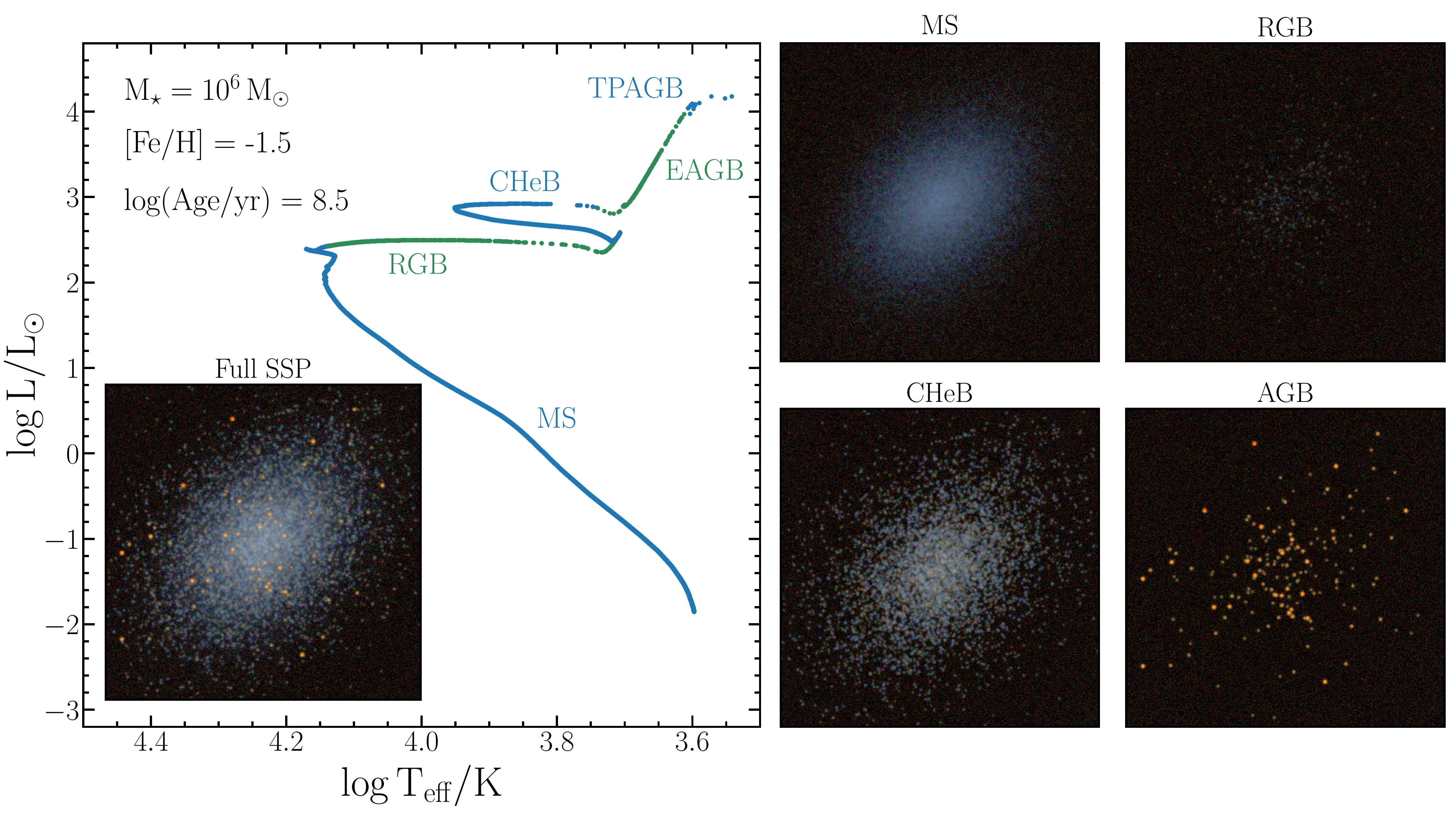}{\textwidth}{}}
	\vspace{-1.1cm}
	\gridline{\fig{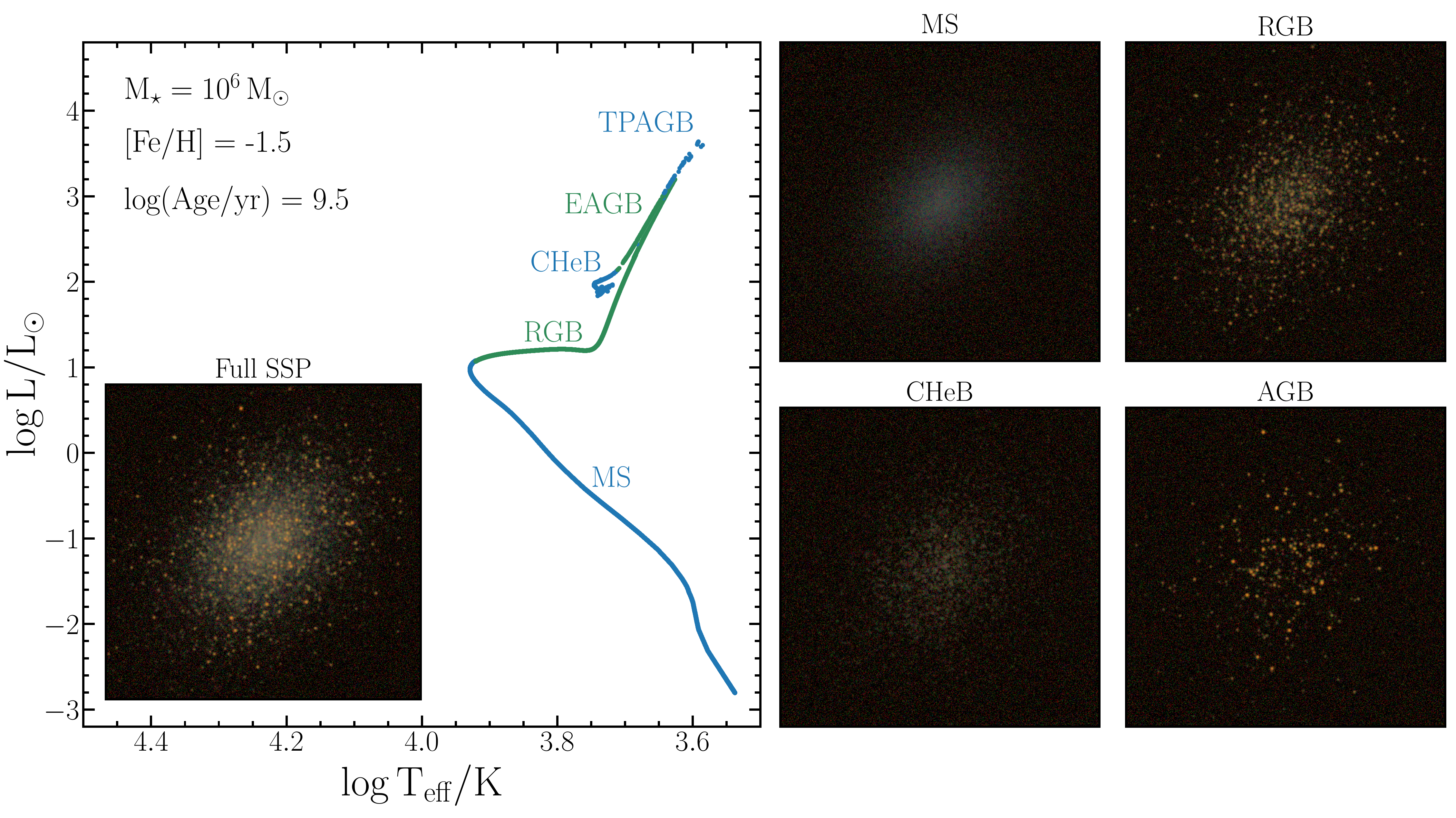}{\textwidth}{}}
	\vspace{-0.5cm}
	\caption{Simulated $gri$-composite images of a dwarf galaxy at 5~Mpc with the indicated stellar mass, metallicity, and ages. The HR-diagrams show the luminosities and temperatures of the stars that were injected into the corresponding mock images. For each dwarf, the four panels on the right deconstruct the full SSP image (inset in left panel) into the four indicated phases of stellar evolution, where the RGB includes the subgiant branch and the AGB includes the early and thermally-pulsating AGB (EAGB and TPAGB, respectively). The evolutionary phases that dominate the fluctuations are clearly visible by eye; CHeB stars dominate the signal of the young population (top), and RGB stars dominate that of the intermediate-age population (bottom). AGB stars contribute non-negligible fluctuations at both ages and are easily isolated in the young galaxy due to the high contrast between their red color and the blue light from main sequence (MS) and CHeB stars. The simulations were tuned to resemble an LSST-like observatory and observing conditions. See Section~\ref{sec:phases} for a note about how we label stellar evolutionary phases.}
	\label{fig:sims-hr}
\end{figure*}

In Figure~\ref{fig:sims-hr}, we show example \code{ArtPop} mock observations of two dwarf galaxies at a distance of 5~Mpc. The dwarf galaxies are both composed of an SSP with stars distributed according to a S\'{e}rsic surface brightness distribution \citep{Sersic:1968aa} with $n=0.8$. All their physical properties are the same except one (top) is very young with $\log(\mathrm{Age/yr}) = 8.5$ and the other (bottom) is of intermediate age with $\log(\mathrm{Age/yr}) = 9.5$. The stellar mass, metallicity, and effective radius of both galaxies are M$_\star = 10^6~M_\odot$, $\mathrm{[Fe/H] = -1.5}$, and $r_\mathrm{eff} = 550$~pc, respectively. The mock observations are tuned to resemble an LSST-like observatory (see Section~\ref{sec:ideal}). The cutouts are $gri$-composite images \citep{Lupton:2004aa} with exposure times of 1.5~hr in $i$ and 45~min in $g$ and $r$.

For each dwarf, the four images on the right deconstruct the full SSP image, which is shown in the inset of left panel, into the four indicated phases of stellar evolution. As throughout this work, the RGB includes the subgiant branch, and the AGB includes the early and thermally-pulsating AGB --- see Section \ref{sec:phases} for a note about how we label stellar evolutionary phases. The Hertzsprung–Russell diagrams on the left show the temperatures and luminosities of the stars that were injected into the corresponding mock images. 

The images in Figure~\ref{fig:sims-hr} are excellent intuition builders for interpreting semi-resolved images of dwarf galaxies. At a glance, we see the smooth S\'{e}rsic distribution of the blue main sequence and SBFs dominated by CHeB stars at early times and the RGB at later times, consistent with our calculations in Figure~\ref{fig:sig-frac}. The AGB is the most luminous and rarest phase in both galaxies and thus contributes non-negligible luminosity fluctuations that are clearly visible in the full SSP images. Intriguingly, AGB stars are easy to identify in the younger galaxy due to the high contrast between their red colors and the blue light from the main sequence and CHeB stars, which suggests ground-based surveys with LSST-quality imaging have the potential to shed light upon this rare phase of stellar evolution using nearby dwarf galaxies.

\subsection{Dwarf Galaxy Physical Parameters}\label{sec:params}
 
In the sections that follow, we simulate thousands of dwarf galaxies with stellar masses ranging from the ultra-faint ($10^4\leq\mathrm{M}_\star \leq 10^5$~M$_\odot$) to low-mass classical (M$_\star\leq 10^7$~M$_\odot$) regimes. The galaxies are assumed to be pure SSPs with S\'{e}rsic surface brightness distributions. For all galaxies, we use an ellipticity of 0.3 and S\'{e}rsic index of $n=0.8$, consistent with the median values of LSB galaxies observed with the Hyper Suprime-Cam Survey \citep{Greco2018cat}. 

For a given stellar mass, we assume empirical scaling relations to derive effective radii and stellar metallicities. For effective radii, we use the mass--size relation compiled by \citet{Danieli:2018aa}. Since we are particularly interested in diffuse stellar systems, we use radii that are larger by 1$\sigma=0.32$~dex than this relation:
\begin{equation}\label{eqn:mass-size}
	\log(r_\mathrm{eff} / \mathrm{kpc}) = 0.23 \log(\mathrm{M_\star / M_\odot}) - 1.93 + 0.32.
\end{equation}
For stellar metallicity, we use the mean mass--metallicity relation for dwarf galaxies from \citet{Kirby:2013aa}:
\begin{equation}\label{eqn:mass-metal}
	[\mathrm{Fe/H}] = -1.69 + 0.3\log\left(\mathrm{\frac{M_\star}{10^6\,M_\odot}}\right).
\end{equation}
We note that using surface brightness to derive stellar metallicity \citep[e.g.,][]{vD-2019-DF44-KCWI} leads to metallicities that are typically higher by ${\sim}0.4$~dex compared to those derived from Equation~(\ref{eqn:mass-metal}) --- we verified that our results are not significantly impacted by this choice. 

\section{Measuring SBFs in practice}\label{sec:sbf-in-pratice}

SBFs from a semi-resolved stellar system are traditionally measured using the azimuthally-averaged power spectrum of its image, as first demonstrated by \citet{Tonry:1988}. The advantage of this method is that it cleanly separates sources of fluctuations that are PSF-convolved---such as Poisson fluctuations in the number of giant stars in each pixel---from those that are due to noise in the detector. Assuming random phases from pixel to pixel, the latter fluctuations produce white noise and thus contribute a constant offset to the power spectrum. In contrast, power from PSF-convolved sources boosts the amplitude of the PSF power spectrum.

A major challenge to this approach is that astrophysical sources such as globular clusters and background galaxies also produce fluctuations on the scale of the PSF. While the power from these sources can be modeled and subtracted under reasonable assumptions, fluctuations arising from sources like star-forming regions and/or dust lanes need to be aggressively masked. There is an extensive literature that addresses the subtleties of measuring SBFs via the power spectrum method \citep[e.g.,][]{Tonry:1988, Blakeslee:1995, Liu:2002, Mei2005sbfsims, Cantiello:2007aa, Jensen:2015}. In this section, we describe the procedure we follow to measure SBFs (Section~\ref{sec:method}), which we then apply to idealized (Section~\ref{sec:ideal}) and realistic (Section~\ref{sec:realistic}) mock dwarf galaxy images. 

\subsection{The Power Spectrum Method} \label{sec:method}

In outline, the major steps for measuring SBFs in a galaxy image are: (1) model and subtract the galaxy's mean surface brightness distribution, and divide the residual image by the square root of the model to normalize the scale of fluctuations; (2) mask sources of fluctuations that are not due to stellar SBFs arising in the galaxy of interest; (3) Fourier transform the masked residual image and calculate the azimuthally-averaged power spectrum, from which the variance of fluctuations may be determined as described below; (4) depending on the image quality, data reduction procedure, and the nature of the target galaxy, it may also be necessary to subtract the additional signal from contaminating sources such as globular clusters, faint background galaxies, and/or correlated noise from resampling.

Assuming steps (1) through (3) have been completed, the azimuthally-averaged power spectrum of the normalized residual image is modeled as
\begin{equation} \label{eqn:Pk}
	P(k) = \sigma_\mathrm{SBF}^2 \times E(k) + \sigma_\mathrm{WN}^2, 
\end{equation}
where $\sigma_\mathrm{SBF}^2$ is the variance from PSF-convolved sources (the desired signal), $\sigma_\mathrm{WN}^2$ is the white (uncorrelated) noise variance associated with the detector, and $k$ is the spatial frequency in units of inverse pixels. The multiplicative factor $E(k)$ is the azimuthally-averaged ``expectation power spectrum'' \citep{Tonry:1990}, which is given by the two-dimensional convolution of the power spectra of the PSF and object mask:  
\begin{equation}\label{eqn:Ek}
	E(k_x,\, k_y) =  \left|\mathrm{PSF}(k_x,\, k_y)\right|^2 \circledast \left|\mathrm{M}(k_x,\, k_y)\right|^2.
\end{equation}
We zero-pad the PSF before calculating its Fourier transform to increase its frequency sampling rate to match that of the object mask. To solve Equation (\ref{eqn:Pk}) for $\sigma_\mathrm{SBF}^2$ and $\sigma_\mathrm{WN}^2$, we use standard non-linear least squares fitting implemented in the \code{scipy} software package.
 
The recovered fluctuation magnitude is sensitive to the range of $k$ values over which the fit is performed \citep[e.g.,][]{Mieske:2003, Cantiello:2005}. At low $k$ (large spatial scales), residual structure from imperfect galaxy subtraction often compromises the expected power spectrum. At large $k$ (small spatial scales), correlated noise between pixels can lead to spurious fits \citep{Mei2005sbfsims}. This issue is usually addressed on a per-object basis, since it depends on the quality of the galaxy subtraction, or by iteratively fitting the power spectrum while varying the range of $k$ values to quantify the statistical uncertainty of the fit \citep[e.g.,][]{Cohen:2018}. 

If astrophysical sources such as globular clusters and/or unresolved background galaxies are also present, then their residual variance $\sigma^2_r$ must be subtracted from $\sigma_\mathrm{SBF}^2$. This may be done analytically by assuming or measuring the form of their luminosity function(s). It is also possible to empirically measure residual variance from contaminating sources\resp{, which are {\it not} associated with the target galaxy,} by measuring the SBF signal in blank fields near the source of interest \citep{Carlsten:2019aa}. The apparent fluctuation magnitude is then given by
\begin{equation}\label{eqn:mbar}
	\overline{\mathrm{m}} =  m_\mathrm{zpt} - 2.5 \log\left(\frac{\sigma_\mathrm{SBF}^2 -\sigma^2_r}{N_\mathrm{pix}}\right), 
\end{equation}
where $N_\mathrm{pix}$ is the number of unmasked pixels, and $m_\mathrm{zpt}$ is the zero point magnitude. Dividing by $N_\mathrm{pix}$ is necessary because $\sigma_\mathrm{SBF}^2$ is equal to the sum of the pixel variances, whereas we are interested in the per-pixel luminosity fluctuations, which we have normalized to have a constant amplitude across the residual image.

\begin{figure*}[th!]
    \centering
    \includegraphics[width=0.95\textwidth]{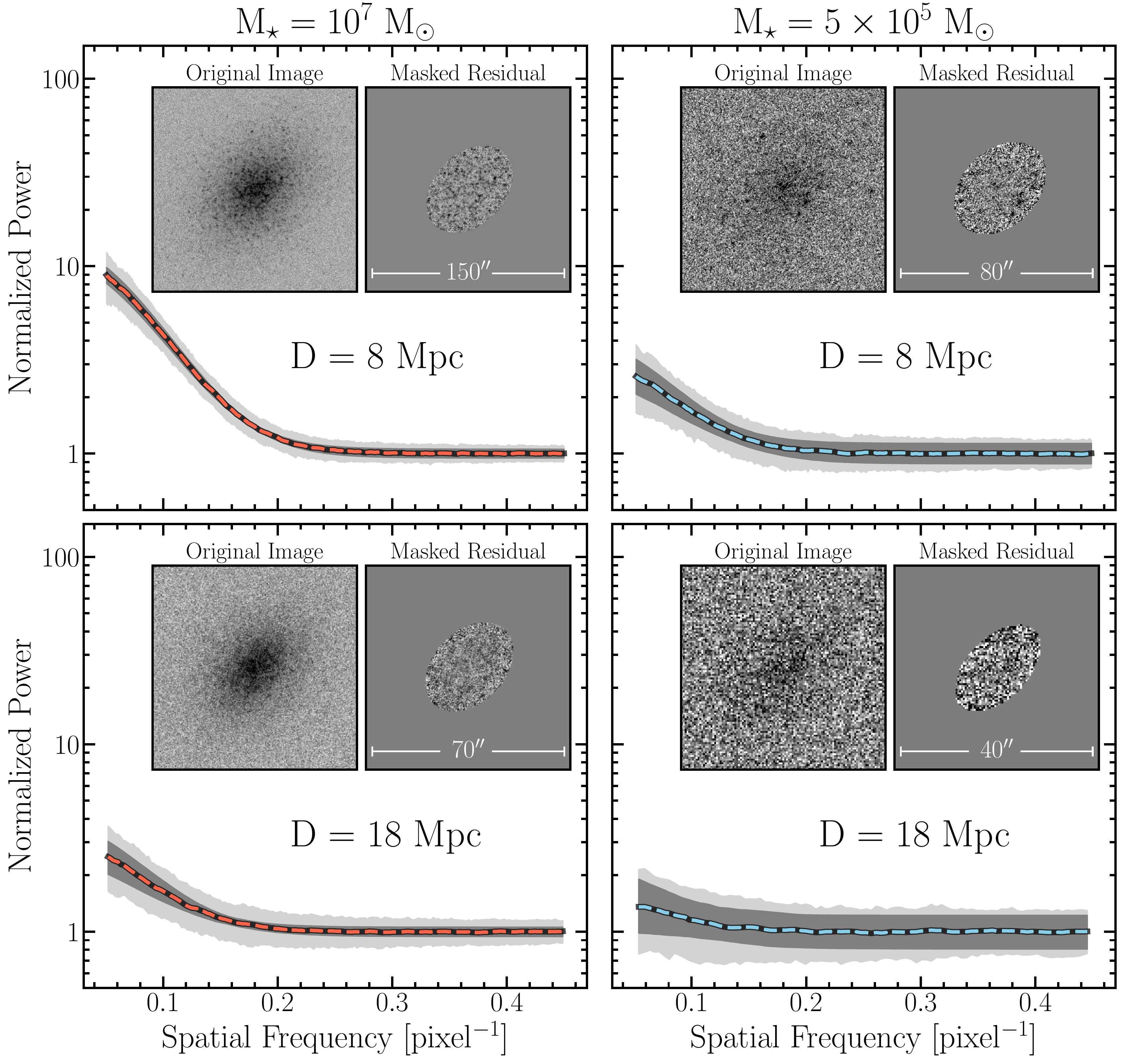}
    \caption{Summary of our fits to the power spectra of mock galaxies with old stellar populations (12.6~Gyr) and stellar masses of $10^7$~M$_\odot$ (left column) and $5\times10^5$~M$_\odot$ (right column). Each panel shows power spectra from 500 realizations of the galaxy shown in the inset images, which show the original image and the masked residual image from which we measure the SBF signal. The light gray shaded regions \johnny{show the 5th - 95th percentile of the ``observed'' data points, and the dark gray regions indicate the 5th - 95th percentile of the best-fit models.} The median best-fit models are indicated by the colored dashed lines. The galaxies are placed at 8 Mpc (top row) and 18 Mpc (bottom row). The mock observation parameters are tuned to resemble 2-year stacks from LSST, as described in Section~\ref{sec:ideal-setup}.}
    \label{fig:power-spectra}   
\end{figure*}

\begin{figure}[th!]
    \centering
    \includegraphics[width=\columnwidth]{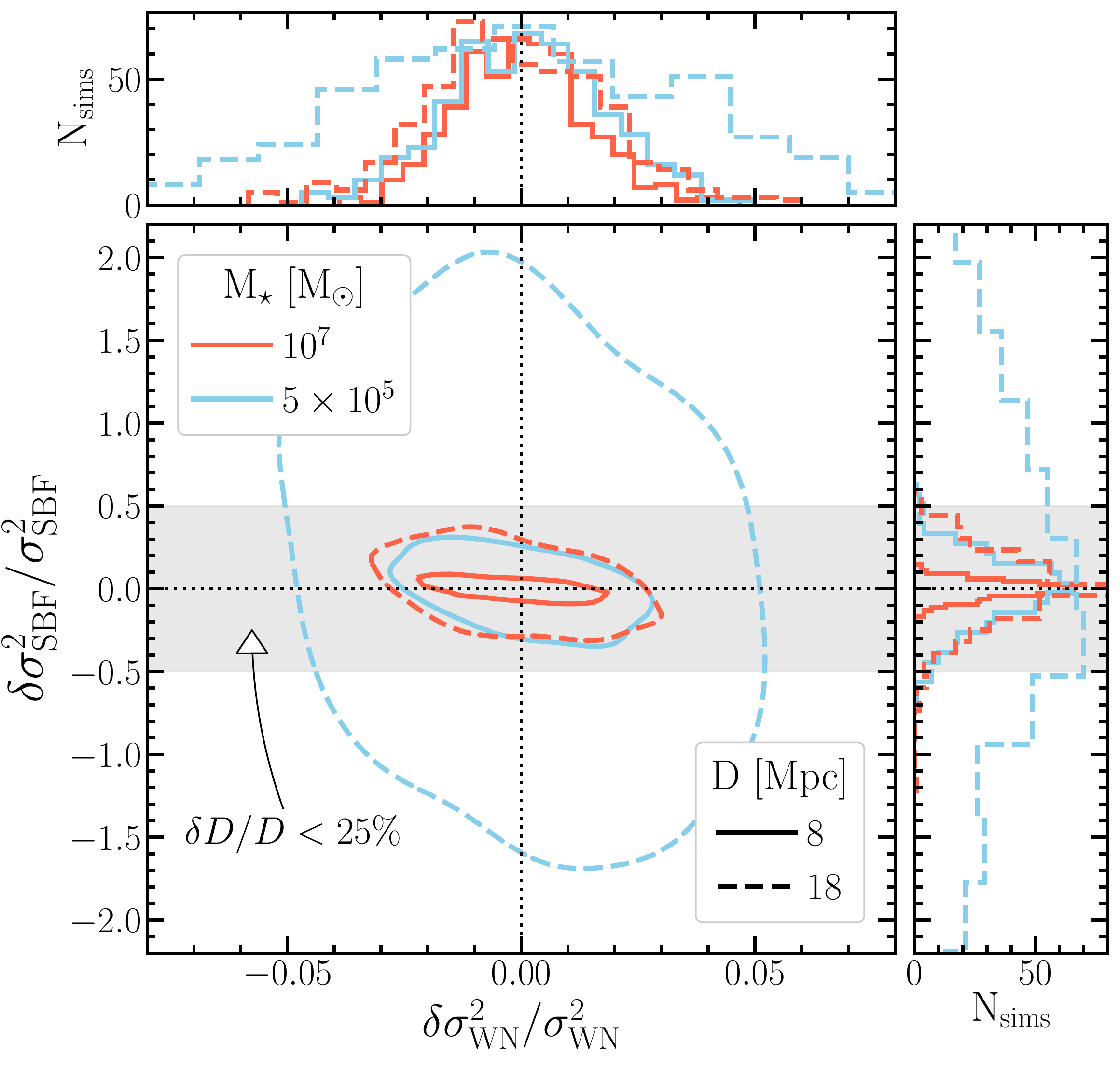}
    \caption{Parameter covariances from the power spectra shown in Figure~\ref{fig:power-spectra}. The parameters are defined in Equation~(\ref{eqn:Pk}), and the differences are defined as $\delta x\equiv x_\mathrm{truth} - x_\mathrm{recovered}$. The lines in the main panel show the 68th percentile contours from 500 simulations for each combination of mass and distance. The histograms on the top and right show the marginalized distributions, and the gray shaded region indicates where the fractional distance error is less than 25\%.}
    \label{fig:delta-params}
\end{figure}

\subsection{SBF Recovery from Idealized Simulations}\label{sec:ideal}

For a known stellar population, how well does $\sigma_\mathrm{SBF}^2$ in Equation (\ref{eqn:Pk}) measure $\mathrm{\overline{L}}$ in Equation (\ref{eqn:Lbar})? \citet{Tonry:1988} explored this question using three simulations of a galaxy at different distances and found a ${\sim}3$-10\% bias in their recovered distances, which they attributed to truncation of the PSF. Here, we revisit this question using idealized \code{ArtPop} simulations, where the target galaxy is the only source in the image and we have {\it perfect} knowledge of both the signal and noise. 

\resp{Throughout this work, we discuss the precision of recovered SBF magnitudes in terms of both magnitude differences and fractional distance errors, which are simply related by $\delta\overline{m} \simeq 2.17\delta \mathrm{D/D}$. Our reported fractional distance errors should therefore be interpreted as a lower limit on the (potentially much larger) total uncertainty.}

\subsubsection{Idealized Simulations: Setup}\label{sec:ideal-setup}

We simulate many realizations of galaxies with old stellar populations (12.6~Gyr) and stellar masses of $10^7$~M$_\odot$ and $5\times10^5$~M$_\odot$, with physical parameters given by the relations in Section~\ref{sec:params}. We inject the galaxies into mock observations that are tuned to resemble an LSST-like observatory with an effective aperture of 6.4~m, pixel scale of 0\farcs2~pixel$^{-1}$, and read noise of 4~electrons \citep{abell2009lsst, LSST-ref-design-2019}. The data are assumed to be sky-noise dominated, with an $i$-band sky brightness of $19.9~\mathrm{mag\ arcsec^{-2}}$, which is typical for dark time near Cerro Pach\'{o}n \citep[e.g.,][]{DES-DR1-2018}. The sky background is constant over the extent of each galaxy, and the sky subtraction is perfect. We model the seeing as a Moffat profile with $\beta=4.765$ \citep{Moffat1969, Trujillo2001moffat} and FWHM of 0\farcs7.

The galaxies are ``observed'' at two distances, 8 Mpc and 18 Mpc, in LSST's $i$-band with an exposure time of 18.4~minutes. These observing parameters produce data with a depth that is similar to what is expected from LSST stacks after 2 years of its deep-wide-fast survey, assuming the calculations and mean number of visits per field from \citet{LSST-ref-design-2019}. 

For each mock galaxy, we \johnny{ignore our knowledge of the true morphology and} fit a 2D S\'{e}rsic function to its mean surface brightness distribution using \code{imfit} \citep{Erwin:2015aa}. \johnny{The best-fit model is used} to generate the normalized residual image, as described in Section~\ref{sec:method}. Since the target galaxy is the only source in the image, step (2) in Section~\ref{sec:method} is not necessary.  

We perform the SBF measurement \johnny{within elliptical apertures} with a range of semi-major axes. There is a trade-off between the area (and hence the number of stars) covered by the measurement and the amount sky noise allowed into the aperture, which is much more dominant in the galaxy outskirts. For these idealized simulations, we find an elliptical aperture with a semi-major axis of $1\times r_\mathrm{eff}$ provides a good a balance between signal and noise. \johnny{The ellipse parameters are derived from the \code{imfit} model.}

Finally, we apply the methods of Section~\ref{sec:method} to recover the apparent fluctuation magnitude from the power spectrum of the normalized residual image. Note that the residual variance from contaminating sources is zero in these simulations. To fit Equation~(\ref{eqn:Pk}), we use a fixed spatial frequency range of $0.05 < k < 0.45$, where the lower bound corresponds to ${\sim}5\times$ the FWHM of the PSF, and the upper bound is near the Nyquist frequency.

\subsubsection{Idealized Simulations: Results}\label{sec:ideal-results}

In Figure~\ref{fig:power-spectra}, we show a summary of our fits to the power spectra. Each panel shows results from 500 realizations of the galaxy shown in the inset images. The light gray shaded region of each power spectrum indicates \johnny{the 5th - 95th percentile of the measured data points, and the dark gray region indicates the 5th - 95th percentile of the best-fit models.} The colored dashed lines indicate the median best-fit models. The top row shows the power spectra for the galaxies at 8~Mpc, and the bottom row shows the power spectra for the galaxies at 18~Mpc. The higher mass ($10^7$~M$_\odot$)  galaxy results are shown in the left column, and the lower mass ($5\times10^5$~M$_\odot$) galaxy results are shown in the right column.

The peak of the power spectrum relative to the constant offset, which we have normalized to unity, is a measure of the signal-to-noise ratio of the measurement. From the higher mass galaxy at 8~Mpc (top left panel) to the lower mass galaxy at 18 Mpc (bottom right panel), the median best-fit power spectrum transitions from sharply peaked to nearly flat. For the former, we typically recover the distance to within a few percent, whereas it is not possible to place any meaningful constraints on the distance for the latter. 

To study the errors on the recovered parameters in more detail, we show the joint parameter covariance distribution in Figure~\ref{fig:delta-params}. The lines in the main panel show the 68th percentile contours from the 500 simulations for each combination of mass and distance, and the histograms on the top and right show the marginalized distributions. The colors of the lines correspond to those in Figure~\ref{fig:power-spectra}. The gray shaded region indicates where the fractional distance error that is less than 25\%. 

At 8~Mpc (solid lines in Figure~\ref{fig:delta-params}), the \resp{apparent SBF magnitudes of the lower and higher mass galaxy are recovered to ${\sim}0.2$ and 0.05~mag ($10\%$ and $2.5\%$ in distance), respectively,} where we calculate the typical \resp{magnitude difference} using the standard deviation of the $\sigma_\mathrm{SBF}^2$ error distribution. In both cases, the recovered SBF signal is unbiased, with a median offset of $\overline{m}_\mathrm{truth} - \overline{m}_\mathrm{recovered}  \equiv \delta\overline{m} < 0.003$~mag, which corresponds to ${<}0.14\%$ in distance. Note that the ``true'' SBF signal is calculated directly from Equation~(\ref{eqn:Lbar}) {\it for each mock galaxy realization}. This means that $\delta\overline{m}$ \resp{does not capture the uncertainty associated with} SBF sampling scatter, as shown in Figure~\ref{fig:m_rs} \resp{and Figure~\ref{fig:absmag-scatter}.}

At 18~Mpc (dashed lines in Figure~\ref{fig:delta-params}), the recovered SBF signal of the higher mass galaxy remains unbiased, but with a lower distance precision of 10\%. In contrast, the average recovered SBF signal of the lower mass galaxy is consistent with zero, resulting in an unconstrained distance and relatively flat $\sigma_\mathrm{SBF}^2$ error distribution. 

Lastly, we note the anticorrelation between the errors of $\sigma_\mathrm{SBF}^2$ and $\sigma_\mathrm{WN}^2$, which is particularly clear for the galaxy of mass $10^7$~M$_\odot$. If the white noise component is underestimated, the SBF signal is likely to be overestimated and vice versa, highlighting the importance of understanding both the signal and the noise when measuring SBFs.

\subsection{SBF Recovery from Images \resp{with Realistic Noise}}\label{sec:realistic}

\begin{table*}[t!]
\begin{center}
\caption{Artificial Galaxy Parameters}\label{tab:params} 
\begin{tabular*}{\textwidth}{@{\extracolsep{\fill}} c c c | c c c c}
\hline\hline
M$_\star$ &  $r_\mathrm{eff}$ & [Fe/H] &  M$_i$ & $\mu_0(i)$ & $\overline{\mathrm{M}}_i$ & $g-i$ \\ [0.2ex] 
\hline
[M$_\odot$] &  [kpc]  &   & [mag] & [mag arcsec$^{-2}$] & [mag] & [mag] \\ [0.4ex] 
\hline    
$1\times10^7$ &  0.93 & -1.4 & -12.7 $\pm$ 0.02 & 24.5 $\pm$ 0.004 & -1.49 $\pm$ 0.02 & 0.77 $\pm$ 0.002 \\
$5\times10^6$ &  0.80 & -1.5 & -11.9 $\pm$ 0.02 & 25.0 $\pm$ 0.007 & -1.47 $\pm$ 0.02 & 0.75 $\pm$ 0.004 \\
$1\times10^6$ &  0.55 & -1.7 & -10.2 $\pm$ 0.08 & 25.9 $\pm$ 0.018 & -1.48 $\pm$ 0.08 & 0.71 $\pm$ 0.009 \\
$5\times10^5$ &  0.47 & -1.8 & -9.5  $\pm$ 0.09 & 26.3 $\pm$ 0.027 & -1.47 $\pm$ 0.09 & 0.69 $\pm$ 0.012 \\
$1\times10^5$ &  0.32 & -2.0 & -7.7  $\pm$ 0.19 & 27.2 $\pm$ 0.059 & -1.47 $\pm$ 0.19 & 0.64 $\pm$ 0.029 \\
$5\times10^4$ &  0.28 & -2.1 & -7.0  $\pm$ 0.23 & 27.6 $\pm$ 0.084 & -1.48 $\pm$ 0.23 & 0.62 $\pm$ 0.038 \\
$1\times10^4$ &  0.19 & -2.3 & -5.4  $\pm$ 0.50 & 28.4 $\pm$ 0.180 & -1.42 $\pm$ 0.50 & 0.58 $\pm$ 0.081 \\
\hline\hline
\end{tabular*}
\end{center}
\vspace{-0.3cm}
\tablecomments{\resp{All artificial galaxies are SSPs of age 12.6~Gyr, with effective radii and metallicities derived from the scaling relations presented in Section~\ref{sec:params}. Additionally, we list the ensemble medians and standard deviations of integrated photometric parameters in the LSST filter system: the $i$-band absolute magnitude, central surface brightness, absolute SBF magnitude, and $g-i$ color.}}
\end{table*}

While the idealized simulations in the previous section are useful for testing and developing our SBF recovery methodology, they neglect several important sources of uncertainty that occur in real observations. They do not include other astrophysical sources of SBFs such as globular clusters and background galaxies. They also assume uncorrelated noise between pixels, which is incorrect if, for example, the data have been resampled, as is often the case for wide-field surveys.

In this section, our goal is to assess the feasibility of measuring SBFs in low-luminosity galaxies using LSST-like data. \resp{Importantly, we emphasize that our aim is to study, as a function of stellar mass, the limiting distance at which SBFs can be recovered from deep ground-based data in the absence of systematic uncertainties such as imperfect sky subtraction, flat-fielding errors, and calibration errors. While such systematics generally dominate the total uncertainty in the low-luminosity regime, they also have the potential to significantly improve over time as observational methods and theoretical models continue to develop and more data are acquired.}

To accomplish this goal, we inject mock galaxies into images from the second data release of the Hyper Suprime-Cam Subaru Strategic Program \citep[HSC-SSP;][]{HSC-DR2}. The data quality and reduction \citep{Bosch:2018aa} of the HSC-SSP are very similar to what is expected for LSST, and with a 5$\sigma$ point-source limit of $i_\mathrm{lim}\simeq26$, the wide-layer depth of HSC-SSP is comparable to the anticipated depth of 2-year LSST stacks. 

By injecting mocks into HSC-SSP images, realistic (correlated) noise and additional astrophysical sources observed with LSST-like depth and resolution are naturally included. \resp{It is important to note that we inject the mock galaxies into post-processed HSC-SSP images and only simulate Poisson white noise associated with the artificial source counts, as described in Section~\ref{sec:artpop}. However, the real noise in the HSC-SSP images is by far the dominant noise component, since our simulated galaxies are faint and have extremely low surface brightnesses.}

\subsubsection{Realistic\resp{-Noise} Simulations: Setup}\label{sec:realistic-setup}

We inject mock galaxies into three quasi-random fields from HSC-SSP's wide layer, avoiding regions that would render SBF measurements exceedingly difficult if not impossible such as near saturated stars or in front of dense galaxy clusters. The FWHM of the $i$-band PSFs in the selected fields ranges from ${\sim}$0\farcs5-0\farcs6, comparable to good-seeing nights at Cerro Pach\'{o}n \citep{LSST-ref-design-2019}. We obtained the PSF for each field using HSC-SSP's PSF Picker tool, which is available on the HSC-SSP data release website\footnote{\url{https://hsc-release.mtk.nao.ac.jp/doc/}}.

For our main suite of simulations, the mock galaxies are composed of old SSPs of age 12.6~Gyr, spanning a stellar mass range of $10^4 \leq \mathrm{M_\star / M_\odot} \leq 10^7$. Table~\ref{tab:params} lists the stellar mass, metallicity, effective radius, and average $i$-band central surface brightness and $g-i$ color of the simulated galaxies. The galaxy parameters are derived from the scaling relations presented in Section~\ref{sec:params}. For each stellar mass, we inject 500 realizations of the galaxy with a random position angle into the $i$-band image of each of the three HSC-SSP fields. Using our idealized simulations as guidance, we place the mock galaxies at various distances to study the limiting SBF distance as a function of stellar mass.

\begin{figure*}[ht!]
    \centering
    \includegraphics[width=\textwidth]{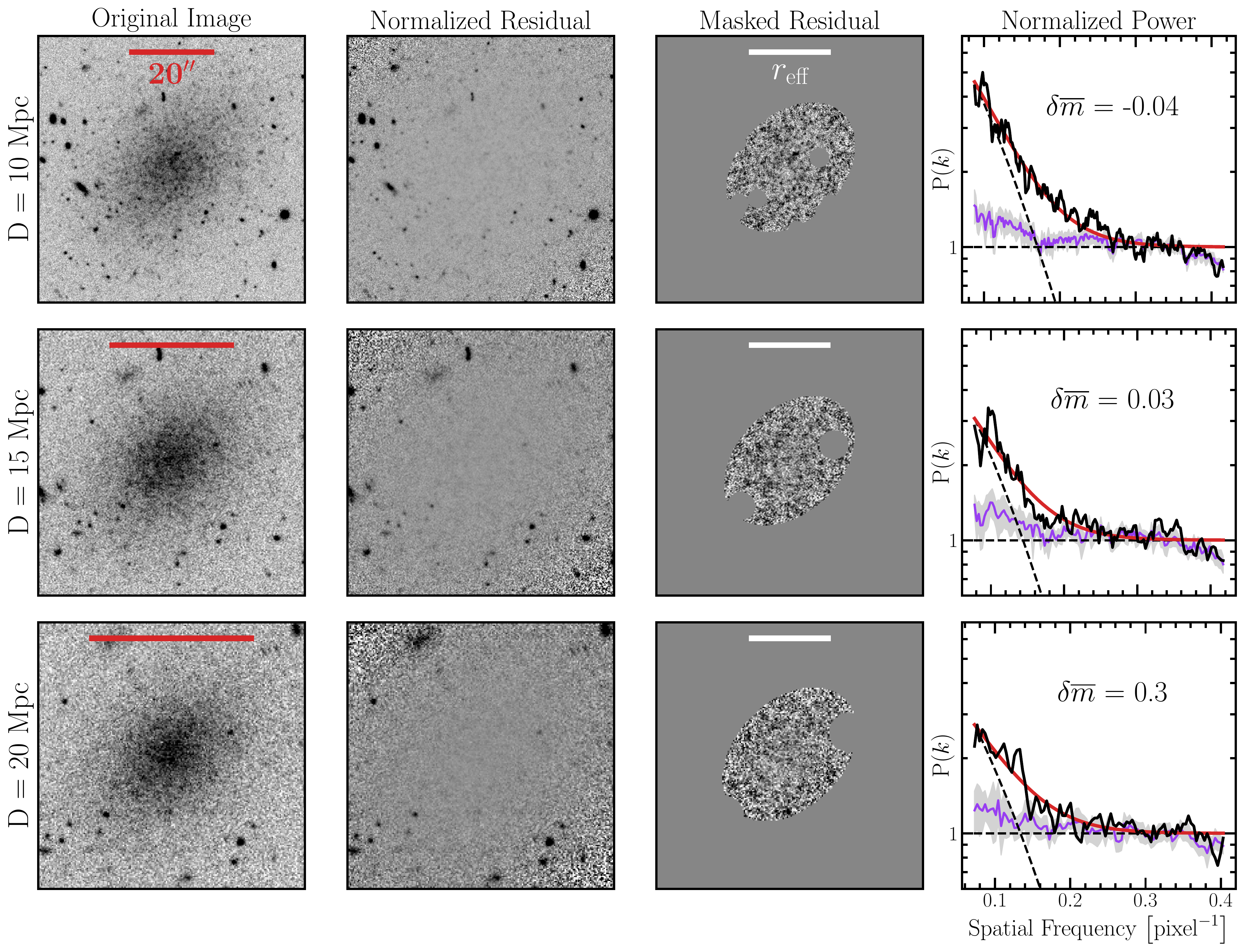}
    \caption{Illustration of the SBF recovery steps for a mock galaxy of stellar mass $10^7$~M$_\odot$ that has been placed at 10~Mpc (top row), 15~Mpc (middle row), and 20~Mpc (bottom row) and injected into an $i$-band image from the wide layer of HSC-SSP. From left to right, the columns show the original images, the normalized residual images, the masked (normalized) residual images, and the azimuthally-averaged normalized power spectra of the masked residual images. In each power spectrum panel, the solid black line shows the measured power spectrum, the solid red line shows the best-fit to Equation~(\ref{eqn:Pk}), and the non-zero-slope dashed black line shows the normalized PSF power spectrum. The solid purple line indicates the median power spectrum of 10 blank fields near the target galaxy, with the gray shaded region indicating the 68th percentile of the distribution. The error on the recovered apparent fluctuation magnitude $\delta\overline{m}\equiv\overline{m}_\mathrm{truth}-\overline{m}_\mathrm{recovered}$ is indicated in each power spectrum panel. \resp{For each row, 20\arcsec\ is indicated by the red line in the first panel and the galaxy's effective radius is indicated by the white line in the third panel.}} 
    \label{fig:hsc-sim-power-spectra}
\end{figure*}

\begin{figure*}[ht!]
    \centering
    \includegraphics[width=\textwidth]{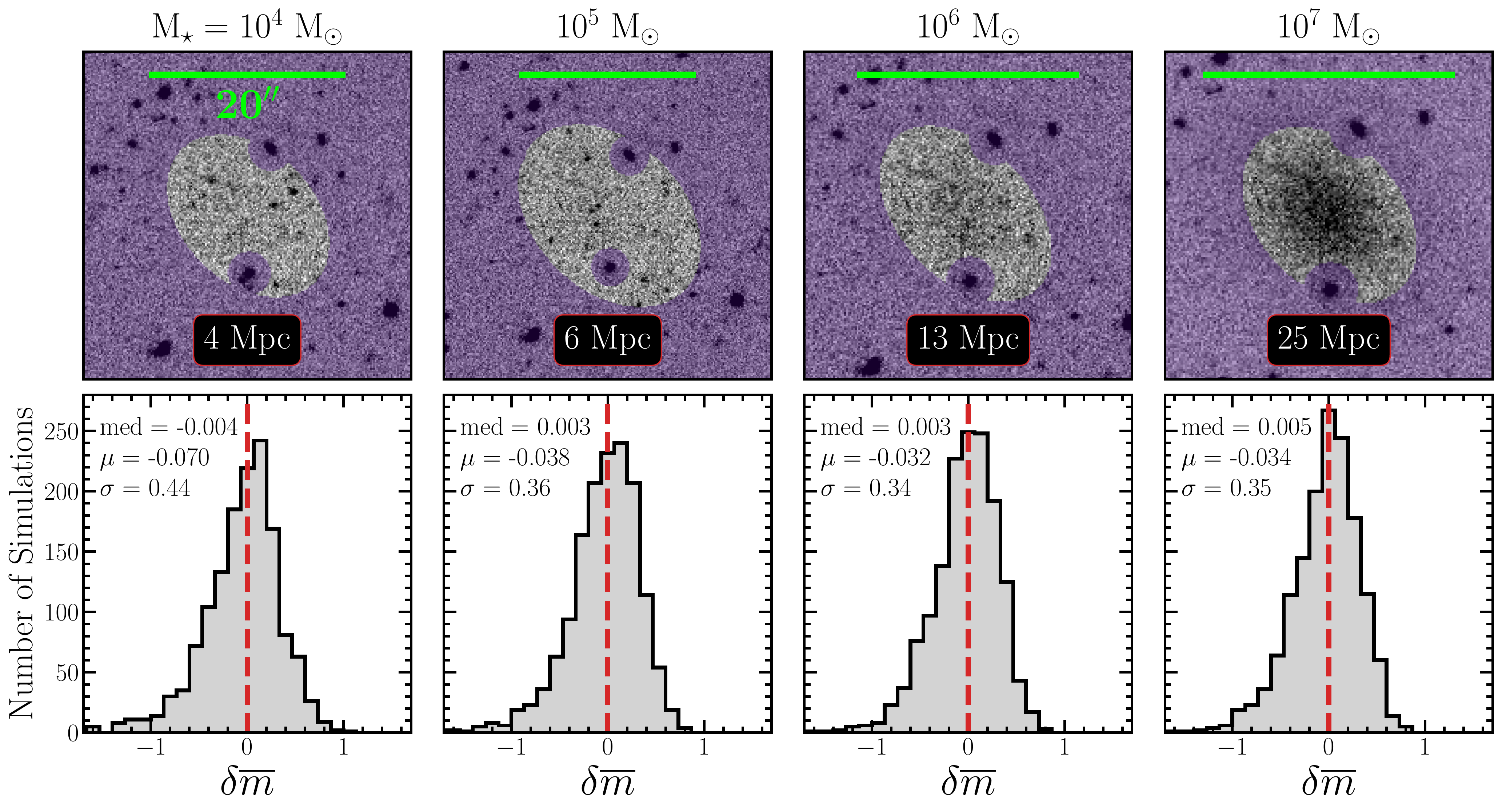}
    \caption{Representative simulation results \resp{from different combinations of stellar mass and distance, for which the apparent SBF magnitudes are typically recovered to within ${\sim}0.3$-0.45 mag (${\sim}15$-20\% in distance).} Each panel in the top row shows an example mock galaxy realization for the indicated combination of stellar mass and distance, and the transparent purple region shows the \resp{SBF measurement} mask. The diffuse stellar component is not visible in images of the two lower mass galaxies on the left, but individual AGB stars (the unmasked point sources) are clearly detected in these galaxies given their relatively nearby distances. Each panel in the bottom row shows the corresponding distribution of $\delta\overline{m}\equiv\overline{m}_\mathrm{truth}-\overline{m}_\mathrm{recovered}$ from the 500 realizations in each of the three HSC-SSP fields (1500 simulations for each mass--distance combination). \resp{We indicate the median (med), mean ($\mu$), and standard deviation ($\sigma$) of the error distributions in each panel}.}
    \label{fig:sbf-recovery}
\end{figure*}

Within each field, we mask any source that---if it were at the same distance of the mock galaxy---would have an absolute $i$-band magnitude brighter than M$_i=-4$, which roughly corresponds to the luminosity of the most luminous giant stars for our assumed SSP parameters. To identify these sources, we select objects in the HSC-SSP catalog based on their \code{cmodel} magnitude, requiring the magnitude uncertainty to be less than 0.2~mag to limit the catalog to high signal to noise ratio detections. At each source position, we apply a circular mask that scales with the object flux. 

 \resp{Our masking algorithm assumes prior knowledge of the galaxy's distance. In real targets, this assumption is generally not satisfied. However, the goal of dwarf galaxy SBF observations is very often to confirm association with a more massive host galaxy or group of galaxies of known distance. In this case, it is reasonable to assume the host distance to generate an object mask \citep[e.g.,][]{Carlsten-2019-M101}. If the target is a potential isolated dwarf galaxy (i.e., there is no host) in the Local Volume, this masking method will need to be modified. For example, a range of distances, and hence object masks, may be used to place constraints on its distance and possibly motivate follow-up observations.}

The final object mask for each field is given by the union of the above source mask and an elliptical mask generated as described in Section~\ref{sec:ideal-setup}. For fixed image quality and depth, the optimal semi-major axis of the ellipse depends on the galaxy's mass and distance, as well as on the density of nearby contaminating sources, since an increasing number of sources will fall within the aperture as the semi-major axis increases. In addition, the impact of dividing by the model is larger in the galaxy outskirts, as the surface brightness approaches zero at large radii. We, therefore, select the aperture sizes on a per-field basis for each combination of galaxy mass and distance that we simulate. For galaxies with M$_\star \leq 10^5$~M$_\odot$, we use semi-major axes of 0.5-0.75$\times r_\mathrm{eff}$. For more massive galaxies, we use somewhat larger semi-major axes of 0.75-1.5$\times r_\mathrm{eff}$

Similar to our idealized simulations, we model each galaxy's surface brightness distribution using \code{imfit}. The main difference in this case is that before performing the fit, we aggressively mask sources that are not associated with the target galaxy using \code{sep}\footnote{\url{https://sep.readthedocs.io}} \citep{Barbary:2016aa}. Note this mask is for fitting purposes and is different from the SBF object mask. For galaxies with masses M$_\star\leq10^5$~M$_\odot$, we fix the centroid and structural parameters to their known values during the fit. Given the extremely low mean surface brightness of such objects, the data lack the necessary surface brightness sensitivity to robustly constrain their average structure. This means, of course, that our fitting results for galaxies with M$_\star\leq10^5$~M$_\odot$ are overly optimistic. However, an analogous scenario may occur in reality, as one can imagine detecting and measuring the structure of an ultra-faint dwarf in the Local Volume with an LSB-optimized, low-resolution survey \citep[e.g.,][]{DF-Wide}, the results of which could serve as a prior for measurements that use data from a higher-resolution survey like HSC-SSP or LSST. 

 We use the best-fit \code{imfit} model to generate the normalized residual image, from which we measure the fluctuation signal following the procedure described in Section~\ref{sec:method}. As we have noted, the best-fit parameters depend strongly on the range of $k$ values used in the fit to Equation~(\ref{eqn:Pk}). To automate the fitting process, we perform the fit 10 times per SBF measurement, where the minimum and maximum $k$ values are drawn from uniform distributions with $0.05 \leq k_\mathrm{min} \leq 0.1$ and $0.35 \leq k_\mathrm{max} \leq 0.45$. We then calculate the SBF signal as the average of the 10 measurements, where we weight each measurement by the $\chi^2$ associated with its fit, similar to the method used in \citet{Cohen:2018}. 
 
We do not include globular clusters in our simulations, but galaxies in the stellar mass range we are studying \resp{generally have} zero to a few globular clusters \citep{Forbes:2018}. We therefore do not need to subtract the fluctuation signal from unmasked globular clusters. We do, however, expect there to be residual variance from faint background galaxies and correlated noise between pixels. Following \citet{Carlsten:2019aa}, we empirically measure this residual variance using blank fields near our target galaxies. For each SBF measurement, we perform identical measurements in 10 nearby blank fields, where the blank fields are normalized using the same model as the target field.

In Figure~\ref{fig:hsc-sim-power-spectra}, we illustrate the above steps for a galaxy of mass $10^7$~M$_\odot$ at 10~Mpc (top row), 15~Mpc (middle row), and 20~Mpc (bottom row). From left to right, the columns show the original images, the normalized residual images, the masked residual images, and the normalized power spectra of the masked residual images. In the leftmost column, we see the galaxy transition from visibly semi-resolved at 10~Mpc to mostly smooth at 20~Mpc. \resp{For each row, the red line in the first panel indicates the scale of 20\arcsec, and the white line in the third panel indicates the size of the galaxy's effective radius.}

The errors on the apparent fluctuation magnitudes for these particular mock galaxy realizations are indicated in the power spectra panels in the right-most column. \resp{The corresponding statistical distance errors} are ${\sim}2\%$ at 10 and 15~Mpc and ${\sim}14\%$ at 20~Mpc. Importantly, this demonstrates that SBFs can be recovered from low-luminosity galaxies even when the galaxy appears smooth to the eye. Note that the \resp{errors do} not necessarily decrease with distance (compare the top and middle rows), since the signal-to-noise ratio depends on a complicated interplay between the strength of the intrinsic signal relative to the sky and detector noise, the size and aggressiveness of the object mask, the quality of the S\'{e}rsic fit, and the residual variance from contaminating sources within the SBF measurement aperture.

\begin{figure}[t!]
    \centering
    \includegraphics[width=\columnwidth]{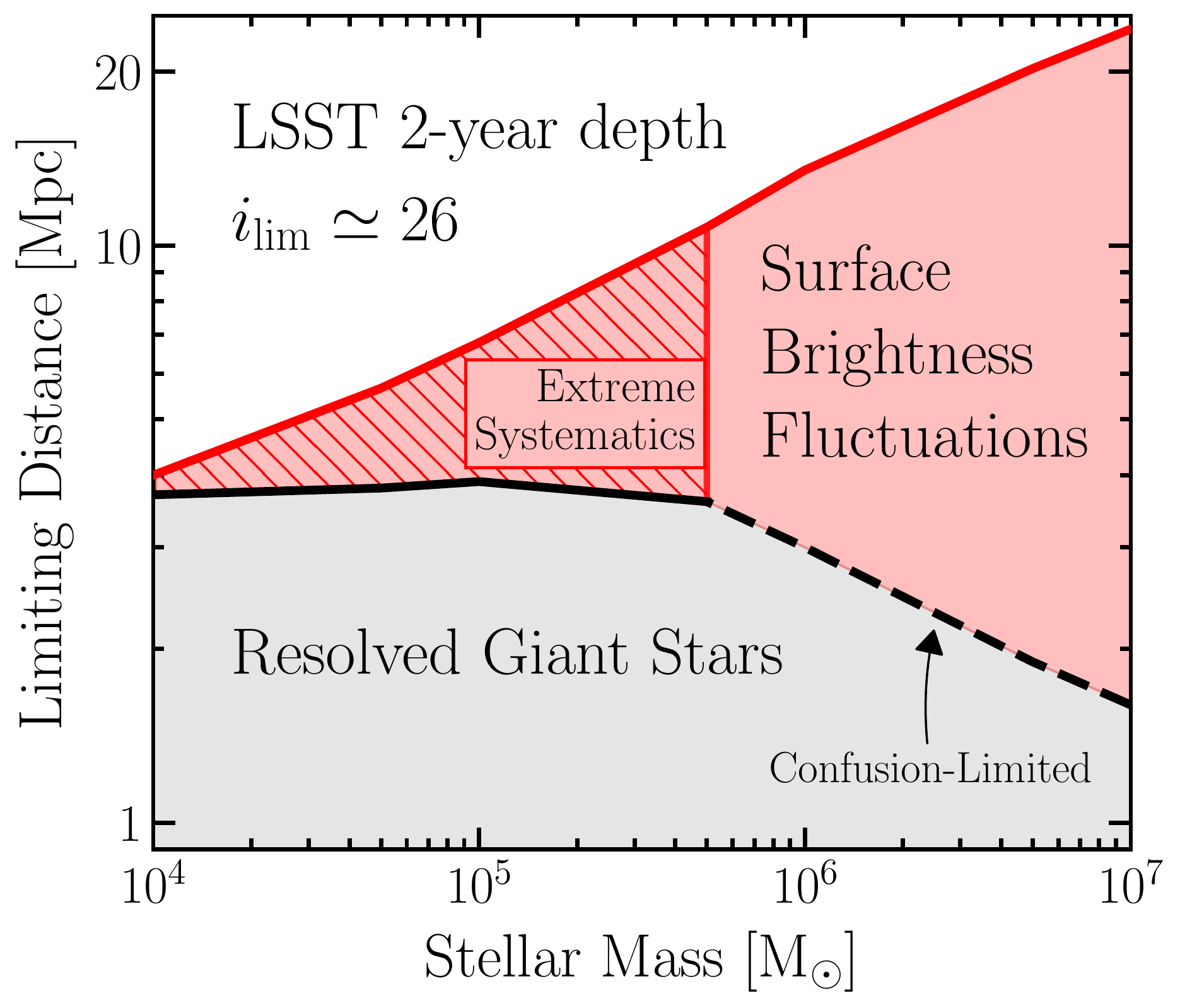}
    \caption{Limiting distance as a function of stellar mass for the SBF method and for detecting galaxies via resolved giant stars. These limits \resp{represent the best-case scenario, ignoring systematic uncertainties, which become very large (${>}0.5$~mag) in the hatched region.} We assume the data quality and depth are similar to what is expected for 2-year LSST stacks. The solid red line indicates where we recover \resp{SBF magnitudes to within 0.3~mag}, assuming old, metal-poor stellar populations. The solid black line shows where at least 30 stars in the target galaxy are brighter than $i_\mathrm{lim}=26$. The dashed black line indicates where  giant stars become confusion-limited with the background of the galaxy's mean stellar distribution.}
    \label{fig:D_lim}
\end{figure}

\subsubsection{Realistic\resp{-Noise} Simulations: Results}

In Figure~\ref{fig:sbf-recovery}, we show representative results \resp{from different combinations of stellar mass and distance, for which the apparent SBF magnitudes are typically recovered to within ${\sim}0.3$-0.45~mag (${\sim}15$-20\% in distance).} In the top row, we show example mock galaxy realizations and their associated object masks (transparent purple regions) for the indicated mass--distance combinations. For the galaxies with M$_\odot \leq 10^5$~M$_\odot$, the diffuse stellar component is more or less invisible, but individual AGB stars are clearly detected, since these galaxies are necessarily at relatively small distances. Nevertheless, it is \resp{in principle possible to measure the fluctuation signal of} such objects from the ground, provided their centroid and structure are independently known. 

The bottom row of Figure~\ref{fig:sbf-recovery} shows the apparent fluctuation magnitude error distributions from the 500 realizations in each of the three HSC-SSP fields (1500 simulations for each mass--distance combination). \resp{In each panel, we list the median, mean, and standard deviations of the distributions. The apparent SBF magnitudes are generally recovered to ${\lesssim}$0.45~mag}. Though the distributions are noisier than our idealized simulations, as expected, the recovered signals in all cases are unbiased, with median offsets that are ${\lesssim}5$~mmag. 

\resp{As indicated by the negative mean values, each distribution has a small tail at $\delta\overline{m}\lesssim-1$, which corresponds to the recovered signal being fainter than the input signal. These are failures of our automated masking and measurement routines, which arise from rare (less than 1-3\% of simulations) but unfortunate combinations of the HSC-SSP field, the mock galaxy realization, and the random position angle of the galaxy relative to nearby sources. Such failures can generally be ameliorated by fine-tuning the masking by hand (e.g., the size of the elliptical aperture and dilation of the object mask), but we choose to keep the simulations fully automated.} 

\section{Limiting \resp{Ground-Based SBF Distances}} \label{sec:D_lim}

The main result from the simulations presented in this paper is summarized in Figure~\ref{fig:D_lim}. This figure shows the \resp{best-case} limiting SBF distance of 2-year LSST stacks, as a function of stellar mass. The solid red line indicates the distance at which we \resp{typically recover SBF magnitudes to within 0.3 mag. For stellar masses $\lesssim 5\times10^5$~M$_\odot$ (red hatched region), we expect theoretical and observational systematics to become very significant (${>}0.5$~mag). This expectation is based in part on the SBF simulations of \citet{Carlsten:2019aa}, who injected mock galaxies into deep CFHT imaging before sky subtraction.}

\resp{At ${\sim}5\times10^5$~M$_\odot$, \citet{Carlsten:2019aa} recovered total luminosities and integrated colors with errors on the order of 0.4 and 0.2~mag, respectively, with a strong dependence on the quality of the sky measurement. Furthermore, SBF sampling scatter becomes non-negligible in this mass range; at the lowest masses, the scatter may be as large as ${\sim}$1~mag. Still, even highly uncertain distance measurements can be useful in certain situations. For example, such measurements can be used to motivate follow-up observations to measure the tip of the RGB, or to determine whether a particular galaxy is likely to be associated with another galaxy or group of known distance.}

The solid black line and gray region show where it is possible to detect the galaxy via resolved stars. To calculate this limit, we count the median number of stars in the galaxy that are brighter than the detection limit of $i_\mathrm{lim} = 26$. We require at least 30 stars to be above the detection limit for the galaxy to be detectable \cite[e.g.,][]{Koposov:2008}. That individual giant stars in the lowest-mass galaxies may be detectable out to a few Mpc does not necessarily mean the distances to such objects can be robustly measured using the tip of the RGB---a galaxy of mass \resp{${\gtrsim}10^4$~M$_\odot$} will likely have fewer than ${\sim}5$ stars within 1~mag from the tip of the RGB. 

\resp{On the other hand, as the left-most panel in Figure~\ref{fig:sbf-recovery} shows, SBF distances may be viable out to ${\sim}$4~Mpc for such low-luminosity systems. This, however, would require that the galaxy's structure and integrated color are well-measured, which is non-trivial in this mass regime. Additionally, it would likely depend on significant progress in stellar population modeling and LSB-optimized observing strategies and reduction methods.} 

For more massive galaxies, the limiting distance for resolving giant stars {\it decreases} with stellar mass, as indicated by the dashed black line. This is due to giant stars becoming confusion-limited with the background of the galaxy's mean stellar distribution. We \johnny{define} the confusion limit as when the surface brightness from the flux of a star at the tip of the RGB, spread over 30 resolution elements, is equal to the mean surface brightness of the galaxy \citep[e.g.,][]{Olsen:2003}.

We assumed old SSPs in our main set of simulations because the agreement between theory and observation is \resp{generally quite good} in this regime (e.g., Figure~\ref{fig:compare-with-obs}). SBF distances to younger populations should in principle be feasible out to larger distances, since at fixed stellar mass, the fluctuation luminosity increases with decreasing SSP age (middle panel of Figure~\ref{fig:lsst-sbf-color}). For a metal poor SSP, the $i$-band fluctuation magnitude fades by ${\sim}1.5$ mag from 300~Myr to 12.6~Gyr, which corresponds to nearly a factor of 4 decrease in the amplitude of the SBF signal. 

Following the procedure presented in Section~\ref{sec:ideal-setup}, we ran a smaller set of simulations for young SSPs of age 300~Myr. Indeed, the distance limits increase by ${\sim}1$~Mpc and ${\sim}20$~Mpc at the low-mass and high-mass end of our mass range of interest, respectively. We do not include these limits in Figure~\ref{fig:D_lim} because SBF distances to young stellar populations are \resp{even more} rife with systematic uncertainty (see below). More empirical and theoretical studies are necessary to fully understand the viability of SBF distances to young stellar populations.

\subsection{\resp{Important Caveats}}\label{sec:caveats}

It is important to note the following caveats about our simulation results. Dwarf galaxies are neither pure SSPs nor perfect S\'{e}rsic galaxies, as we have assumed in this work. While these assumptions are reasonable approximations for dwarf elliptical galaxies, they are inconsistent with the stellar populations and morphologies of dwarf irregular galaxies. Additionally, star forming regions and dust within galaxies that have recently undergone star formation are a particularly concerning source of SBFs, which we have not attempted to model.

\resp{In addition,} we have quantified the uncertainties using the fluctuation magnitude offset directly, whereas in practice one must use a theoretically or empirically calibrated fluctuation-color relation, which currently introduces systematic uncertainties on the order of ${\sim}$0.2 mag or ${\sim}$9\% in distance. This also means that our predicted uncertainties do not include the potentially significant sampling scatter expected for ultra-faint systems. \resp{Finally, since we inject the artificial galaxies into post-processed HSC-SSP images, our simulations do not include the impact of imperfect sky subtraction on the measured galaxy properties such as its structure and integrated color, which can contribute ${\gtrsim}0.2$~mag to the SBF magnitude uncertainty, depending on the stellar mass.} 

Notwithstanding the above caveats, our simulation results suggest that the SBF method will play an essential role in confirming and studying low-luminosity galaxies with the next generation of wide-field imaging surveys. LSST's deep-wide-fast survey mode will uniformly observe at least 18,000~deg$^2$ with similar image quality and depth to our simulations after 2 years of data collection. At the end of its main 10-year survey, \johnny{the stacked images will be approximately 1 magnitude deeper than the 2-year stacks.} This will undoubtedly lead to larger limiting SBF distances, but in practice, the measurement will be more difficult because of the larger density of background sources. Stated succinctly, LSST will deliver SBF distances to dwarf galaxies out to ${\lesssim}25$~Mpc within the first few years of its deep-wide-fast survey.   

\section{Summary and Conclusion} \label{sec:summary}

In this work, we have presented a detailed study of the SBF method \resp{using stellar population synthesis modeling and realistic image simulations.} With an eye towards the next generation of wide-field imaging surveys, we have placed an emphasis on low-luminosity galaxies. Our primary results may be summarized as follows.

\begin{enumerate}
	\item Using the MIST isochrones, we provide new predictions for absolute fluctuation magnitudes for single-age stellar populations in the LSST, HST, and proposed RST filter systems. 
	\item \resp{Choice of isochrone model remains a major source of uncertainty in calculating SBF magnitudes, particularly for young to intermediate-age stellar populations in red bandpasses, where the predictions from different models can differ by as much as ${\sim}$2~mag (Figure~\ref{fig:fsps-compare-sbf}).}
	\item SBFs in very low-luminosity galaxies ($\mathrm{M}_\star < 10^6$~M$_\odot$) have a \johnny{sampling scatter} of ${\sim}0.2$-1.2~mag, depending on stellar mass and bandpass, which is due to incomplete sampling of the stellar mass function (Figure~\ref{fig:absmag-scatter}).
	\item Consistent with previous work, we find that the predictions for simple stellar populations are in excellent agreement with the observed fluctuation-color relation for metal-poor, old stellar populations with colors $g-i \gtrsim 0.5$ (Figure~\ref{fig:compare-with-obs}).
	\item For bluer and younger galaxies, composite \johnny{double-burst stellar populations} provide better agreement with the observations than single-age populations, particularly in the color range $0.25 \lesssim g-i \lesssim 0.65$ (Figure~\ref{fig:compare-with-obs}). This is also the color range where \resp{MIST predicts} AGB stars begin to dominate the \resp{luminosity variance} in metal poor systems (Figure~\ref{fig:phase-sbf-color}).
	\item \resp{Through comparisons with observed SBF-color relations, we identify two apparent shortcomings of the MIST version 1.2 models: i) at high metallicity ($\mathrm{[Fe/H]}\approx0$), the MIST $i$-band SBF magnitudes in old SSPs are ${\sim}0.3$~mag fainter than other widely-used models (Figure~\ref{fig:old-pops-compare}), and ii) the near-infrared colors of intermediate-age SSPs are ${\sim}0.05$-$0.1$~mag bluer than that of observed star clusters (Figure~\ref{fig:compare-IR}). In Section~\ref{sec:compare-with-obs}, we discuss potential sources of these discrepancies, which will be addressed in a future version of MIST.}
	\item Based on recovered SBF distances to thousands of artificial galaxies that were injected into HSC-SSP imaging data, LSST will provide data of sufficient quality and depth to \resp{measure SBF magnitudes (to within ${\sim}$0.2-0.5~mag) in old dwarf galaxies} in the ultra-faint $\left(\mathrm{10^4 \leq M_\star/M_\odot \leq 10^5}\right)$ and low-mass classical (M$_\star\leq10^7$~M$_\odot$) regimes out to ${\sim}4$~Mpc and ${\sim}25$~Mpc, respectively, within the first few years of its deep-wide-fast survey (Figure~\ref{fig:D_lim}). \resp{See Section~\ref{sec:caveats} for important caveats about our simulation results.}
\end{enumerate}

While there are \resp{significant practical challenges and systematic uncertainties} that will make it difficult (or in some cases impossible) to fully exploit the SBF method in dwarf galaxy research, this work has shown that we will soon have tens of thousands of square degrees over which SBFs have the potential to play an essential role in confirming and studying low-luminosity galaxies in the nearby universe.

\acknowledgments
\resp{The authors thank the anonymous referee for their detailed report, which greatly improved this paper.} We are grateful to Roberto Abraham for carefully reading an earlier version of this work and providing very useful feedback. J.P.G. thanks Annika Peter for many useful conversations and words of encouragement, as well as Rachael Beaton for answering lots of questions related to measuring extragalactic distances and Ben Cook for his help with \code{PCMDPy}. J.P.G. is supported by an NSF Astronomy and Astrophysics Postdoctoral Fellowship under award AST-1801921. The computations in this paper were run on the CCAPP condos of the Ruby Cluster and the Pitzer Cluster at the Ohio Supercomputer Center \citep{OhioSupercomputerCenter1987}. This paper is based upon work supported by the National Aeronautics and Space Administration (NASA) under Contract No. NNG16PJ26C issued through the WFIRST Science Investigation Teams Program. 

The Hyper Suprime-Cam (HSC) collaboration includes the astronomical communities of Japan and Taiwan, and Princeton University. The HSC instrumentation and software were developed by the National Astronomical Observatory of Japan (NAOJ), the Kavli Institute for the Physics and Mathematics of the Universe (Kavli IPMU), the University of Tokyo, the High Energy Accelerator Research Organization (KEK), the Academia Sinica Institute for Astronomy and Astrophysics in Taiwan (ASIAA), and Princeton University. Funding was contributed by the FIRST program from Japanese Cabinet Office, the Ministry of Education, Culture, Sports, Science and Technology (MEXT), the Japan Society for the Promotion of Science (JSPS), Japan Science and Technology Agency (JST), the Toray Science Foundation, NAOJ, Kavli IPMU, KEK, ASIAA, and Princeton University. 

This paper makes use of software developed for the Large Synoptic Survey Telescope. We thank the LSST Project for making their code available as free software at \url{http://dm.lsst.org}

This paper is based in part on data collected at the Subaru Telescope and retrieved from the HSC data archive system, which is operated by Subaru Telescope and Astronomy Data Center at National Astronomical Observatory of Japan. Data analysis was in part carried out with the cooperation of Center for Computational Astrophysics, National Astronomical Observatory of Japan.

\software{
  \code{ArtPop} (J. Greco \& S. Danieli, in preparation), 
  \code{astropy} \citep{Astropy-Collaboration:2013aa}, 
  \code{imfit} \citep{Erwin:2015aa}, 
  \code{FSPS} \citep{Conroy:2009aa, Conroy-Gunn-2010},
  \code{matplotlib} \citep{Hunter:2007aa}, 
  \code{numpy} \citep{Van-der-Walt:2011aa}, 
  \code{photutils} \citep{photutils},
  \code{PCMDPy} \citep{Cook:2019},
  \code{sep} \citep{Barbary:2016aa},
  \code{scipy} (\url{https://www.scipy.org}),
  \code{schwimmbad} \citep{schwimmbad}
}


\startlongtable
\begin{deluxetable*}{ccrrrrrrrrrrrr}\label{tab:lsst}
\tablecaption{LSST SSP Fluctuation and Mean Stellar Magnitudes}
\tablecolumns{13}
\tablewidth{\textwidth}
\tablehead{
\colhead{$\log(\mathrm{Age/yr})$} &
\colhead{[Fe/H]} &
\colhead{$u$} &
\colhead{$g$} &
\colhead{$r$} &
\colhead{$i$} &
\colhead{$z$} &
\colhead{$y$} &
\colhead{\Mbar$_u$} &
\colhead{\Mbar$_g$} &
\colhead{\Mbar$_r$} &
\colhead{\Mbar$_i$} &
\colhead{\Mbar$_z$} &
\colhead{\Mbar$_y$} 
}
\startdata
8.0 & -2.5 & 3.60 & 3.40 & 3.68 & 3.88 & 4.02 & 4.09 & -2.00 & -2.40 & -2.99 & -3.66 & -4.08 & -4.33\\
8.0 & -2.0 & 3.45 & 3.19 & 3.44 & 3.63 & 3.74 & 3.80 & -2.30 & -2.80 & -3.20 & -3.78 & -4.17 & -4.41\\
8.0 & -1.5 & 3.41 & 3.05 & 3.26 & 3.41 & 3.51 & 3.55 & -2.45 & -3.07 & -3.24 & -3.59 & -3.87 & -4.06\\
8.0 & -1.0 & 3.48 & 2.99 & 3.09 & 3.18 & 3.22 & 3.23 & -2.31 & -3.21 & -3.36 & -3.63 & -3.85 & -4.01\\
8.0 & -0.5 & 3.79 & 3.26 & 3.15 & 3.09 & 3.07 & 3.03 & -0.75 & -1.79 & -2.97 & -3.54 & -3.85 & -4.04\\
8.0 & 0.0 & 3.78 & 3.24 & 3.21 & 3.19 & 3.17 & 3.13 & -0.94 & -1.66 & -2.53 & -3.13 & -3.59 & -4.00\\
8.5 & -2.5 & 4.44 & 3.94 & 4.10 & 4.22 & 4.29 & 4.31 & -0.57 & -1.41 & -2.43 & -3.25 & -3.70 & -3.96\\
8.5 & -2.0 & 4.39 & 3.83 & 3.97 & 4.08 & 4.14 & 4.16 & -0.65 & -1.47 & -2.24 & -2.99 & -3.42 & -3.68\\
8.5 & -1.5 & 4.41 & 3.76 & 3.83 & 3.90 & 3.94 & 3.94 & -0.59 & -1.60 & -2.18 & -2.81 & -3.21 & -3.45\\
8.5 & -1.0 & 4.61 & 3.90 & 3.82 & 3.80 & 3.80 & 3.79 & 0.21 & -1.01 & -1.96 & -2.60 & -3.00 & -3.27\\
8.5 & -0.5 & 4.76 & 3.99 & 3.91 & 3.88 & 3.88 & 3.85 & 0.73 & -0.36 & -1.38 & -2.15 & -2.69 & -3.09\\
8.5 & 0.0 & 4.95 & 4.11 & 4.07 & 4.07 & 4.06 & 4.01 & 0.78 & 0.00 & -0.59 & -1.40 & -2.14 & -2.85\\
9.0 & -2.5 & 5.36 & 4.60 & 4.68 & 4.76 & 4.81 & 4.81 & 0.55 & -0.50 & -1.08 & -1.72 & -2.17 & -2.47\\
9.0 & -2.0 & 5.38 & 4.58 & 4.50 & 4.49 & 4.49 & 4.48 & 0.82 & -0.38 & -1.28 & -1.93 & -2.33 & -2.58\\
9.0 & -1.5 & 5.56 & 4.67 & 4.44 & 4.34 & 4.31 & 4.27 & 1.28 & -0.24 & -1.40 & -2.03 & -2.40 & -2.63\\
9.0 & -1.0 & 5.67 & 4.66 & 4.36 & 4.23 & 4.17 & 4.12 & 1.82 & 0.10 & -1.21 & -1.94 & -2.40 & -2.70\\
9.0 & -0.5 & 6.01 & 4.91 & 4.61 & 4.47 & 4.40 & 4.34 & 2.51 & 0.71 & -0.76 & -1.73 & -2.35 & -2.76\\
9.0 & 0.0 & 6.38 & 5.23 & 4.88 & 4.74 & 4.67 & 4.59 & 2.82 & 1.38 & 0.31 & -0.79 & -1.66 & -2.43\\
9.5 & -2.5 & 6.67 & 5.81 & 5.65 & 5.59 & 5.58 & 5.56 & 1.44 & 0.19 & -0.56 & -0.95 & -1.15 & -1.25\\
9.5 & -2.0 & 6.54 & 5.62 & 5.33 & 5.21 & 5.16 & 5.12 & 1.67 & 0.04 & -1.07 & -1.65 & -1.97 & -2.18\\
9.5 & -1.5 & 6.68 & 5.69 & 5.33 & 5.18 & 5.11 & 5.06 & 2.08 & 0.10 & -1.21 & -1.83 & -2.15 & -2.35\\
9.5 & -1.0 & 6.93 & 5.83 & 5.40 & 5.21 & 5.12 & 5.06 & 2.73 & 0.40 & -1.00 & -1.65 & -2.01 & -2.24\\
9.5 & -0.5 & 7.33 & 6.09 & 5.55 & 5.32 & 5.20 & 5.12 & 3.48 & 0.93 & -0.58 & -1.39 & -1.90 & -2.25\\
9.5 & 0.0 & 7.94 & 6.50 & 5.88 & 5.58 & 5.41 & 5.28 & 4.28 & 1.72 & 0.29 & -0.78 & -1.52 & -2.13\\
10.0 & -2.5 & 7.17 & 6.29 & 5.98 & 5.83 & 5.77 & 5.73 & 1.45 & -0.06 & -0.98 & -1.42 & -1.63 & -1.75\\
10.0 & -2.0 & 7.63 & 6.67 & 6.24 & 6.04 & 5.96 & 5.91 & 2.03 & 0.19 & -0.91 & -1.45 & -1.74 & -1.92\\
10.0 & -1.5 & 7.75 & 6.70 & 6.20 & 5.97 & 5.88 & 5.82 & 2.41 & 0.31 & -0.90 & -1.47 & -1.79 & -2.00\\
10.0 & -1.0 & 8.06 & 6.85 & 6.27 & 6.02 & 5.90 & 5.83 & 3.02 & 0.61 & -0.79 & -1.43 & -1.78 & -2.01\\
10.0 & -0.5 & 8.57 & 7.13 & 6.47 & 6.17 & 6.02 & 5.92 & 3.90 & 1.26 & -0.22 & -1.09 & -1.63 & -1.98\\
10.0 & 0.0 & 9.31 & 7.57 & 6.83 & 6.47 & 6.27 & 6.11 & 4.99 & 2.10 & 0.75 & -0.32 & -1.10 & -1.81\\
10.1 & -2.5 & 7.30 & 6.41 & 6.05 & 5.89 & 5.81 & 5.77 & 1.49 & -0.10 & -1.05 & -1.50 & -1.71 & -1.83\\
10.1 & -2.0 & 7.75 & 6.77 & 6.33 & 6.12 & 6.04 & 5.98 & 1.98 & 0.16 & -0.93 & -1.47 & -1.74 & -1.91\\
10.1 & -1.5 & 7.98 & 6.91 & 6.39 & 6.16 & 6.06 & 6.00 & 2.45 & 0.33 & -0.90 & -1.46 & -1.75 & -1.94\\
10.1 & -1.0 & 8.27 & 7.04 & 6.45 & 6.20 & 6.08 & 6.00 & 2.89 & 0.65 & -0.74 & -1.38 & -1.74 & -1.97\\
10.1 & -0.5 & 8.83 & 7.32 & 6.63 & 6.32 & 6.16 & 6.05 & 3.74 & 1.28 & -0.18 & -1.05 & -1.59 & -1.94\\
10.1 & 0.0 & 9.62 & 7.79 & 7.02 & 6.64 & 6.43 & 6.26 & 3.49 & 2.07 & 0.83 & -0.23 & -1.02 & -1.75\\
10.15 & -2.5 & 7.37 & 6.47 & 6.10 & 5.92 & 5.84 & 5.80 & 1.52 & -0.11 & -1.08 & -1.53 & -1.74 & -1.86\\
10.15 & -2.0 & 7.82 & 6.82 & 6.38 & 6.17 & 6.08 & 6.03 & 1.96 & 0.15 & -0.95 & -1.48 & -1.75 & -1.92\\
10.15 & -1.5 & 8.04 & 6.96 & 6.44 & 6.21 & 6.11 & 6.05 & 2.36 & 0.33 & -0.89 & -1.45 & -1.75 & -1.93\\
10.15 & -1.0 & 8.34 & 7.10 & 6.50 & 6.24 & 6.12 & 6.04 & 2.85 & 0.65 & -0.75 & -1.39 & -1.75 & -1.98\\
10.15 & -0.5 & 8.99 & 7.48 & 6.79 & 6.48 & 6.31 & 6.21 & 3.85 & 1.36 & -0.10 & -0.98 & -1.53 & -1.88\\
10.15 & 0.0 & 9.77 & 7.88 & 7.09 & 6.70 & 6.48 & 6.31 & 4.92 & 2.16 & 0.84 & -0.21 & -1.00 & -1.74\\
\enddata
\tablecomments{MIST predictions for SSP absolute SBF magnitudes and mean magnitudes for calculating colors in the LSST $ugrizy$ filter system. Calculations assume the \citet{Kroupa:2001} IMF, and magnitudes are in the AB magnitude system.}
\end{deluxetable*}

\startlongtable
\begin{deluxetable*}{ccrrrrrrrrrr}\label{tab:hst}
\centering
\tablecaption{HST SSP Fluctuation and Mean Stellar Magnitudes}
\tablecolumns{12}
\tablewidth{\textwidth}
\tablehead{
\colhead{$\log(\mathrm{Age/yr})$} &
\colhead{[Fe/H]} &
\colhead{F$_\mathrm{435W}$} &
\colhead{F$_\mathrm{475W}$} &
\colhead{F$_\mathrm{555W}$} &
\colhead{F$_\mathrm{606W}$} &
\colhead{F$_\mathrm{814W}$} &
\colhead{\Mbar$_\mathrm{435W}$} &
\colhead{\Mbar$_\mathrm{475W}$} &
\colhead{\Mbar$_\mathrm{555W}$} &
\colhead{\Mbar$_\mathrm{606W}$} &
\colhead{\Mbar$_\mathrm{814W}$} 
}
\startdata
8.0 & -2.5 & 3.35 & 3.42 & 3.57 & 3.68 & 4.12 & -2.22 & -2.14 & -1.98 & -1.85 & -1.39\\
8.0 & -2.0 & 3.13 & 3.18 & 3.29 & 3.38 & 3.68 & -2.79 & -2.80 & -2.90 & -3.08 & -3.95\\
8.0 & -1.5 & 3.01 & 3.05 & 3.14 & 3.21 & 3.45 & -3.06 & -3.07 & -3.09 & -3.17 & -3.71\\
8.0 & -1.0 & 2.99 & 2.99 & 3.03 & 3.07 & 3.20 & -3.19 & -3.21 & -3.24 & -3.30 & -3.72\\
8.0 & -0.5 & 3.32 & 3.26 & 3.21 & 3.17 & 3.08 & -1.32 & -1.73 & -2.35 & -2.74 & -3.68\\
8.0 & 0.0 & 3.28 & 3.25 & 3.23 & 3.22 & 3.18 & -1.41 & -1.62 & -2.03 & -2.33 & -3.33\\
8.5 & -2.5 & 3.95 & 3.99 & 4.10 & 4.19 & 4.55 & -1.02 & -0.97 & -0.83 & -0.72 & -0.33\\
8.5 & -2.0 & 3.83 & 3.83 & 3.89 & 3.94 & 4.11 & -1.37 & -1.46 & -1.73 & -2.05 & -3.19\\
8.5 & -1.5 & 3.78 & 3.76 & 3.78 & 3.81 & 3.92 & -1.50 & -1.59 & -1.79 & -2.03 & -2.99\\
8.5 & -1.0 & 3.98 & 3.91 & 3.85 & 3.84 & 3.80 & -0.67 & -0.96 & -1.41 & -1.76 & -2.78\\
8.5 & -0.5 & 4.07 & 4.00 & 3.94 & 3.92 & 3.88 & -0.03 & -0.31 & -0.79 & -1.16 & -2.40\\
8.5 & 0.0 & 4.18 & 4.11 & 4.08 & 4.08 & 4.07 & 0.08 & 0.02 & -0.18 & -0.42 & -1.75\\
9.0 & -2.5 & 4.72 & 4.67 & 4.66 & 4.67 & 4.74 & 0.00 & -0.07 & -0.14 & -0.18 & -0.32\\
9.0 & -2.0 & 4.67 & 4.59 & 4.53 & 4.51 & 4.49 & -0.09 & -0.37 & -0.83 & -1.17 & -2.11\\
9.0 & -1.5 & 4.82 & 4.68 & 4.55 & 4.48 & 4.33 & 0.29 & -0.21 & -0.85 & -1.25 & -2.20\\
9.0 & -1.0 & 4.84 & 4.67 & 4.50 & 4.42 & 4.21 & 0.70 & 0.14 & -0.59 & -1.06 & -2.15\\
9.0 & -0.5 & 5.11 & 4.94 & 4.76 & 4.67 & 4.44 & 1.32 & 0.79 & 0.04 & -0.50 & -2.03\\
9.0 & 0.0 & 5.46 & 5.26 & 5.05 & 4.95 & 4.71 & 1.80 & 1.44 & 0.91 & 0.51 & -1.22\\
9.5 & -2.5 & 5.76 & 5.64 & 5.53 & 5.48 & 5.38 & 0.63 & 0.30 & -0.13 & -0.39 & -1.04\\
9.5 & -2.0 & 5.79 & 5.63 & 5.47 & 5.39 & 5.18 & 0.61 & 0.09 & -0.54 & -0.92 & -1.80\\
9.5 & -1.5 & 5.90 & 5.71 & 5.51 & 5.40 & 5.15 & 0.82 & 0.18 & -0.54 & -0.99 & -1.97\\
9.5 & -1.0 & 6.10 & 5.87 & 5.61 & 5.48 & 5.17 & 1.21 & 0.49 & -0.29 & -0.76 & -1.81\\
9.5 & -0.5 & 6.42 & 6.13 & 5.82 & 5.65 & 5.26 & 1.79 & 1.03 & 0.19 & -0.33 & -1.63\\
9.5 & 0.0 & 6.88 & 6.54 & 6.18 & 5.99 & 5.50 & 2.56 & 1.82 & 1.02 & 0.51 & -1.15\\
10.0 & -2.5 & 6.84 & 6.65 & 6.45 & 6.34 & 6.07 & 0.72 & 0.28 & -0.25 & -0.57 & -1.31\\
10.0 & -2.0 & 6.93 & 6.70 & 6.46 & 6.32 & 6.00 & 0.80 & 0.25 & -0.37 & -0.75 & -1.58\\
10.0 & -1.5 & 6.99 & 6.73 & 6.44 & 6.29 & 5.93 & 1.03 & 0.39 & -0.31 & -0.72 & -1.61\\
10.0 & -1.0 & 7.19 & 6.89 & 6.55 & 6.37 & 5.97 & 1.43 & 0.70 & -0.08 & -0.56 & -1.59\\
10.0 & -0.5 & 7.54 & 7.18 & 6.80 & 6.59 & 6.10 & 2.12 & 1.36 & 0.54 & 0.02 & -1.35\\
10.0 & 0.0 & 8.04 & 7.62 & 7.19 & 6.96 & 6.37 & 2.95 & 2.20 & 1.43 & 0.96 & -0.71\\
10.1 & -2.5 & 7.02 & 6.83 & 6.61 & 6.49 & 6.21 & 0.73 & 0.28 & -0.26 & -0.59 & -1.34\\
10.1 & -2.0 & 7.03 & 6.80 & 6.55 & 6.41 & 6.09 & 0.74 & 0.21 & -0.40 & -0.77 & -1.59\\
10.1 & -1.5 & 7.21 & 6.95 & 6.65 & 6.49 & 6.12 & 1.05 & 0.40 & -0.30 & -0.71 & -1.60\\
10.1 & -1.0 & 7.40 & 7.09 & 6.75 & 6.56 & 6.15 & 1.45 & 0.74 & -0.03 & -0.51 & -1.55\\
10.1 & -0.5 & 7.75 & 7.37 & 6.97 & 6.75 & 6.24 & 2.12 & 1.38 & 0.58 & 0.06 & -1.31\\
10.1 & 0.0 & 8.28 & 7.84 & 7.40 & 7.15 & 6.54 & 2.65 & 2.15 & 1.48 & 1.03 & -0.63\\
10.15 & -2.5 & 7.09 & 6.89 & 6.67 & 6.55 & 6.26 & 0.72 & 0.26 & -0.28 & -0.61 & -1.36\\
10.15 & -2.0 & 7.08 & 6.85 & 6.60 & 6.46 & 6.14 & 0.71 & 0.20 & -0.41 & -0.78 & -1.61\\
10.15 & -1.5 & 7.26 & 6.99 & 6.70 & 6.54 & 6.17 & 1.03 & 0.40 & -0.29 & -0.70 & -1.59\\
10.15 & -1.0 & 7.46 & 7.14 & 6.80 & 6.61 & 6.19 & 1.44 & 0.74 & -0.04 & -0.52 & -1.56\\
10.15 & -0.5 & 7.91 & 7.53 & 7.13 & 6.91 & 6.40 & 2.19 & 1.45 & 0.65 & 0.13 & -1.25\\
10.15 & 0.0 & 8.38 & 7.94 & 7.48 & 7.23 & 6.60 & 3.00 & 2.26 & 1.51 & 1.04 & -0.61\\
\enddata
\tablecomments{MIST predictions for SSP absolute SBF magnitudes and mean magnitudes for calculating colors in the proposed HST filter system. Calculations assume the \citet{Kroupa:2001} IMF, and magnitudes are in the AB magnitude system.}
\end{deluxetable*}

\startlongtable
\begin{deluxetable*}{ccrrrrrrrrrrrr}\label{tab:wfirst}
\tablecaption{RST SSP Fluctuation and Mean Stellar Magnitudes}
\tablecolumns{13}
\tablewidth{\textwidth}
\tablehead{
\colhead{$\log(\mathrm{Age/yr})$} &
\colhead{[Fe/H]} &
\colhead{R$_{062}$} &
\colhead{Z$_{087}$} &
\colhead{Y$_{106}$} &
\colhead{J$_{129}$} &
\colhead{H$_{158}$} &
\colhead{F$_{184}$} &
\colhead{\Mbar$_\mathrm{R062}$} &
\colhead{\Mbar$_\mathrm{Z087}$} &
\colhead{\Mbar$_\mathrm{Y106}$} &
\colhead{\Mbar$_\mathrm{J129}$} &
\colhead{\Mbar$_\mathrm{H158}$} &
\colhead{\Mbar$_\mathrm{F184}$} 
}
\startdata
8.0 & -2.5 & 3.66 & 4.01 & 4.18 & 4.38 & 4.55 & 4.76 & -2.62 & -3.90 & -4.20 & -4.59 & -6.46 & -6.71\\
8.0 & -2.0 & 3.43 & 3.73 & 3.88 & 4.07 & 4.22 & 4.42 & -2.84 & -3.99 & -4.28 & -4.68 & -6.58 & -6.84\\
8.0 & -1.5 & 3.25 & 3.50 & 3.62 & 3.79 & 3.93 & 4.12 & -2.88 & -3.69 & -3.90 & -4.27 & -6.17 & -6.46\\
8.0 & -1.0 & 3.09 & 3.21 & 3.27 & 3.37 & 3.43 & 3.60 & -3.00 & -3.68 & -3.83 & -4.13 & -6.00 & -6.30\\
8.0 & -0.5 & 3.16 & 3.06 & 3.02 & 2.98 & 2.93 & 3.04 & -2.59 & -3.67 & -3.89 & -4.20 & -6.04 & -6.38\\
8.0 & 0.0 & 3.21 & 3.16 & 3.11 & 3.07 & 3.01 & 3.09 & -2.15 & -3.43 & -3.98 & -4.53 & -6.57 & -7.05\\
8.5 & -2.5 & 4.10 & 4.28 & 4.37 & 4.50 & 4.60 & 4.77 & -2.06 & -3.52 & -3.85 & -4.27 & -6.22 & -6.51\\
8.5 & -2.0 & 3.97 & 4.13 & 4.21 & 4.35 & 4.45 & 4.63 & -1.88 & -3.25 & -3.57 & -4.01 & -5.98 & -6.28\\
8.5 & -1.5 & 3.83 & 3.93 & 3.98 & 4.09 & 4.17 & 4.34 & -1.82 & -3.03 & -3.35 & -3.81 & -5.84 & -6.18\\
8.5 & -1.0 & 3.83 & 3.80 & 3.80 & 3.83 & 3.86 & 4.01 & -1.59 & -2.83 & -3.18 & -3.66 & -5.74 & -6.17\\
8.5 & -0.5 & 3.91 & 3.87 & 3.85 & 3.85 & 3.84 & 3.97 & -1.02 & -2.53 & -3.08 & -3.69 & -5.88 & -6.39\\
8.5 & 0.0 & 4.07 & 4.05 & 4.00 & 3.98 & 3.95 & 4.04 & -0.25 & -2.01 & -3.09 & -3.98 & -6.20 & -6.73\\
9.0 & -2.5 & 4.68 & 4.80 & 4.86 & 4.99 & 5.11 & 5.29 & -0.73 & -2.00 & -2.41 & -2.96 & -5.19 & -5.60\\
9.0 & -2.0 & 4.51 & 4.48 & 4.50 & 4.56 & 4.61 & 4.78 & -0.92 & -2.15 & -2.47 & -2.92 & -5.00 & -5.37\\
9.0 & -1.5 & 4.45 & 4.30 & 4.27 & 4.27 & 4.26 & 4.41 & -1.04 & -2.22 & -2.51 & -2.91 & -4.94 & -5.32\\
9.0 & -1.0 & 4.38 & 4.17 & 4.11 & 4.07 & 4.01 & 4.13 & -0.85 & -2.23 & -2.62 & -3.11 & -5.22 & -5.66\\
9.0 & -0.5 & 4.62 & 4.40 & 4.31 & 4.24 & 4.15 & 4.24 & -0.44 & -2.19 & -2.75 & -3.32 & -5.46 & -5.93\\
9.0 & 0.0 & 4.90 & 4.66 & 4.54 & 4.45 & 4.37 & 4.44 & 0.58 & -1.54 & -2.64 & -3.45 & -5.63 & -6.14\\
9.5 & -2.5 & 5.66 & 5.57 & 5.58 & 5.63 & 5.70 & 5.88 & -0.19 & -0.98 & -1.02 & -1.16 & -2.80 & -2.95\\
9.5 & -2.0 & 5.34 & 5.15 & 5.11 & 5.11 & 5.10 & 5.25 & -0.71 & -1.80 & -2.02 & -2.36 & -4.33 & -4.68\\
9.5 & -1.5 & 5.34 & 5.11 & 5.05 & 5.02 & 4.99 & 5.13 & -0.85 & -1.97 & -2.18 & -2.50 & -4.36 & -4.65\\
9.5 & -1.0 & 5.41 & 5.12 & 5.03 & 4.96 & 4.89 & 5.01 & -0.64 & -1.84 & -2.11 & -2.49 & -4.45 & -4.81\\
9.5 & -0.5 & 5.57 & 5.20 & 5.06 & 4.95 & 4.82 & 4.89 & -0.25 & -1.74 & -2.18 & -2.66 & -4.71 & -5.14\\
9.5 & 0.0 & 5.89 & 5.41 & 5.20 & 5.02 & 4.86 & 4.89 & 0.55 & -1.39 & -2.22 & -2.86 & -4.93 & -5.39\\
10.0 & -2.5 & 5.99 & 5.77 & 5.72 & 5.72 & 5.73 & 5.89 & -0.61 & -1.46 & -1.52 & -1.66 & -3.31 & -3.47\\
10.0 & -2.0 & 6.25 & 5.96 & 5.89 & 5.85 & 5.82 & 5.96 & -0.56 & -1.57 & -1.74 & -2.01 & -3.88 & -4.16\\
10.0 & -1.5 & 6.21 & 5.88 & 5.79 & 5.72 & 5.66 & 5.78 & -0.54 & -1.62 & -1.84 & -2.15 & -4.08 & -4.39\\
10.0 & -1.0 & 6.28 & 5.90 & 5.78 & 5.68 & 5.58 & 5.68 & -0.43 & -1.61 & -1.87 & -2.24 & -4.18 & -4.52\\
10.0 & -0.5 & 6.48 & 6.02 & 5.86 & 5.71 & 5.55 & 5.61 & 0.09 & -1.47 & -1.90 & -2.37 & -4.42 & -4.86\\
10.0 & 0.0 & 6.84 & 6.26 & 6.00 & 5.80 & 5.61 & 5.63 & 1.01 & -0.99 & -1.94 & -2.58 & -4.65 & -5.11\\
10.1 & -2.5 & 6.06 & 5.81 & 5.76 & 5.74 & 5.74 & 5.90 & -0.68 & -1.54 & -1.60 & -1.74 & -3.39 & -3.55\\
10.1 & -2.0 & 6.34 & 6.04 & 5.96 & 5.92 & 5.89 & 6.03 & -0.57 & -1.57 & -1.72 & -1.96 & -3.77 & -4.02\\
10.1 & -1.5 & 6.41 & 6.06 & 5.97 & 5.91 & 5.84 & 5.97 & -0.54 & -1.58 & -1.77 & -2.05 & -3.90 & -4.17\\
10.1 & -1.0 & 6.47 & 6.08 & 5.96 & 5.86 & 5.75 & 5.85 & -0.39 & -1.57 & -1.83 & -2.20 & -4.14 & -4.49\\
10.1 & -0.5 & 6.64 & 6.16 & 5.99 & 5.83 & 5.67 & 5.73 & 0.13 & -1.43 & -1.87 & -2.33 & -4.38 & -4.81\\
10.1 & 0.0 & 7.02 & 6.42 & 6.15 & 5.94 & 5.74 & 5.77 & 1.09 & -0.91 & -1.89 & -2.54 & -4.60 & -5.07\\
10.15 & -2.5 & 6.11 & 5.84 & 5.79 & 5.76 & 5.76 & 5.91 & -0.71 & -1.57 & -1.63 & -1.77 & -3.42 & -3.59\\
10.15 & -2.0 & 6.39 & 6.08 & 6.01 & 5.96 & 5.93 & 6.07 & -0.59 & -1.58 & -1.72 & -1.96 & -3.75 & -3.99\\
10.15 & -1.5 & 6.45 & 6.11 & 6.01 & 5.95 & 5.88 & 6.01 & -0.53 & -1.58 & -1.76 & -2.05 & -3.89 & -4.17\\
10.15 & -1.0 & 6.52 & 6.12 & 6.00 & 5.89 & 5.78 & 5.88 & -0.39 & -1.58 & -1.85 & -2.22 & -4.16 & -4.50\\
10.15 & -0.5 & 6.80 & 6.31 & 6.14 & 5.99 & 5.83 & 5.89 & 0.20 & -1.37 & -1.81 & -2.28 & -4.33 & -4.77\\
10.15 & 0.0 & 7.10 & 6.48 & 6.20 & 5.98 & 5.78 & 5.80 & 1.10 & -0.90 & -1.89 & -2.54 & -4.61 & -5.07\\
\enddata
\tablecomments{MIST predictions for SSP absolute SBF magnitudes and mean magnitudes for calculating colors in the proposed RST filter system. Calculations assume the \citet{Kroupa:2001} IMF, and magnitudes are in the AB magnitude system.}
\end{deluxetable*}

\bibliography{ref}

\begin{thebibliography}{}
\expandafter\ifx\csname natexlab\endcsname\relax\def\natexlab#1{#1}\fi

\bibitem[{{Abbott} {et~al.}(2018){Abbott}, {Abdalla}, {Allam}, {Amara},
  {Annis}, {Asorey}, {Avila}, {Ballester}, {Banerji}, {Barkhouse}, {Baruah},
  {Baumer}, {Bechtol}, {Becker}, {Benoit-L{\'e}vy}, {Bernstein}, {Bertin},
  {Blazek}, {Bocquet}, {Brooks}, {Brout}, {Buckley-Geer}, {Burke}, {Busti},
  {Campisano}, {Cardiel-Sas}, {Carnero Rosell}, {Carrasco Kind}, {Carretero},
  {Castander}, {Cawthon}, {Chang}, {Chen}, {Conselice}, {Costa}, {Crocce},
  {Cunha}, {D'Andrea}, {da Costa}, {Das}, {Daues}, {Davis}, {Davis}, {De
  Vicente}, {DePoy}, {DeRose}, {Desai}, {Diehl}, {Dietrich}, {Dodelson},
  {Doel}, {Drlica-Wagner}, {Eifler}, {Elliott}, {Evrard}, {Farahi}, {Fausti
  Neto}, {Fernandez}, {Finley}, {Flaugher}, {Foley}, {Fosalba}, {Friedel},
  {Frieman}, {Garc{\'\i}a-Bellido}, {Gaztanaga}, {Gerdes}, {Giannantonio},
  {Gill}, {Glazebrook}, {Goldstein}, {Gower}, {Gruen}, {Gruendl}, {Gschwend},
  {Gupta}, {Gutierrez}, {Hamilton}, {Hartley}, {Hinton}, {Hislop}, {Hollowood},
  {Honscheid}, {Hoyle}, {Huterer}, {Jain}, {James}, {Jeltema}, {Johnson},
  {Johnson}, {Kacprzak}, {Kent}, {Khullar}, {Klein}, {Kovacs}, {Koziol},
  {Krause}, {Kremin}, {Kron}, {Kuehn}, {Kuhlmann}, {Kuropatkin}, {Lahav},
  {Lasker}, {Li}, {Li}, {Liddle}, {Lima}, {Lin}, {L{\'o}pez-Reyes}, {MacCrann},
  {Maia}, {Maloney}, {Manera}, {March}, {Marriner}, {Marshall}, {Martini},
  {McClintock}, {McKay}, {McMahon}, {Melchior}, {Menanteau}, {Miller},
  {Miquel}, {Mohr}, {Morganson}, {Mould}, {Neilsen}, {Nichol}, {Nogueira},
  {Nord}, {Nugent}, {Nunes}, {Ogand o}, {Old}, {Pace}, {Palmese},
  {Paz-Chinch{\'o}n}, {Peiris}, {Percival}, {Petravick}, {Plazas}, {Poh},
  {Pond}, {Porredon}, {Pujol}, {Refregier}, {Reil}, {Ricker}, {Rollins},
  {Romer}, {Roodman}, {Rooney}, {Ross}, {Rykoff}, {Sako}, {Sanchez}, {Sanchez},
  {Santiago}, {Saro}, {Scarpine}, {Scolnic}, {Serrano}, {Sevilla-Noarbe},
  {Sheldon}, {Shipp}, {Silveira}, {Smith}, {Smith}, {Smith}, {Soares-Santos},
  {Sobreira}, {Song}, {Stebbins}, {Suchyta}, {Sullivan}, {Swanson}, {Tarle},
  {Thaler}, {Thomas}, {Thomas}, {Troxel}, {Tucker}, {Vikram}, {Vivas},
  {Walker}, {Wechsler}, {Weller}, {Wester}, {Wolf}, {Wu}, {Yanny}, {Zenteno},
  {Zhang}, {Zuntz}, {DES Collaboration}, {Juneau}, {Fitzpatrick}, {Nikutta},
  {Nidever}, {Olsen}, {Scott}, \& {NOAO Data Lab}}]{DES-DR1-2018}
{Abbott}, T.~M.~C., {Abdalla}, F.~B., {Allam}, S., {et~al.} 2018, \apjs, 239,
  18

\bibitem[{Abell {et~al.}(2009)Abell, Allison, Anderson, Andrew, Angel, Armus,
  Arnett, Asztalos, Axelrod, Bailey, {et~al.}}]{abell2009lsst}
Abell, P.~A., Allison, J., Anderson, S.~F., {et~al.} 2009, arXiv:0912.0201

\bibitem[{{Aihara} {et~al.}(2018){Aihara}, {Arimoto}, {Armstrong}, {Arnouts},
  {Bahcall}, {Bickerton}, {Bosch}, {Bundy}, {Capak}, {Chan}, {Chiba}, {Coupon},
  {Egami}, {Enoki}, {Finet}, {Fujimori}, {Fujimoto}, {Furusawa}, {Furusawa},
  {Goto}, {Goulding}, {Greco}, {Greene}, {Gunn}, {Hamana}, {Harikane},
  {Hashimoto}, {Hattori}, {Hayashi}, {Hayashi}, {He{\l}miniak}, {Higuchi},
  {Hikage}, {Ho}, {Hsieh}, {Huang}, {Huang}, {Ikeda}, {Imanishi}, {Inoue},
  {Iwasawa}, {Iwata}, {Jaelani}, {Jian}, {Kamata}, {Karoji}, {Kashikawa},
  {Katayama}, {Kawanomoto}, {Kayo}, {Koda}, {Koike}, {Kojima}, {Komiyama},
  {Konno}, {Koshida}, {Koyama}, {Kusakabe}, {Leauthaud}, {Lee}, {Lin}, {Lin},
  {Lupton}, {Mandelbaum}, {Matsuoka}, {Medezinski}, {Mineo}, {Miyama},
  {Miyatake}, {Miyazaki}, {Momose}, {More}, {More}, {Moritani}, {Moriya},
  {Morokuma}, {Mukae}, {Murata}, {Murayama}, {Nagao}, {Nakata}, {Niida},
  {Niikura}, {Nishizawa}, {Obuchi}, {Oguri}, {Oishi}, {Okabe}, {Okamoto},
  {Okura}, {Ono}, {Onodera}, {Onoue}, {Osato}, {Ouchi}, {Price}, {Pyo}, {Sako},
  {Sawicki}, {Shibuya}, {Shimasaku}, {Shimono}, {Shirasaki}, {Silverman},
  {Simet}, {Speagle}, {Spergel}, {Strauss}, {Sugahara}, {Sugiyama}, {Suto},
  {Suyu}, {Suzuki}, {Tait}, {Takada}, {Takata}, {Tamura}, {Tanaka}, {Tanaka},
  {Tanaka}, {Tanaka}, {Terai}, {Terashima}, {Toba}, {Tominaga}, {Toshikawa},
  {Turner}, {Uchida}, {Uchiyama}, {Umetsu}, {Uraguchi}, {Urata}, {Usuda},
  {Utsumi}, {Wang}, {Wang}, {Wong}, {Yabe}, {Yamada}, {Yamanoi}, {Yasuda},
  {Yeh}, {Yonehara}, \& {Yuma}}]{Aihara:2018aa}
{Aihara}, H., {Arimoto}, N., {Armstrong}, R., {et~al.} 2018, \pasj, 70, S4

\bibitem[{{Aihara} {et~al.}(2019){Aihara}, {AlSayyad}, {Ando}, {Armstrong},
  {Bosch}, {Egami}, {Furusawa}, {Furusawa}, {Goulding}, {Harikane}, {Hikage},
  {Ho}, {Hsieh}, {Huang}, {Ikeda}, {Imanishi}, {Ito}, {Iwata}, {Jaelani},
  {Kakuma}, {Kawana}, {Kikuta}, {Kobayashi}, {Koike}, {Komiyama}, {Li},
  {Liang}, {Lin}, {Luo}, {Lupton}, {Lust}, {MacArthur}, {Matsuoka}, {Mineo},
  {Miyatake}, {Miyazaki}, {More}, {Murata}, {Namiki}, {Nishizawa}, {Oguri},
  {Okabe}, {Okamoto}, {Okura}, {Ono}, {Onodera}, {Onoue}, {Osato}, {Ouchi},
  {Shibuya}, {Strauss}, {Sugiyama}, {Suto}, {Takada}, {Takagi}, {Takata},
  {Takita}, {Tanaka}, {Terai}, {Toba}, {Uchiyama}, {Utsumi}, {Wang}, {Wang}, \&
  {Yamada}}]{HSC-DR2}
{Aihara}, H., {AlSayyad}, Y., {Ando}, M., {et~al.} 2019, \pasj, 71, 114

\bibitem[{{Astropy Collaboration} {et~al.}(2013){Astropy Collaboration},
  {Robitaille}, {Tollerud}, {Greenfield}, {Droettboom}, {Bray}, {Aldcroft},
  {Davis}, {Ginsburg}, {Price-Whelan}, {Kerzendorf}, {Conley}, {Crighton},
  {Barbary}, {Muna}, {Ferguson}, {Grollier}, {Parikh}, {Nair}, {Unther},
  {Deil}, {Woillez}, {Conseil}, {Kramer}, {Turner}, {Singer}, {Fox}, {Weaver},
  {Zabalza}, {Edwards}, {Azalee Bostroem}, {Burke}, {Casey}, {Crawford},
  {Dencheva}, {Ely}, {Jenness}, {Labrie}, {Lim}, {Pierfederici}, {Pontzen},
  {Ptak}, {Refsdal}, {Servillat}, \&
  {Streicher}}]{Astropy-Collaboration:2013aa}
{Astropy Collaboration}, {Robitaille}, T.~P., {Tollerud}, E.~J., {et~al.} 2013,
  \aap, 558, A33

\bibitem[{{Bahcall} \& {Schneider}(1988)}]{Bahcall:1988}
{Bahcall}, J.~N., \& {Schneider}, D.~P. 1988, in IAU Symposium, Vol. 126, The
  Harlow-Shapley Symposium on Globular Cluster Systems in Galaxies, ed. J.~E.
  {Grindlay} \& A.~G.~D. {Philip}, 455--463

\bibitem[{Barbary(2016)}]{Barbary:2016aa}
Barbary, K. 2016, The Journal of Open Source Software, 1,
  doi:10.21105/joss.00058

\bibitem[{{Beaton} {et~al.}(2018){Beaton}, {Bono}, {Braga}, {Dall'Ora},
  {Fiorentino}, {Jang}, {Mart{\'\i}nez-V{\'a}zquez}, {Matsunaga}, {Monelli},
  {Neeley}, \& {Salaris}}]{Beaton2018-old-pops}
{Beaton}, R.~L., {Bono}, G., {Braga}, V.~F., {et~al.} 2018, \ssr, 214, 113

\bibitem[{{Bennet} {et~al.}(2017){Bennet}, {Sand}, {Crnojevi{\'c}}, {Spekkens},
  {Zaritsky}, \& {Karunakaran}}]{Bennet-2017}
{Bennet}, P., {Sand}, D.~J., {Crnojevi{\'c}}, D., {et~al.} 2017, \apj, 850, 109

\bibitem[{{Bessell} \& {Murphy}(2012)}]{Bessell:2012}
{Bessell}, M., \& {Murphy}, S. 2012, \pasp, 124, 140

\bibitem[{{Blakeslee} \& {Cantiello}(2018)}]{Blakeslee-DF2:2018}
{Blakeslee}, J.~P., \& {Cantiello}, M. 2018, Research Notes of the American
  Astronomical Society, 2, 146

\bibitem[{{Blakeslee} \& {Tonry}(1995)}]{Blakeslee:1995}
{Blakeslee}, J.~P., \& {Tonry}, J.~L. 1995, \apj, 442, 579

\bibitem[{{Blakeslee} {et~al.}(2001){Blakeslee}, {Vazdekis}, \&
  {Ajhar}}]{Blakeslee:2001aa}
{Blakeslee}, J.~P., {Vazdekis}, A., \& {Ajhar}, E.~A. 2001, \mnras, 320, 193

\bibitem[{{Blakeslee} {et~al.}(2009){Blakeslee}, {Jord{\'a}n}, {Mei},
  {C{\^o}t{\'e}}, {Ferrarese}, {Infante}, {Peng}, {Tonry}, \&
  {West}}]{Blakeslee:2009}
{Blakeslee}, J.~P., {Jord{\'a}n}, A., {Mei}, S., {et~al.} 2009, \apj, 694, 556

\bibitem[{{Blakeslee} {et~al.}(2010){Blakeslee}, {Cantiello}, {Mei},
  {C{\^o}t{\'e}}, {Barber DeGraaff}, {Ferrarese}, {Jord{\'a}n}, {Peng},
  {Tonry}, \& {Worthey}}]{Blakeslee:2010aa}
{Blakeslee}, J.~P., {Cantiello}, M., {Mei}, S., {et~al.} 2010, \apj, 724, 657

\bibitem[{{Bloecker}(1995)}]{Bloecker:1995}
{Bloecker}, T. 1995, \aap, 297, 727

\bibitem[{{Bosch} {et~al.}(2018){Bosch}, {Armstrong}, {Bickerton}, {Furusawa},
  {Ikeda}, {Koike}, {Lupton}, {Mineo}, {Price}, {Takata}, {Tanaka}, {Yasuda},
  {AlSayyad}, {Becker}, {Coulton}, {Coupon}, {Garmilla}, {Huang}, {Krughoff},
  {Lang}, {Leauthaud}, {Lim}, {Lust}, {MacArthur}, {Mandelbaum}, {Miyatake},
  {Miyazaki}, {Murata}, {More}, {Okura}, {Owen}, {Swinbank}, {Strauss},
  {Yamada}, \& {Yamanoi}}]{Bosch:2018aa}
{Bosch}, J., {Armstrong}, R., {Bickerton}, S., {et~al.} 2018, \pasj, 70, S5

\bibitem[{Bradley {et~al.}(2017)Bradley, Sipocz, Robitaille, Vin{\'\i}cius,
  Tollerud, Deil, Barbary, G{\"u}nther, Cara, Busko, Droettboom, Bostroem,
  Bray, Bratholm, Pickering, Craig, Barentsen, Pascual, Conseil, adonath,
  Greco, Kerzendorf, de~Val-Borro, StuartLittlefair, Ogaz, Lim, Ferreira,
  D'Eugenio, \& Weaver}]{photutils}
Bradley, L., Sipocz, B., Robitaille, T., {et~al.} 2017, astropy/photutils:
  v0.4, , , doi:10.5281/zenodo.1039309

\bibitem[{{Bressan} {et~al.}(2012){Bressan}, {Marigo}, {Girardi}, {Salasnich},
  {Dal Cero}, {Rubele}, \& {Nanni}}]{PARSEC-2012}
{Bressan}, A., {Marigo}, P., {Girardi}, L., {et~al.} 2012, \mnras, 427, 127

\bibitem[{{Buzzoni}(1993)}]{Buzzoni:1993}
{Buzzoni}, A. 1993, \aap, 275, 433

\bibitem[{{Cantiello} {et~al.}(2007){Cantiello}, {Blakeslee}, {Raimondo},
  {Brocato}, \& {Capaccioli}}]{Cantiello:2007aa}
{Cantiello}, M., {Blakeslee}, J., {Raimondo}, G., {Brocato}, E., \&
  {Capaccioli}, M. 2007, \apj, 668, 130

\bibitem[{{Cantiello} {et~al.}(2005){Cantiello}, {Blakeslee}, {Raimondo},
  {Mei}, {Brocato}, \& {Capaccioli}}]{Cantiello:2005}
{Cantiello}, M., {Blakeslee}, J.~P., {Raimondo}, G., {et~al.} 2005, \apj, 634,
  239

\bibitem[{{Cantiello} {et~al.}(2003){Cantiello}, {Raimondo}, {Brocato}, \&
  {Capaccioli}}]{Cantiello:2003aa}
{Cantiello}, M., {Raimondo}, G., {Brocato}, E., \& {Capaccioli}, M. 2003, \aj,
  125, 2783

\bibitem[{{Cantiello} {et~al.}(2018{\natexlab{a}}){Cantiello}, {Jensen},
  {Blakeslee}, {Berger}, {Levan}, {Tanvir}, {Raimondo}, {Brocato}, {Alexander},
  {Blanchard}, {Branchesi}, {Cano}, {Chornock}, {Covino}, {Cowperthwaite},
  {D'Avanzo}, {Eftekhari}, {Fong}, {Fruchter}, {Grado}, {Hjorth}, {Holz},
  {Lyman}, {Mandel}, {Margutti}, {Nicholl}, {Villar}, \&
  {Williams}}]{Cantiello-GW-source:2018}
{Cantiello}, M., {Jensen}, J.~B., {Blakeslee}, J.~P., {et~al.}
  2018{\natexlab{a}}, \apjl, 854, L31

\bibitem[{{Cantiello} {et~al.}(2018{\natexlab{b}}){Cantiello}, {Blakeslee},
  {Ferrarese}, {C{\^o}t{\'e}}, {Roediger}, {Raimondo}, {Peng}, {Gwyn},
  {Durrell}, \& {Cuillandre}}]{Cantiello:2018aa}
{Cantiello}, M., {Blakeslee}, J.~P., {Ferrarese}, L., {et~al.}
  2018{\natexlab{b}}, \apj, 856, 126

\bibitem[{{Carlsten} {et~al.}(2019{\natexlab{a}}){Carlsten}, {Beaton}, {Greco},
  \& {Greene}}]{Carlsten:2019aa}
{Carlsten}, S.~G., {Beaton}, R.~L., {Greco}, J.~P., \& {Greene}, J.~E.
  2019{\natexlab{a}}, \apj, 879, 13

\bibitem[{{Carlsten} {et~al.}(2019{\natexlab{b}}){Carlsten}, {Beaton}, {Greco},
  \& {Greene}}]{Carlsten-2019-M101}
---. 2019{\natexlab{b}}, \apjl, 878, L16

\bibitem[{{Carlsten} {et~al.}(2019{\natexlab{c}}){Carlsten}, {Greco}, {Beaton},
  \& {Greene}}]{Carlsten2019-survey}
{Carlsten}, S.~G., {Greco}, J.~P., {Beaton}, R.~L., \& {Greene}, J.~E.
  2019{\natexlab{c}}, arXiv e-prints, arXiv:1909.07389

\bibitem[{{Choi} {et~al.}(2016){Choi}, {Dotter}, {Conroy}, {Cantiello},
  {Paxton}, \& {Johnson}}]{Choi:2016}
{Choi}, J., {Dotter}, A., {Conroy}, C., {et~al.} 2016, \apj, 823, 102

\bibitem[{{Cohen}(1982)}]{Cohen:1982}
{Cohen}, J.~G. 1982, \apj, 258, 143

\bibitem[{{Cohen} {et~al.}(2003){Cohen}, {Wheaton}, \&
  {Megeath}}]{Cohen-2MASS:2003}
{Cohen}, M., {Wheaton}, W.~A., \& {Megeath}, S.~T. 2003, \aj, 126, 1090

\bibitem[{{Cohen} {et~al.}(2018){Cohen}, {van Dokkum}, {Danieli}, {Romanowsky},
  {Abraham}, {Merritt}, {Zhang}, {Mowla}, {Kruijssen}, {Conroy}, \&
  {Wasserman}}]{Cohen:2018}
{Cohen}, Y., {van Dokkum}, P., {Danieli}, S., {et~al.} 2018, \apj, 868, 96

\bibitem[{{Conroy}(2013)}]{Conroy-SPS-review-2013}
{Conroy}, C. 2013, \araa, 51, 393

\bibitem[{{Conroy} \& {Gunn}(2010)}]{Conroy-Gunn-2010}
{Conroy}, C., \& {Gunn}, J.~E. 2010, \apj, 712, 833

\bibitem[{{Conroy} {et~al.}(2009){Conroy}, {Gunn}, \& {White}}]{Conroy:2009aa}
{Conroy}, C., {Gunn}, J.~E., \& {White}, M. 2009, \apj, 699, 486

\bibitem[{{Conroy} \& {van Dokkum}(2016)}]{Conroy-2016-pcmd}
{Conroy}, C., \& {van Dokkum}, P.~G. 2016, \apj, 827, 9

\bibitem[{{Cook} {et~al.}(2019){Cook}, {Conroy}, {van Dokkum}, \&
  {Speagle}}]{Cook:2019}
{Cook}, B.~A., {Conroy}, C., {van Dokkum}, P., \& {Speagle}, J.~S. 2019, \apj,
  876, 78

\bibitem[{{Cordier} {et~al.}(2007){Cordier}, {Pietrinferni}, {Cassisi}, \&
  {Salaris}}]{Cordier-2007-BaSTI}
{Cordier}, D., {Pietrinferni}, A., {Cassisi}, S., \& {Salaris}, M. 2007, \aj,
  133, 468

\bibitem[{{C{\^o}t{\'e}} {et~al.}(2004){C{\^o}t{\'e}}, {Blakeslee},
  {Ferrarese}, {Jord{\'a}n}, {Mei}, {Merritt}, {Milosavljevi{\'c}}, {Peng},
  {Tonry}, \& {West}}]{Cote:2004}
{C{\^o}t{\'e}}, P., {Blakeslee}, J.~P., {Ferrarese}, L., {et~al.} 2004, \apjs,
  153, 223

\bibitem[{{Danieli} {et~al.}(2019{\natexlab{a}}){Danieli}, {van Dokkum},
  {Abraham}, {Conroy}, {Dolphin}, \& {Romanowsky}}]{Danieli2019-DF4}
{Danieli}, S., {van Dokkum}, P., {Abraham}, R., {et~al.} 2019{\natexlab{a}},
  arXiv e-prints, arXiv:1910.07529

\bibitem[{{Danieli} {et~al.}(2018){Danieli}, {van Dokkum}, \&
  {Conroy}}]{Danieli:2018aa}
{Danieli}, S., {van Dokkum}, P., \& {Conroy}, C. 2018, \apj, 856, 69

\bibitem[{{Danieli} {et~al.}(2017){Danieli}, {van Dokkum}, {Merritt},
  {Abraham}, {Zhang}, {Karachentsev}, \& {Makarova}}]{Danieli:2017}
{Danieli}, S., {van Dokkum}, P., {Merritt}, A., {et~al.} 2017, \apj, 837, 136

\bibitem[{{Danieli} {et~al.}(2019{\natexlab{b}}){Danieli}, {Lokhorst}, {Zhang},
  {Merritt}, {van Dokkum}, {Abraham}, {Conroy}, {Gilhuly}, {Greco}, {Janssens},
  {Li}, {Liu}, {Miller}, \& {Mowla}}]{DF-Wide}
{Danieli}, S., {Lokhorst}, D., {Zhang}, J., {et~al.} 2019{\natexlab{b}}, arXiv
  e-prints, arXiv:1910.14045

\bibitem[{{Dark Energy Survey Collaboration} {et~al.}(2016){Dark Energy Survey
  Collaboration}, {Abbott}, {Abdalla}, {Aleksi{\'c}}, {Allam}, {Amara},
  {Bacon}, {Balbinot}, {Banerji}, {Bechtol}, {Benoit-L{\'e}vy}, {Bernstein},
  {Bertin}, {Blazek}, {Bonnett}, {Bridle}, {Brooks}, {Brunner}, {Buckley-Geer},
  {Burke}, {Caminha}, {Capozzi}, {Carlsen}, {Carnero-Rosell}, {Carollo},
  {Carrasco-Kind}, {Carretero}, {Castander}, {Clerkin}, {Collett}, {Conselice},
  {Crocce}, {Cunha}, {D'Andrea}, {da Costa}, {Davis}, {Desai}, {Diehl},
  {Dietrich}, {Dodelson}, {Doel}, {Drlica-Wagner}, {Estrada}, {Etherington},
  {Evrard}, {Fabbri}, {Finley}, {Flaugher}, {Foley}, {Fosalba}, {Frieman},
  {Garc{\'{\i}}a-Bellido}, {Gaztanaga}, {Gerdes}, {Giannantonio}, {Goldstein},
  {Gruen}, {Gruendl}, {Guarnieri}, {Gutierrez}, {Hartley}, {Honscheid}, {Jain},
  {James}, {Jeltema}, {Jouvel}, {Kessler}, {King}, {Kirk}, {Kron}, {Kuehn},
  {Kuropatkin}, {Lahav}, {Li}, {Lima}, {Lin}, {Maia}, {Makler}, {Manera},
  {Maraston}, {Marshall}, {Martini}, {McMahon}, {Melchior}, {Merson}, {Miller},
  {Miquel}, {Mohr}, {Morice-Atkinson}, {Naidoo}, {Neilsen}, {Nichol}, {Nord},
  {Ogando}, {Ostrovski}, {Palmese}, {Papadopoulos}, {Peiris}, {Peoples},
  {Percival}, {Plazas}, {Reed}, {Refregier}, {Romer}, {Roodman}, {Ross},
  {Rozo}, {Rykoff}, {Sadeh}, {Sako}, {S{\'a}nchez}, {Sanchez}, {Santiago},
  {Scarpine}, {Schubnell}, {Sevilla-Noarbe}, {Sheldon}, {Smith}, {Smith},
  {Soares-Santos}, {Sobreira}, {Soumagnac}, {Suchyta}, {Sullivan}, {Swanson},
  {Tarle}, {Thaler}, {Thomas}, {Thomas}, {Tucker}, {Vieira}, {Vikram},
  {Walker}, {Wechsler}, {Weller}, {Wester}, {Whiteway}, {Wilcox}, {Yanny},
  {Zhang}, \& {Zuntz}}]{DES:2016aa}
{Dark Energy Survey Collaboration}, {Abbott}, T., {Abdalla}, F.~B., {et~al.}
  2016, \mnras, 460, 1270

\bibitem[{{Dotter}(2016)}]{Dotter:2016}
{Dotter}, A. 2016, \apjs, 222, 8

\bibitem[{{Erwin}(2015)}]{Erwin:2015aa}
{Erwin}, P. 2015, \apj, 799, 226

\bibitem[{{Forbes} {et~al.}(2018){Forbes}, {Read}, {Gieles}, \&
  {Collins}}]{Forbes:2018}
{Forbes}, D.~A., {Read}, J.~I., {Gieles}, M., \& {Collins}, M. L.~M. 2018,
  \mnras, 481, 5592

\bibitem[{{Freedman} {et~al.}(2001){Freedman}, {Madore}, {Gibson}, {Ferrarese},
  {Kelson}, {Sakai}, {Mould}, {Kennicutt}, {Ford}, {Graham}, {Huchra},
  {Hughes}, {Illingworth}, {Macri}, \& {Stetson}}]{Freedman2001}
{Freedman}, W.~L., {Madore}, B.~F., {Gibson}, B.~K., {et~al.} 2001, \apj, 553,
  47

\bibitem[{{Gonz{\'a}lez} {et~al.}(2004){Gonz{\'a}lez}, {Liu}, \& {Bruzual
  A.}}]{Gonzalez:2004}
{Gonz{\'a}lez}, R.~A., {Liu}, M.~C., \& {Bruzual A.}, G. 2004, \apj, 611, 270

\bibitem[{{Gonz{\'a}lez-L{\'o}pezlira}
  {et~al.}(2005){Gonz{\'a}lez-L{\'o}pezlira}, {Albarr{\'a}n}, {Mouhcine},
  {Liu}, {Bruzual-A.}, \& {de Batz}}]{Gonzalez:2005}
{Gonz{\'a}lez-L{\'o}pezlira}, R.~A., {Albarr{\'a}n}, M.~Y., {Mouhcine}, M.,
  {et~al.} 2005, \mnras, 363, 1279

\bibitem[{{Gonz{\'a}lez-L{\'o}pezlira}
  {et~al.}(2010){Gonz{\'a}lez-L{\'o}pezlira}, {Bruzual-A.}, {Charlot},
  {Ballesteros-Paredes}, \& {Loinard}}]{2010MNRAS.403.1213G}
{Gonz{\'a}lez-L{\'o}pezlira}, R.~A., {Bruzual-A.}, G., {Charlot}, S.,
  {Ballesteros-Paredes}, J., \& {Loinard}, L. 2010, \mnras, 403, 1213

\bibitem[{{Greco} {et~al.}(2018){Greco}, {Greene}, {Strauss}, {Macarthur},
  {Flowers}, {Goulding}, {Huang}, {Kim}, {Komiyama}, {Leauthaud}, {Leisman},
  {Lupton}, {Sif{\'o}n}, \& {Wang}}]{Greco2018cat}
{Greco}, J.~P., {Greene}, J.~E., {Strauss}, M.~A., {et~al.} 2018, \apj, 857,
  104

\bibitem[{{Gu} {et~al.}(2018){Gu}, {Conroy}, {Law}, {van Dokkum}, {Yan},
  {Wake}, {Bundy}, {Merritt}, {Abraham}, {Zhang}, {Bershady}, {Bizyaev},
  {Brinkmann}, {Drory}, {Grabowski}, {Masters}, {Pan}, {Parejko}, {Weijmans},
  \& {Zhang}}]{Gu-2018}
{Gu}, M., {Conroy}, C., {Law}, D., {et~al.} 2018, \apj, 859, 37

\bibitem[{Hunter(2007)}]{Hunter:2007aa}
Hunter, J.~D. 2007, Computing In Science \& Engineering, 9, 90

\bibitem[{Ivezi\'c {et~al.}(2008)Ivezi\'c, Tyson, Acosta, Allsman, Anderson,
  Andrew, Angel, Axelrod, Barr, Becker, {et~al.}}]{ivezic2008lsst}
Ivezi\'c, v., Tyson, J.~A., Acosta, E., {et~al.} 2008, arXiv:0805.2366v4

\bibitem[{{Ivezi{\'c}} {et~al.}(2019){Ivezi{\'c}}, {Kahn}, {Tyson}, {Abel},
  {Acosta}, {Allsman}, {Alonso}, {AlSayyad}, {Anderson}, {Andrew}, {Angel},
  {Angeli}, {Ansari}, {Antilogus}, {Araujo}, {Armstrong}, {Arndt}, {Astier},
  {Aubourg}, {Auza}, {Axelrod}, {Bard}, {Barr}, {Barrau}, {Bartlett}, {Bauer},
  {Bauman}, {Baumont}, {Bechtol}, {Bechtol}, {Becker}, {Becla}, {Beldica},
  {Bellavia}, {Bianco}, {Biswas}, {Blanc}, {Blazek}, {Bland ford}, {Bloom},
  {Bogart}, {Bond}, {Booth}, {Borgland}, {Borne}, {Bosch}, {Boutigny},
  {Brackett}, {Bradshaw}, {Brand t}, {Brown}, {Bullock}, {Burchat}, {Burke},
  {Cagnoli}, {Calabrese}, {Callahan}, {Callen}, {Carlin}, {Carlson}, {Chand
  rasekharan}, {Charles-Emerson}, {Chesley}, {Cheu}, {Chiang}, {Chiang},
  {Chirino}, {Chow}, {Ciardi}, {Claver}, {Cohen-Tanugi}, {Cockrum}, {Coles},
  {Connolly}, {Cook}, {Cooray}, {Covey}, {Cribbs}, {Cui}, {Cutri}, {Daly},
  {Daniel}, {Daruich}, {Daubard}, {Daues}, {Dawson}, {Delgado}, {Dellapenna},
  {de Peyster}, {de Val-Borro}, {Digel}, {Doherty}, {Dubois},
  {Dubois-Felsmann}, {Durech}, {Economou}, {Eifler}, {Eracleous}, {Emmons},
  {Fausti Neto}, {Ferguson}, {Figueroa}, {Fisher-Levine}, {Focke}, {Foss},
  {Frank}, {Freemon}, {Gangler}, {Gawiser}, {Geary}, {Gee}, {Geha}, {Gessner},
  {Gibson}, {Gilmore}, {Glanzman}, {Glick}, {Goldina}, {Goldstein}, {Goodenow},
  {Graham}, {Gressler}, {Gris}, {Guy}, {Guyonnet}, {Haller}, {Harris},
  {Hascall}, {Haupt}, {Hernand ez}, {Herrmann}, {Hileman}, {Hoblitt},
  {Hodgson}, {Hogan}, {Howard}, {Huang}, {Huffer}, {Ingraham}, {Innes},
  {Jacoby}, {Jain}, {Jammes}, {Jee}, {Jenness}, {Jernigan}, {Jevremovi{\'c}},
  {Johns}, {Johnson}, {Johnson}, {Jones}, {Juramy-Gilles}, {Juri{\'c}},
  {Kalirai}, {Kallivayalil}, {Kalmbach}, {Kantor}, {Karst}, {Kasliwal},
  {Kelly}, {Kessler}, {Kinnison}, {Kirkby}, {Knox}, {Kotov}, {Krabbendam},
  {Krughoff}, {Kub{\'a}nek}, {Kuczewski}, {Kulkarni}, {Ku}, {Kurita}, {Lage},
  {Lambert}, {Lange}, {Langton}, {Le Guillou}, {Levine}, {Liang}, {Lim},
  {Lintott}, {Long}, {Lopez}, {Lotz}, {Lupton}, {Lust}, {MacArthur}, {Mahabal},
  {Mand elbaum}, {Markiewicz}, {Marsh}, {Marshall}, {Marshall}, {May},
  {McKercher}, {McQueen}, {Meyers}, {Migliore}, {Miller}, {Mills}, {Miraval},
  {Moeyens}, {Moolekamp}, {Monet}, {Moniez}, {Monkewitz}, {Montgomery},
  {Morrison}, {Mueller}, {Muller}, {Mu{\~n}oz Arancibia}, {Neill}, {Newbry},
  {Nief}, {Nomerotski}, {Nordby}, {O'Connor}, {Oliver}, {Olivier}, {Olsen},
  {O'Mullane}, {Ortiz}, {Osier}, {Owen}, {Pain}, {Palecek}, {Parejko},
  {Parsons}, {Pease}, {Peterson}, {Peterson}, {Petravick}, {Libby Petrick},
  {Petry}, {Pierfederici}, {Pietrowicz}, {Pike}, {Pinto}, {Plante}, {Plate},
  {Plutchak}, {Price}, {Prouza}, {Radeka}, {Rajagopal}, {Rasmussen},
  {Regnault}, {Reil}, {Reiss}, {Reuter}, {Ridgway}, {Riot}, {Ritz}, {Robinson},
  {Roby}, {Roodman}, {Rosing}, {Roucelle}, {Rumore}, {Russo}, {Saha},
  {Sassolas}, {Schalk}, {Schellart}, {Schindler}, {Schmidt}, {Schneider},
  {Schneider}, {Schoening}, {Schumacher}, {Schwamb}, {Sebag}, {Selvy},
  {Sembroski}, {Seppala}, {Serio}, {Serrano}, {Shaw}, {Shipsey}, {Sick},
  {Silvestri}, {Slater}, {Smith}, {Smith}, {Sobhani}, {Soldahl},
  {Storrie-Lombardi}, {Stover}, {Strauss}, {Street}, {Stubbs}, {Sullivan},
  {Sweeney}, {Swinbank}, {Szalay}, {Takacs}, {Tether}, {Thaler}, {Thayer},
  {Thomas}, {Thornton}, {Thukral}, {Tice}, {Trilling}, {Turri}, {Van Berg},
  {Vanden Berk}, {Vetter}, {Virieux}, {Vucina}, {Wahl}, {Walkowicz}, {Walsh},
  {Walter}, {Wang}, {Wang}, {Warner}, {Wiecha}, {Willman}, {Winters},
  {Wittman}, {Wolff}, {Wood-Vasey}, {Wu}, {Xin}, {Yoachim}, \&
  {Zhan}}]{LSST-ref-design-2019}
{Ivezi{\'c}}, {\v{Z}}., {Kahn}, S.~M., {Tyson}, J.~A., {et~al.} 2019, \apj,
  873, 111

\bibitem[{{Jacoby} {et~al.}(1992){Jacoby}, {Branch}, {Ciardullo}, {Davies},
  {Harris}, {Pierce}, {Pritchet}, {Tonry}, \& {Welch}}]{Jacoby-1992-distances}
{Jacoby}, G.~H., {Branch}, D., {Ciardullo}, R., {et~al.} 1992, \pasp, 104, 599

\bibitem[{{Jang} {et~al.}(2018){Jang}, {Hatt}, {Beaton}, {Lee}, {Freedman},
  {Madore}, {Hoyt}, {Monson}, {Rich}, {Scowcroft}, \& {Seibert}}]{Jang:2018}
{Jang}, I.~S., {Hatt}, D., {Beaton}, R.~L., {et~al.} 2018, \apj, 852, 60

\bibitem[{{Jensen} {et~al.}(2015){Jensen}, {Blakeslee}, {Gibson}, {Lee},
  {Cantiello}, {Raimondo}, {Boyer}, \& {Cho}}]{Jensen:2015}
{Jensen}, J.~B., {Blakeslee}, J.~P., {Gibson}, Z., {et~al.} 2015, \apj, 808, 91

\bibitem[{{Jensen} {et~al.}(2003){Jensen}, {Tonry}, {Barris}, {Thompson},
  {Liu}, {Rieke}, {Ajhar}, \& {Blakeslee}}]{Jensen:2003}
{Jensen}, J.~B., {Tonry}, J.~L., {Barris}, B.~J., {et~al.} 2003, \apj, 583, 712

\bibitem[{{Jensen} {et~al.}(1998){Jensen}, {Tonry}, \& {Luppino}}]{Jensen:1998}
{Jensen}, J.~B., {Tonry}, J.~L., \& {Luppino}, G.~A. 1998, \apj, 505, 111

\bibitem[{{Jerjen} {et~al.}(1998){Jerjen}, {Freeman}, \&
  {Binggeli}}]{Jerjen:1998}
{Jerjen}, H., {Freeman}, K.~C., \& {Binggeli}, B. 1998, \aj, 116, 2873

\bibitem[{{Jerjen} {et~al.}(2000){Jerjen}, {Freeman}, \&
  {Binggeli}}]{Jerjen:2000a}
---. 2000, \aj, 119, 166

\bibitem[{{Jerjen} {et~al.}(2001){Jerjen}, {Rekola}, {Takalo}, {Coleman}, \&
  {Valtonen}}]{Jerjen:2001}
{Jerjen}, H., {Rekola}, R., {Takalo}, L., {Coleman}, M., \& {Valtonen}, M.
  2001, \aap, 380, 90

\bibitem[{{Kadowaki} {et~al.}(2017){Kadowaki}, {Zaritsky}, \&
  {Donnerstein}}]{Kadowaki:2017aa}
{Kadowaki}, J., {Zaritsky}, D., \& {Donnerstein}, R.~L. 2017, \apjl, 838, L21

\bibitem[{{Karachentsev} {et~al.}(2013){Karachentsev}, {Makarov}, \&
  {Kaisina}}]{Karachentsev:2013}
{Karachentsev}, I.~D., {Makarov}, D.~I., \& {Kaisina}, E.~I. 2013, \aj, 145,
  101

\bibitem[{{Kirby} {et~al.}(2013){Kirby}, {Cohen}, {Guhathakurta}, {Cheng},
  {Bullock}, \& {Gallazzi}}]{Kirby:2013aa}
{Kirby}, E.~N., {Cohen}, J.~G., {Guhathakurta}, P., {et~al.} 2013, \apj, 779,
  102

\bibitem[{{Koposov} {et~al.}(2008){Koposov}, {Belokurov}, {Evans}, {Hewett},
  {Irwin}, {Gilmore}, {Zucker}, {Rix}, {Fellhauer}, {Bell}, \&
  {Glushkova}}]{Koposov:2008}
{Koposov}, S., {Belokurov}, V., {Evans}, N.~W., {et~al.} 2008, \apj, 686, 279

\bibitem[{{Kroupa}(2001)}]{Kroupa:2001}
{Kroupa}, P. 2001, \mnras, 322, 231

\bibitem[{{Leavitt} \& {Pickering}(1912)}]{Leavitt-Law-1912}
{Leavitt}, H.~S., \& {Pickering}, E.~C. 1912, Harvard College Observatory
  Circular, 173, 1

\bibitem[{{Lee} {et~al.}(2010){Lee}, {Worthey}, \&
  {Blakeslee}}]{Lee-Worthey-Blakeslee-2010}
{Lee}, H.-c., {Worthey}, G., \& {Blakeslee}, J.~P. 2010, \apj, 710, 421

\bibitem[{{Lee} {et~al.}(1993){Lee}, {Freedman}, \& {Madore}}]{Lee-1993-TRGB}
{Lee}, M.~G., {Freedman}, W.~L., \& {Madore}, B.~F. 1993, \apj, 417, 553

\bibitem[{{Liu} {et~al.}(2000){Liu}, {Charlot}, \& {Graham}}]{Liu:2000}
{Liu}, M.~C., {Charlot}, S., \& {Graham}, J.~R. 2000, \apj, 543, 644

\bibitem[{{Liu} {et~al.}(2002){Liu}, {Graham}, \& {Charlot}}]{Liu:2002}
{Liu}, M.~C., {Graham}, J.~R., \& {Charlot}, S. 2002, \apj, 564, 216

\bibitem[{{Lupton} {et~al.}(2004){Lupton}, {Blanton}, {Fekete}, {Hogg},
  {O'Mullane}, {Szalay}, \& {Wherry}}]{Lupton:2004aa}
{Lupton}, R., {Blanton}, M.~R., {Fekete}, G., {et~al.} 2004, \pasp, 116, 133

\bibitem[{{Madore} \& {Freedman}(1995)}]{Madore-Freedman-1995}
{Madore}, B.~F., \& {Freedman}, W.~L. 1995, \aj, 109, 1645

\bibitem[{{Mart{\'\i}nez-Delgado} {et~al.}(2018){Mart{\'\i}nez-Delgado},
  {Grebel}, {Javanmardi}, {Boschin}, {Longeard}, {Carballo-Bello}, {Makarov},
  {Beasley}, {Donatiello}, {Haynes}, {Forbes}, \&
  {Romanowsky}}]{Martinez-Delgado:2018}
{Mart{\'\i}nez-Delgado}, D., {Grebel}, E.~K., {Javanmardi}, B., {et~al.} 2018,
  \aap, 620, A126

\bibitem[{{McQuinn} {et~al.}(2010{\natexlab{a}}){McQuinn}, {Skillman},
  {Cannon}, {Dalcanton}, {Dolphin}, {Hidalgo-Rodr{\'\i}guez}, {Holtzman},
  {Stark}, {Weisz}, \& {Williams}}]{McQuinn:2010-I}
{McQuinn}, K. B.~W., {Skillman}, E.~D., {Cannon}, J.~M., {et~al.}
  2010{\natexlab{a}}, \apj, 721, 297

\bibitem[{{McQuinn} {et~al.}(2010{\natexlab{b}}){McQuinn}, {Skillman},
  {Cannon}, {Dalcanton}, {Dolphin}, {Hidalgo-Rodr{\'\i}guez}, {Holtzman},
  {Stark}, {Weisz}, \& {Williams}}]{McQuinn:2010}
---. 2010{\natexlab{b}}, \apj, 724, 49

\bibitem[{{Mei} {et~al.}(2001){Mei}, {Quinn}, \& {Silva}}]{Mei:2001sims}
{Mei}, S., {Quinn}, P.~J., \& {Silva}, D.~R. 2001, \aap, 371, 779

\bibitem[{{Mei} {et~al.}(2005){Mei}, {Blakeslee}, {Tonry}, {Jord{\'a}n},
  {Peng}, {C{\^o}t{\'e}}, {Ferrarese}, {Merritt}, {Milosavljevi{\'c}}, \&
  {West}}]{Mei2005sbfsims}
{Mei}, S., {Blakeslee}, J.~P., {Tonry}, J.~L., {et~al.} 2005, \apjs, 156, 113

\bibitem[{{Mei} {et~al.}(2007){Mei}, {Blakeslee}, {C{\^o}t{\'e}}, {Tonry},
  {West}, {Ferrarese}, {Jord{\'a}n}, {Peng}, {Anthony}, \&
  {Merritt}}]{Mei:2007}
{Mei}, S., {Blakeslee}, J.~P., {C{\^o}t{\'e}}, P., {et~al.} 2007, \apj, 655,
  144

\bibitem[{{Mieske} {et~al.}(2003){Mieske}, {Hilker}, \&
  {Infante}}]{Mieske:2003}
{Mieske}, S., {Hilker}, M., \& {Infante}, L. 2003, \aap, 403, 43

\bibitem[{{Mitzkus} {et~al.}(2018){Mitzkus}, {Walcher}, {Roth}, {Coelho},
  {Cioni}, {Raimondo}, \& {Rejkuba}}]{Mitzkus:2018}
{Mitzkus}, M., {Walcher}, C.~J., {Roth}, M.~M., {et~al.} 2018, \mnras, 480, 629

\bibitem[{{Moffat}(1969)}]{Moffat1969}
{Moffat}, A.~F.~J. 1969, \aap, 3, 455

\bibitem[{{M{\"u}ller} {et~al.}(2017){M{\"u}ller}, {Jerjen}, \&
  {Binggeli}}]{Muller-2017-lsb-Centaurus}
{M{\"u}ller}, O., {Jerjen}, H., \& {Binggeli}, B. 2017, \aap, 597, A7

\bibitem[{{Ohio Supercomputer Center}(1987)}]{OhioSupercomputerCenter1987}
{Ohio Supercomputer Center}. 1987, Ohio Supercomputer Center, ,

\bibitem[{{Oke} \& {Gunn}(1983)}]{Oke:1983aa}
{Oke}, J.~B., \& {Gunn}, J.~E. 1983, \apj, 266, 713

\bibitem[{{Olsen} {et~al.}(2003){Olsen}, {Blum}, \& {Rigaut}}]{Olsen:2003}
{Olsen}, K. A.~G., {Blum}, R.~D., \& {Rigaut}, F. 2003, \aj, 126, 452

\bibitem[{{Paxton} {et~al.}(2011){Paxton}, {Bildsten}, {Dotter}, {Herwig},
  {Lesaffre}, \& {Timmes}}]{Paxton:2011}
{Paxton}, B., {Bildsten}, L., {Dotter}, A., {et~al.} 2011, \apjs, 192, 3

\bibitem[{{Paxton} {et~al.}(2013){Paxton}, {Cantiello}, {Arras}, {Bildsten},
  {Brown}, {Dotter}, {Mankovich}, {Montgomery}, {Stello}, {Timmes}, \&
  {Townsend}}]{Paxton:2013}
{Paxton}, B., {Cantiello}, M., {Arras}, P., {et~al.} 2013, \apjs, 208, 4

\bibitem[{{Paxton} {et~al.}(2015){Paxton}, {Marchant}, {Schwab}, {Bauer},
  {Bildsten}, {Cantiello}, {Dessart}, {Farmer}, {Hu}, {Langer}, {Townsend},
  {Townsley}, \& {Timmes}}]{Paxton:2015}
{Paxton}, B., {Marchant}, P., {Schwab}, J., {et~al.} 2015, \apjs, 220, 15

\bibitem[{{Pietrinferni} {et~al.}(2004){Pietrinferni}, {Cassisi}, {Salaris}, \&
  {Castelli}}]{BaSTI-2004}
{Pietrinferni}, A., {Cassisi}, S., {Salaris}, M., \& {Castelli}, F. 2004, \apj,
  612, 168

\bibitem[{Price-Whelan \& Foreman-Mackey(2017)}]{schwimmbad}
Price-Whelan, A.~M., \& Foreman-Mackey, D. 2017, The Journal of Open Source
  Software, 2, doi:10.21105/joss.00357

\bibitem[{{Prole} {et~al.}(2019){Prole}, {van der Burg}, {Hilker}, \&
  {Davies}}]{Prole2019}
{Prole}, D.~J., {van der Burg}, R. F.~J., {Hilker}, M., \& {Davies}, J.~I.
  2019, arXiv e-prints, arXiv:1910.14057

\bibitem[{{Raimondo}(2009)}]{Raimondo:2009}
{Raimondo}, G. 2009, \apj, 700, 1247

\bibitem[{{Raimondo} {et~al.}(2005){Raimondo}, {Brocato}, {Cantiello}, \&
  {Capaccioli}}]{Raimondo:2005}
{Raimondo}, G., {Brocato}, E., {Cantiello}, M., \& {Capaccioli}, M. 2005, \aj,
  130, 2625

\bibitem[{{Reimers}(1975)}]{Reimers:1975}
{Reimers}, D. 1975, Memoires of the Societe Royale des Sciences de Liege, 8,
  369

\bibitem[{{Rekola} {et~al.}(2005){Rekola}, {Jerjen}, \& {Flynn}}]{Rekola:2005}
{Rekola}, R., {Jerjen}, H., \& {Flynn}, C. 2005, \aap, 437, 823

\bibitem[{{Santos} \& {Frogel}(1997)}]{Santos:1997}
{Santos}, Jo{\~a}o F.~C., J., \& {Frogel}, J.~A. 1997, \apj, 479, 764

\bibitem[{{Schlafly} \& {Finkbeiner}(2011)}]{Schlafly:2011aa}
{Schlafly}, E.~F., \& {Finkbeiner}, D.~P. 2011, \apj, 737, 103

\bibitem[{{Schlegel} {et~al.}(1998){Schlegel}, {Finkbeiner}, \&
  {Davis}}]{Schlegel:1998aa}
{Schlegel}, D.~J., {Finkbeiner}, D.~P., \& {Davis}, M. 1998, \apj, 500, 525

\bibitem[{{S\'{e}rsic}(1968)}]{Sersic:1968aa}
{S\'{e}rsic}, J.~L. 1968, {Atlas de galaxias australes}

\bibitem[{{Spergel} {et~al.}(2015){Spergel}, {Gehrels}, {Baltay}, {Bennett},
  {Breckinridge}, {Donahue}, {Dressler}, {Gaudi}, {Greene}, {Guyon}, {Hirata},
  {Kalirai}, {Kasdin}, {Macintosh}, {Moos}, {Perlmutter}, {Postman},
  {Rauscher}, {Rhodes}, {Wang}, {Weinberg}, {Benford}, {Hudson}, {Jeong},
  {Mellier}, {Traub}, {Yamada}, {Capak}, {Colbert}, {Masters}, {Penny},
  {Savransky}, {Stern}, {Zimmerman}, {Barry}, {Bartusek}, {Carpenter}, {Cheng},
  {Content}, {Dekens}, {Demers}, {Grady}, {Jackson}, {Kuan}, {Kruk}, {Melton},
  {Nemati}, {Parvin}, {Poberezhskiy}, {Peddie}, {Ruffa}, {Wallace}, {Whipple},
  {Wollack}, \& {Zhao}}]{Spergel:2015}
{Spergel}, D., {Gehrels}, N., {Baltay}, C., {et~al.} 2015, arXiv e-prints,
  arXiv:1503.03757

\bibitem[{{Tonry} \& {Schneider}(1988)}]{Tonry:1988}
{Tonry}, J., \& {Schneider}, D.~P. 1988, \aj, 96, 807

\bibitem[{{Tonry}(1991)}]{Tonry:1991a}
{Tonry}, J.~L. 1991, \apjl, 373, L1

\bibitem[{{Tonry} {et~al.}(1989){Tonry}, {Ajhar}, \& {Luppino}}]{Tonry-1989}
{Tonry}, J.~L., {Ajhar}, E.~A., \& {Luppino}, G.~A. 1989, \apjl, 346, L57

\bibitem[{{Tonry} {et~al.}(1990{\natexlab{a}}){Tonry}, {Ajhar}, \&
  {Luppino}}]{Tonry:1990virgo}
---. 1990{\natexlab{a}}, \aj, 100, 1416

\bibitem[{{Tonry} {et~al.}(1990{\natexlab{b}}){Tonry}, {Ajhar}, \&
  {Luppino}}]{Tonry:1990}
---. 1990{\natexlab{b}}, \aj, 100, 1416

\bibitem[{{Tonry} {et~al.}(1997){Tonry}, {Blakeslee}, {Ajhar}, \&
  {Dressler}}]{Tonry:1997}
{Tonry}, J.~L., {Blakeslee}, J.~P., {Ajhar}, E.~A., \& {Dressler}, A. 1997,
  \apj, 475, 399

\bibitem[{{Tonry} {et~al.}(2001){Tonry}, {Dressler}, {Blakeslee}, {Ajhar},
  {Fletcher}, {Luppino}, {Metzger}, \& {Moore}}]{Tonry:2001aa}
{Tonry}, J.~L., {Dressler}, A., {Blakeslee}, J.~P., {et~al.} 2001, \apj, 546,
  681

\bibitem[{{Trujillo} {et~al.}(2001){Trujillo}, {Aguerri}, {Cepa}, \&
  {Guti{\'e}rrez}}]{Trujillo2001moffat}
{Trujillo}, I., {Aguerri}, J.~A.~L., {Cepa}, J., \& {Guti{\'e}rrez}, C.~M.
  2001, \mnras, 328, 977

\bibitem[{{Trujillo} {et~al.}(2019){Trujillo}, {Beasley}, {Borlaff},
  {Carrasco}, {Di Cintio}, {Filho}, {Monelli}, {Montes}, {Rom{\'a}n},
  {Ruiz-Lara}, {S{\'a}nchez Almeida}, {Valls-Gabaud}, \&
  {Vazdekis}}]{Trujillo:2019}
{Trujillo}, I., {Beasley}, M.~A., {Borlaff}, A., {et~al.} 2019, \mnras, 486,
  1192

\bibitem[{{Van der Walt} {et~al.}(2011){Van der Walt}, Colbert, \&
  Varoquaux}]{Van-der-Walt:2011aa}
{Van der Walt}, S., Colbert, S.~C., \& Varoquaux, G. 2011, {Computing in
  Science \& Engineering}, 13, 22

\bibitem[{{van Dokkum} {et~al.}(2018){van Dokkum}, {Danieli}, {Cohen},
  {Romanowsky}, \& {Conroy}}]{vD2018-DF2-distance}
{van Dokkum}, P., {Danieli}, S., {Cohen}, Y., {Romanowsky}, A.~J., \& {Conroy},
  C. 2018, \apjl, 864, L18

\bibitem[{{van Dokkum} {et~al.}(2019){van Dokkum}, {Wasserman}, {Danieli},
  {Abraham}, {Brodie}, {Conroy}, {Forbes}, {Martin}, {Matuszewski},
  {Romanowsky}, \& {Villaume}}]{vD-2019-DF44-KCWI}
{van Dokkum}, P., {Wasserman}, A., {Danieli}, S., {et~al.} 2019, \apj, 880, 91

\bibitem[{{van Dokkum} \& {Conroy}(2014)}]{vD:2014}
{van Dokkum}, P.~G., \& {Conroy}, C. 2014, \apj, 797, 56

\bibitem[{{van Dokkum} {et~al.}(2015){van Dokkum}, {Romanowsky}, {Abraham},
  {Brodie}, {Conroy}, {Geha}, {Merritt}, {Villaume}, \&
  {Zhang}}]{van-Dokkum:2015ab}
{van Dokkum}, P.~G., {Romanowsky}, A.~J., {Abraham}, R., {et~al.} 2015, \apjl,
  804, L26

\bibitem[{{Weisz} {et~al.}(2014){Weisz}, {Dolphin}, {Skillman}, {Holtzman},
  {Gilbert}, {Dalcanton}, \& {Williams}}]{Weisz:2014}
{Weisz}, D.~R., {Dolphin}, A.~E., {Skillman}, E.~D., {et~al.} 2014, \apj, 789,
  147

\bibitem[{{Weisz} {et~al.}(2011){Weisz}, {Dalcanton}, {Williams}, {Gilbert},
  {Skillman}, {Seth}, {Dolphin}, {McQuinn}, {Gogarten}, {Holtzman}, {Rosema},
  {Cole}, {Karachentsev}, \& {Zaritsky}}]{Weisz:2011}
{Weisz}, D.~R., {Dalcanton}, J.~J., {Williams}, B.~F., {et~al.} 2011, \apj,
  739, 5

\bibitem[{{Worthey}(1993)}]{Worthey:1993}
{Worthey}, G. 1993, \apj, 409, 530

\bibitem[{{Zaritsky} {et~al.}(2019){Zaritsky}, {Donnerstein}, {Dey},
  {Kadowaki}, {Zhang}, {Karunakaran}, {Mart{\'\i}nez-Delgado}, {Rahman}, \&
  {Spekkens}}]{Zaritsky2019}
{Zaritsky}, D., {Donnerstein}, R., {Dey}, A., {et~al.} 2019, \apjs, 240, 1

\end{thebibliography}

\end{document}